\documentclass[a4paper,fleqn,usenatbib]{mnras}

\usepackage[T1]{fontenc}
\usepackage{ae,aecompl}

\usepackage{graphicx}	
\usepackage{amsmath}	
\usepackage{amssymb}	
\usepackage{xspace}

\usepackage{natbib}
\usepackage[usenames]{color}

\newcommand{\bd}{\begin{displaymath}}
\newcommand{\ed}{\end{displaymath}}
\newcommand{\be}{\begin{equation}}
\newcommand{\ee}{\end{equation}}
\newcommand{\beaa}{\begin{eqnarray*}}
\newcommand{\eeaa}{\end{eqnarray*}}
\newcommand{\bea}{\begin{eqnarray}}
\newcommand{\eea}{\end{eqnarray}}

\def\hequad{HE\,0435$-$1223}
\def\hedouble{HE\,1104$-$1805}
\def\wfilens{WFI2033$-$4723}
\def\blens{B1608$+$656}
\def\rxjlens{RXJ1131$-$1231}

\def\Ok{\Omega_{\rm k}}
\def\Ode{\Omega_{\rm DE}}
\def\Om{\Omega_{\rm m}}
\def\OL{\Omega_{\Lambda}}
\def\Neff{N_{\rm eff}}

\def\tdist{D_{\Delta t}}

\def\Dd{D_{\rm d}}
\def\Dds{D_{\rm ds}}
\def\Ds{D_{\rm s}}

\def\kext{\kappa_{\rm ext}}

\def\zd{z_{\rm d}}
\def\zs{z_{\rm s}}

\def\hst{\textit{HST}}

\def\GLEE{{\sc Glee}\xspace}

\def\kmsMpc {\rm km\,s^{-1}\,Mpc^{-1}}

\newcommand{\sref}[1]{Section~\ref{#1}}
\newcommand{\fref}[1]{Figure~\ref{#1}}
\newcommand{\tref}[1]{Table~\ref{#1}}

\title[H0LiCOW program overview]{H0LiCOW I. $H_0$ Lenses in
  COSMOGRAIL's Wellspring: Program Overview}

\author[S.~H.~Suyu et al.]{\parbox{\textwidth}{
S.~H.~Suyu,$^{1,2,3}$\thanks{E-mail: suyu@mpa-garching.mpg.de}
V.~Bonvin,$^{4}$
F.~Courbin,$^{4}$
C.~D.~Fassnacht,$^{5}$
C.~E.~Rusu,$^{5}$
D.~Sluse,$^{6}$
T.~Treu,$^{7}$
K.~C.~Wong,$^{8,2}$
M.~W.~Auger,$^{9}$
X.~Ding,$^{7,10}$
S.~Hilbert,$^{11,12}$
P.~J.~Marshall,$^{13}$
N.~Rumbaugh,$^{5}$
A.~Sonnenfeld,$^{14,7,15}$
M.~Tewes,$^{16}$
O.~Tihhonova,$^{4}$
A.~Agnello,$^{17}$
R.~D.~Blandford,$^{13}$
G.~C.-F.~Chen,$^{5,2}$
T.~Collett,$^{18}$
L.~V.~E.~Koopmans,$^{19}$
K.~Liao,$^{7}$
G.~Meylan,$^{4}$
C.~Spiniello$^{1}$}
\\
\\
\parbox{\textwidth}{
$^{1}$Max-Planck-Institut f{\"u}r Astrophysik, Karl-Schwarzschild-Str.~1, 85748 Garching, Germany\\
$^{2}$Institute of Astronomy and Astrophysics, Academia Sinica, P.O.~Box 23-141, Taipei 10617, Taiwan\\
$^{3}$Physik-Department, Technische Universit\"at M\"unchen, James-Franck-Stra\ss{}e~1, 85748 Garching, Germany\\
$^{4}$Laboratoire d'Astrophysique, Ecole Polytechnique
F{\'e}d{\'e}rale de Lausanne (EPFL), Observatoire de Sauverny, CH-1290
Versoix, Switzerland\\
$^{5}$Department of Physics, University of California, Davis, CA
95616, USA\\
$^{6}$STAR Institute, Quartier Agora - All\'ee du six Ao\^ut, 19c
B-4000 Li\`ege, Belgium\\
$^{7}$Department of Physics and Astronomy, University of California,
Los Angeles, CA 90095, USA \\
$^{8}$National Astronomical Observatory of Japan, 2-21-1 Osawa,
Mitaka, Tokyo 181-8588, Japan\\
$^{9}$Institute of Astronomy, University of Cambridge, Madingley Road,
Cambridge CB3 0HA, UK \\
$^{10}$Department of Astronomy, Beijing Normal University, Beijing
100875, China \\
$^{11}$Exzellenzcluster Universe, Boltzmannstr. 2, 85748 Garching, Germany \\
$^{12}$Ludwig-Maximilians-Universit{\"a}t, Universit{\"a}ts-Sternwarte, Scheinerstr. 1, 81679 M{\"u}nchen, Germany \\
$^{13}$Kavli Institute for Particle Astrophysics and Cosmology,
Stanford University, 452 Lomita Mall, Stanford, CA 94035, USA\\
$^{14}$Kavli IPMU (WPI), UTIAS, The University of Tokyo, Kashiwa, Chiba 277-8583, Japan \\
$^{15}$Physics Department, University of California, Santa Barbara, CA,
93106, USA \\
$^{16}$Argelander-Institut f\"ur Astronomie, Auf dem H\"ugel 71,
D-53121 Bonn, Germany\\
$^{17}$ESO-European Southern Observatory, D-85748 Garching bei
M\"unchen, Germany \\
$^{18}$Institute of Cosmology and Gravitation, University of
Portsmouth, Burnaby Rd, Portsmouth PO1 3FX, UK \\
$^{19}$Kapteyn Astronomical Institute, University of Groningen,
P.O. Box 800, 9700-AV Groningen, The Netherlands \\
}}

\date{Accepted XXX. Received YYY; in original form ZZZ}

\pubyear{2016}

\begin{document}
\label{firstpage}
\pagerange{\pageref{firstpage}--\pageref{lastpage}}
\maketitle

\begin{abstract}
Strong gravitational lens systems with time delays between the
multiple images allow measurements of time-delay distances, which are primarily sensitive to the 
Hubble constant that is key to probing dark energy, neutrino
physics, and the spatial curvature of the Universe, as well as discovering new physics.
We present H0LiCOW ($H_0$ Lenses in COSMOGRAIL's Wellspring), a
program that aims to measure $H_0$ with $<3.5\%$ uncertainty from five lens systems (\blens, \rxjlens,
\hequad, \wfilens\ and \hedouble).
  We have been acquiring
(1) time delays
through COSMOGRAIL and Very Large Array monitoring, (2)
high-resolution {\it Hubble Space Telescope} imaging for the lens
mass modeling, (3) wide-field imaging and spectroscopy to
characterize the lens environment, and (4) moderate-resolution
spectroscopy to obtain the stellar velocity dispersion of the
lenses for mass modeling.  In cosmological models with one-parameter extension to flat $\Lambda$CDM, we
expect to measure $H_0$ to $<3.5\%$ in most models, spatial curvature $\Ok$ to 0.004, $w$ to 0.14, and the
effective number of neutrino species to 0.2 (1$\sigma$ uncertainties)
when combined with current CMB experiments.  These are, respectively, a factor of
$\sim15$, $\sim2$, and $\sim1.5$ tighter than CMB alone.  Our data set
will further enable us to study the stellar initial mass function of the lens galaxies, and
the co-evolution of supermassive black holes and their host
galaxies.  This program will provide a foundation for extracting
cosmological distances from the hundreds of time-delay lenses
that are expected to be discovered in current and future surveys.

\end{abstract}



\section{Introduction}
\label{sec:intro}

In the past decade, the so-called ``flat $\Lambda$CDM'' cosmological
model consisting of dark energy (with density characterized by a
cosmological constant
$\Lambda$) and cold dark matter (CDM) in a spatially flat Universe has
emerged as the standard cosmological model.  This simple model has
provided excellent fit to various cosmological observations including
the temperature anisotropies in the cosmic microwave background (CMB)
and galaxy density correlations in baryon acoustic oscillations (BAO).
Recent CMB experiments, particularly the Wilkinson Microwave
Anisotropy Probe \citep[WMAP;][]{KomatsuEtal11,HinshawEtal12} and the Planck
satellite \citep{Planck2013P16, Planck2015P13}, and BAO surveys
\citep[e.g.,][]{AndersonEtal14, RossEtal15, KazinEtal14}, have yielded
stringent constraints
with unprecedented precision on cosmological parameters in the
spatially-flat $\Lambda$CDM model.

An interesting result from Planck is its predicted value of the Hubble
constant ($H_0$), a key cosmological parameter that sets the
present-day expansion rate as well as the age, size, and critical
density of the Universe.  Planck does not directly measure $H_0$, but
rather enables its indirect inference through measurements of combinations of
cosmological parameters given assumptions of the background cosmological
model.  Intriguingly, Planck's value of $H_0 = 67.8\pm0.9\,\kmsMpc$ \citep{Planck2015P13}, from
Planck temperature data and Planck lensing under the flat $\Lambda$CDM
model, is lower than
recent direct measurements based on the distance ladder, of
$73.24\pm1.74\,\kmsMpc$ from the SH0ES program \citep{RiessEtal16} and of
$74.3\pm2.1\,\kmsMpc$ \citep{FreedmanEtal12} from the Carnegie-Chicago
Hubble Program \citep{BeatonEtal16}.  On the other hand, Planck's
$H_0$ value is similar to the results of some of the megamaser
measurements 
(e.g., $H_0 = 68.9\pm7.1\,\kmsMpc$ from \citeauthor{ReidEtal13}
\citeyear{ReidEtal13}, $H_0=73^{+26}_{-22}\,\kmsMpc$ from \citeauthor{KuoEtal15}
\citeyear{KuoEtal15}, and
$H_0=66.0\pm6.0\,\kmsMpc$ from \citeauthor{GaoEtal16}
\citeyear{GaoEtal16}),
although the uncertainties of these
maser $H_0$ measurements are still substantial relative to that of Planck.
A 1\% {\it direct} measurement of the Hubble constant is highly needed:
such 1\% measurements of $H_0$ would address the
possible tension with the CMB value which, if significant, would
point towards deviations from the standard flat $\Lambda$CDM and
new physics.  In fact, when one relaxes, for example, the flatness or
$\Lambda$ assumption in the CMB analysis, strong parameter
degeneracies between $H_0$ and other cosmological parameters appear,
and the degenerate $H_0$ values from the CMB become compatible with
the local $H_0$ measurements from the distance ladder
\citep{Planck2015P13, RiessEtal16, FreedmanEtal12}.
Thus, a 1\% measurement of $H_0$ is crucial for
understanding the nature of dark
energy, neutrino physics, the spatial curvature of the Universe and the
validity of General Relativity \citep[e.g.,][]{Hu05, SuyuEtal12b,
  WeinbergEtal13}.  In particular, the dark energy figure of merit of
any survey that does not directly measure $H_0$ improves by
$\sim40\%$ if $H_0$ is known to 1\%.  Furthermore, independent methods
to measure $H_0$
are necessary to overcome
systematic effects, such as the known unknowns (e.g., the effects of
crowding or metallicity dependence
in the cosmic
distance ladder) and the unknown unknowns
in order to
robustly verify or rule out the standard cosmological paradigm.

Strong gravitational lenses with measured time delays between the
multiple images provide a competitive approach to measuring the Hubble
constant, completely independent of the local distance ladder: we
have demonstrated that we can constrain $H_0$ to $\sim7-8\%$ precision
from a single time-delay lens system with ancillary data
\citep{SuyuEtal10, SuyuEtal14}.  The time-delay
method was first proposed by \citet{Refsdal64} even before the
discovery of the first strong gravitational lens system
\citep{WalshEtal79}, consisting of a foreground mass distribution that
is located close along the line of sight to a background source
\citep[see][for a recent review]{TreuMarshall16}.  The
light from the background source is deflected by the foreground
``lens'' mass distribution; such light bending produces distorted
and, in rare cases of ``strong lensing'', multiple and often
spectacular images of the background source (e.g.,
Figure~\ref{fig:lens_sample}).

When the background source is one that varies in its luminosity, such
as an active galactic nucleus \citep[AGN; e.g.,][]{VanderriestEtal89,
  SchechterEtal97, FassnachtEtal99, FassnachtEtal02, KochanekEtal06,
  CourbinEtal11} or a supernova \citep[SN; e.g.,][]{QuimbyEtal14,
  KellyEtal15, KellyEtal16, GrilloEtal16, KawamataEtal16, TreuEtal16,
  GoobarEtal16, MoreEtal16b},
the variability is manifest in each
of the multiple images, but delayed in time relative to each other due
to the different light paths.  This time delay ($\Delta t$) thus depends on the
``time-delay distance'' ($\tdist$) and the lens mass distribution.
Specifically,
$\Delta t = \tdist \Delta \phi / c$,
 where $\Delta \phi$ is the
Fermat potential difference that is determined by the lens mass distribution and
$c$ is the speed of light.
Therefore, by measuring the time delay from photometric light
curves of the quasar images and modeling the
lens mass distribution, one can determine the time-delay distance to
the lens system and use the distance-redshift relation to constrain
cosmological models.

More precisely, the time-delay distance is
\be
\tdist \equiv (1+\zd) \frac{\Dd \Ds}{\Dds}
\ee
\citep{Refsdal64, SuyuEtal10}, where $\zd$ is the redshift of the
foreground deflector (also referred to as the strong lens), $\Dd$ is the angular diameter distance to
the deflector, $\Ds$ is the angular diameter distance to the source, and
$\Dds$ is the angular diameter distance between the deflector and the
source.  This time-delay distance is for a single strong lens plane,
with other line-of-sight mass distributions only weakly perturbing the
strong lens system and characterized via external shear and
convergence.  For cases where there are massive line-of-sight mass
distributions at a different redshift from the strong lens galaxy yet
close in projection to it such that these massive structures cannot be
well approximated by an external shear/convergence, it is necessary to
use the multi-plane lensing formalism
\citep[e.g.,][]{BlandfordNarayan86, SchneiderEtal92}.  In
general, multi-lens plane ray tracing does not yield a
single time-delay distance but rather several combinations of
distances.  Nonetheless, even in some of these cases, we can derive an
effective time-delay distance.

As a result of the unique combination of these three angular diameter
distances, the time-delay distance $\tdist$ is primarily sensitive to
the Hubble constant, in contrast to other non-local distance probes
such as supernova (SN) that probe relative luminosity distances \citep[e.g.,][]{RiessEtal98,
  PerlmutterEtal99, ConleyEtal11, SuzukiEtal12,BetouleEtal14} and BAO
\citep[e.g.,][]{EisensteinEtal05, PercivalEtal10, BlakeEtal11,
  AndersonEtal14} that yield absolute angular diameter distances.
 We note though that BAO, together with the CMB, can be used to
  calibrate the absolute magnitude of SN; assuming that the
  absolute magnitude of SN does not evolve with redshift,
  this combination of BAO and SN provides an ``inverse-distance
  ladder'' for the Hubble constant that is insensitive to
  assumptions on dark energy properties and spatial curvature
  \citep[e.g.,][]{HeavensEtal14, AubourgEtal15}.  While BAO and the
  time-delay method both provide angular diameter distance
  measurements, the distinction is that BAO gives angular diameter
  distances at specific redshifts whereas
  the time-delay method yields time-delay distances ($\tdist$) which are each a
  combination of three angular diameter distances.  One could in fact
  determine the angular diameter distance to the lens $\Dd$ in
  addition to $\tdist$ for time-delay lenses that have stellar velocity
  dispersion measurements of the foreground lens galaxy
  \citep{ParaficzHjorth09, JeeEtal15}.
Without time delays, lenses with stellar velocity
dispersion measurements can still offer a way to determine the cosomolgical
matter and dark-energy density parameters via a ratio of angular
diameter distances \citep[e.g.,][]{FutamaseHamana99,
  FutamaseYoshida01, GrilloEtal08}.
Recently \citet{JeeEtal16} have shown that measurements of $\tdist$ and
$\Dd$ from a modest sample of time-delay lenses with lens velocity
dispersion measurements yield competitive
constraints on cosmological models. In practice, both distances appear
as intermediate quantities between the sought after cosmological
parameters and the observed quantities.

In order to measure distances precisely and accurately from time-delay
lenses, we need four key ingredients in addition to the spectroscopic
redshifts of the lens and the source:
(1) time delays,
(2) high-resolution and high signal-to-noise ratio images of the
  lens systems,
(3) characterization of the lens environment, and
(4) stellar velocity dispersion of the lens galaxy.
These can be obtained via imaging and spectroscopy from
\textit{Hubble Space Telescope} (\hst) and ground-based observatories. In
\sref{sec:ingredients}, we detail each of these requirements.

We initiated the H0LiCOW ($H_0$ Lenses in COSMOGRAIL's Wellspring) program
with the aim of measuring the Hubble constant with better than $3.5\%$
precision and accuracy (in most background cosmological models),
through a sample of five time-delay lenses.  We
obtain the key ingredients to each of the lenses through observational
follow-ups and novel analysis techniques.  In particular, we have
high-quality lensed quasar light curves, primarily obtained via
optical monitoring by the COSMOGRAIL
\citep[COSmological MOnitoring of GRAvItational Lenses;
e.g.,][]{CourbinEtal05, VuissozEtal08, CourbinEtal11, TewesEtal13b}
and \citet{KochanekEtal06} teams but also via radio-wavelength
monitoring \citep{FassnachtEtal02}. COSMOGRAIL has been
monitoring more than 20 lensed quasars for more than a decade.  The
unprecendented quality of the light curves combined with new curve
shifting algorithms \citep{TewesEtal13a} lead to time delays with
typically $\sim$3\% accuracy \citep{FassnachtEtal02, CourbinEtal11, TewesEtal13b}.
In addition, we obtain \hst\ imaging that reveal the ``Einstein ring'' of the lens systems in
high resolution, and develop state-of-the-art lens modeling techniques
\citep[][Suyu et al.~in preparation]{SuyuEtal09, SuyuHalkola10, SuyuEtal12a} and
kinematic modeling methods \citep{AugerEtal10, SonnenfeldEtal12} to
obtain the lens mass distribution with a few percent uncertainty
\citep[e.g.,][]{SuyuEtal13, SuyuEtal14}.  We further obtain wide-field
imaging and spectroscopy to characterize the environment of the field,
as well as the spectroscopy
of the lens galaxy to obtain the stellar velocity dispersion.  The
exquisite follow-up data set that we have acquired allow us not only
to constrain cosmology but also to study lens galaxy and source
properties for understanding galaxy evolution, including the dark
matter distribution in galaxies, the stellar initial mass function of
galaxies and the co-evolution between supermassive black holes and
their host galaxies.

A crucial aspect of our program is the use of blind analysis
\citep[e.g.,][]{ConleyEtal06, SuzukiEtal12, SuyuEtal13,
  vonderLindenEtal14} to test for residual systematics and avoid
subconscious experimenter
bias.  In particular, we have developed core analysis
techniques for the first lens whose dissection was not blinded
\citep[\blens;][]{SuyuEtal10}; we
subsequently build upon these techniques and
perform blind analysis on the other lenses in the sample.
In the blind analysis, the idea is not to blind all the model
parameters being inferred,
but rather just
the cosmological parameters that we aim to measure (as well as any
derived parameters or summary statistics
from which we could infer the cosmological parameters).  We therefore
blind the time-delay distance and all cosmological parameters in our
analysis.  Specifically, throughout the analysis, we only ever plot
these blinded parameters offset by their posterior median value.  We can then
still use the parameter correlations and the uncertainties to cross
check our analysis, since the temptation to stop investigating systematic
errors when the ``right answer'' has been obtained has been removed.
Only when the
collaboration deems the
analysis to be final and complete do we ``open the box'' to reveal the
median values of the parameters, and then publish these results without
modifications.

This paper (hereafter, H0LiCOW Paper I) is the first of the series,
and gives an overview of the program.  There are four more papers
that detail the data sets and analysis of the H0LiCOW lens system
\hequad.  In particular, \citet[][hereafter H0LiCOW~Paper~II]{SluseEtal16} 
present the spectroscopic follow-up of
the strong lens field to
measure redshifts of massive and nearby objects close in projection to
the strong lens system and identify galaxy groups along the line of sight.
\citet[][hereafter H0LiCOW~Paper~III]{RusuEtal16b} 
use our
multi-band wide-field imaging to characterize the lens
environment in combination with ray tracing with numerical
simulations.    \citet[][hereafter H0LiCOW Paper IV]{WongEtal16} 
 perform the lens mass
modeling of the strong lens
system incorporating the time delays, high-resolution imaging and
lens stellar kinematics data sets to infer the distance to the lens
via blind analysis.  \citet[][hereafter
H0LiCOW~Paper~V]{BonvinEtal16b} present the time-delay 
measurements from COSMOGRAIL lens monitoring
and the cosmological inference based on the previous three papers.

The outline of this paper is as follows.  We describe the key
ingredients for time-delay cosmography in \sref{sec:ingredients},
present the five H0LiCOW lens systems in \sref{sec:lenses}, and
describe our observational campaign in \sref{sec:followup} .  The key
components of the four analysis papers introduced above are summarized
in \sref{sec:papers}.  We show the forecasted cosmographic constraints
from the H0LiCOW sample in \sref{sec:forecast}.  We summarize in
\sref{sec:summary} with an outlook for the program.


\section{Observational requirements of the time-delay method}
\label{sec:ingredients}

In this section, we describe the observational requirements of the four
ingredients for accurate and precise distance measurements from
time-delay lenses.

\begin{enumerate}
\item {\it Time delays}.  Monitoring campaigns to map out the
  variability of the multiple lensed images over time have been
  carried out both in the radio and optical wavelengths
  \citep[e.g.,][]{VanderriestEtal89, SchechterEtal97, BurudEtal02,
    HjorthEtal02, FassnachtEtal02, VuissozEtal07,
    KochanekEtal06, RumbaughEtal15}.  Regular and frequent
  observations, at least once every few days, are necessary so that
  the variability pattern of the background source can be observed in
  each of the multiple images and be matched up to obtain the time
  delays.  Monitoring in the optical requires a long baseline or high
  photometric precision to overcome systematic variations due to
  microlensing by stars in the lensing galaxy that could be mistaken
  as the background source intrinsic variability
  \citep[e.g.,][]{TewesEtal13b, SluseTewes14}.  Curve-shifting methods have been
  developed to measure the time delays from the light curves
  \citep[e.g.,][]{PressEtal92b, PeltEtal96, FassnachtEtal02,
    HarvaEtal08, HirvEtal11, MorganEtal08, TewesEtal13a,
    HojjatiEtal13}.  A recent time-delay
  challenge showed that some of the methods can recover accurately the
  time delays in a blind test \citep{DoblerEtal15,LiaoEtal15},
  particularly the methods we use from the COSMOGRAIL collaboration
  \citep[e.g.,][]{TewesEtal13a, BonvinEtal16}.
\item {\it Well-resolved lensed images}.  The strong lensing
  information, such as the multiple image positions of the background
  source, is needed to obtain the foreground lens mass distribution
  for converting the time delays into distances.  Deep and
  high-resolution imaging of the strong lens system reveal the
  ``Einstein rings'' that are the spatially extended and
  lensed images of the background source, such as the host galaxy of
  the AGN.  In the past decade, methods have been developed to take
  advantage of the thousands of intensity pixels of the extended
  images to constrain precisely within a few percent
  the lens potential at the location of the multiple images
  \citep[e.g.,][]{KochanekEtal01, WarrenDye03, TreuKoopmans04, Koopmans05, DyeEtal08,
    VegettiKoopmans09,
    SuyuEtal09, SuyuEtal13, BirrerEtal15, ChenEtal16}.
  The time-delay distance is particularly sensitive to the radial
  profile of the lens galaxy mass distribution
  \citep[e.g.,][]{Kochanek02, Wucknitz02, WucknitzEtal04, Suyu12}.
  Imaging with high signal-to-noise-ratio and high angular resolution
  of the Einstein ring helps to constrain the lens radial profile in
  the region of the ring, and hence the time-delay distance, up to a
  mass-sheet transformation (described below).
\item {\it The lens environment}.  The distribution of mass external to the lens
  galaxy, such as that associated with galaxies which are close in projection
  to the lens system along the line of sight, affects the time
  delays between the multiple images and hence our cosmological
  distance measurements.
  An external convergence $\kext$ can be absorbed by the lens and source
  model leaving the fit to the lensed images unchanged, but the predicted
  time delays altered by a factor of $(1-\kext)$.
  To break this ``mass-sheet degeneracy'' \citep{FalcoEtal85}, one
  can study the environment of the lens system to constrain $\kext$
  within a few percent\footnote{in terms of its impact on $\tdist$}
  through spectroscopic/photometric observations of local galaxy
  groups and line-of-sight structures \citep[e.g.,][]{MomchevaEtal06,
    FassnachtEtal06, MomchevaEtal15} in combination with ray-tracing
  through numerical N-body simulations \citep[e.g.,][]{HilbertEtal07,
    HilbertEtal09, SuyuEtal10, GreeneEtal13, CollettEtal13}. Furthermore,
  \citet{McCullyEtal14, McCullyEtal16} developed a new
  framework to model line-of-sight mass distributions efficiently and
  quantified the environment effects through realistic simulations of
  lens fields.  By reconstructing the 3-dimensional mass distribution
  of strong-lens sightlines, \citet{McCullyEtal16} can obtain 
  constraints on $\kext$ that are consistent with but tighter than those
  from the aforementioned statistical approach of combining galaxy
  number density observations with N-body simulations
  (see also \citeauthor{CollettEtal13}
  \citeyear{CollettEtal13} whose sightline mass reconstruction
  also produces 
  tighter constraints on $\kext$ 
  than the statistical approach).
  Recently, \citet{CollettCunnington16} have pointed out
  that the external convergence over an ensemble of lenses usually
  does not average to zero -- lenses, like typical massive galaxies,
  preferentially live in locally over-dense regions
  \citep{HolderSchechter03, TreuEtal09,
  FassnachtEtal11} and are therefore slightly easier to detect and
  monitor.  Nonetheless, this bias in detection and/or selection that
  is due to overdensity is
  expected to have currently negligible impact on $\tdist$ ($<1\%$ impact).
  In contrast,
  measurements of $\Dd$ that come from combining delays with the lens velocity
  dispersion are impervious to $\kext$ \citep{JeeEtal15}.
\item {\it The lens galaxy stellar velocity dispersion}.  The combination
  of lensing and stellar kinematics is a powerful probe of the lens
  galaxy mass distribution \citep[e.g.,][]{RomanowskyKochanek99, TreuKoopmans02,
    KoopmansEtal03, BarnabeEtal09, BarnabeEtal11, SonnenfeldEtal12}
  since the combination breaks degeneracies that are inherent in each
  approach, and in particular the mass-sheet degeneracy in lensing.
  \citet{SchneiderSluse13} pointed out that the mass-sheet degeneracy
  can manifest as a lens mass profile degeneracy, which
  \citet{XuEtal16} investigated using simulated galaxies.
  Moreover, the mass-sheet
  degeneracy is in fact a special case of a more general
  ``source-position transformation'' \citep{SchneiderSluse14,
    UnruhEtal16}, although this latter
  transformation typically does not leave the multiple time delays
  invariant.
  To break such lensing degeneracies, information from the lens galaxy
  stellar kinematics is crucial: \citet{SuyuEtal14} showed that the
  lens velocity dispersion substantially reduced the dependence of the
  time-delay distance on lens mass profile assumptions.  The lens
  velocity dispersion is also a
  key ingredient for measuring $\Dd$, which is more sensitive to
  dark energy properties than $\tdist$ \citep{JeeEtal15, JeeEtal16}.
\end{enumerate}


\section{H0LiCOW Sample of Lenses}
\label{sec:lenses}

\begin{figure*}

\flushleft \hspace{0.1\columnwidth} \blens\hspace{0.20\columnwidth}\rxjlens\hspace{0.17\columnwidth} \hequad\hspace{0.16\columnwidth}
\wfilens\hspace{0.16\columnwidth} \hedouble \\
\includegraphics[width=0.40\columnwidth]{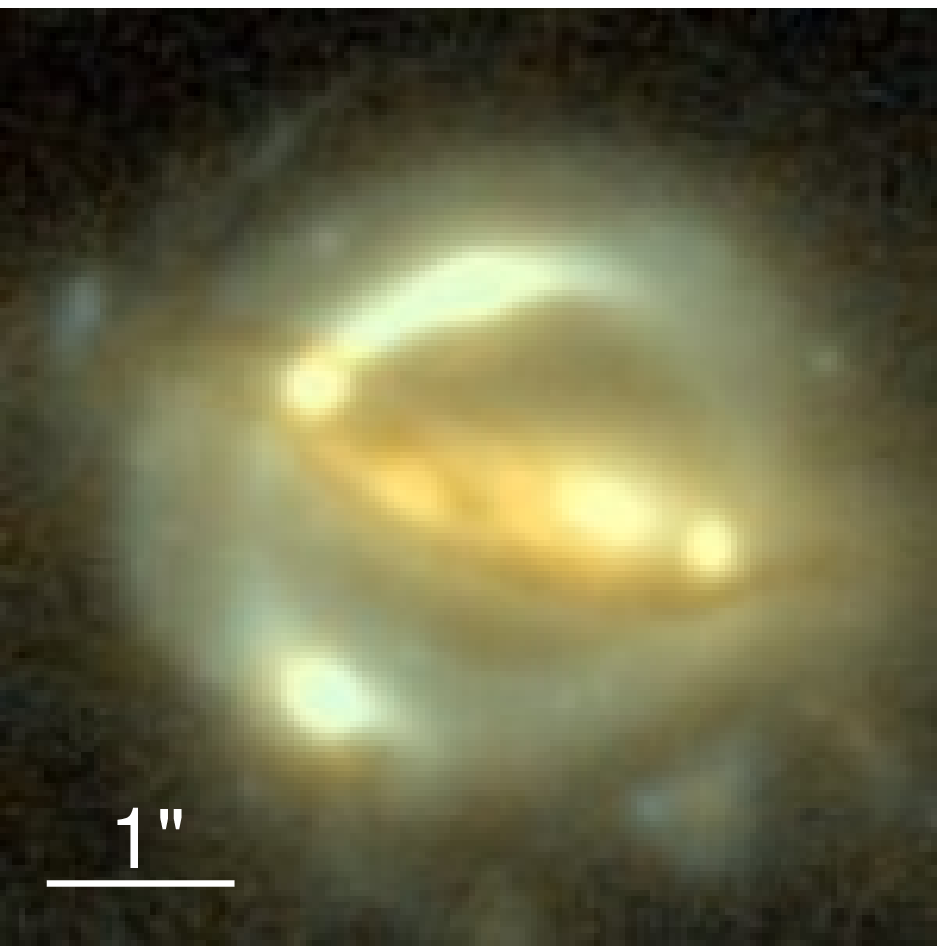}
\includegraphics[width=0.40\columnwidth]{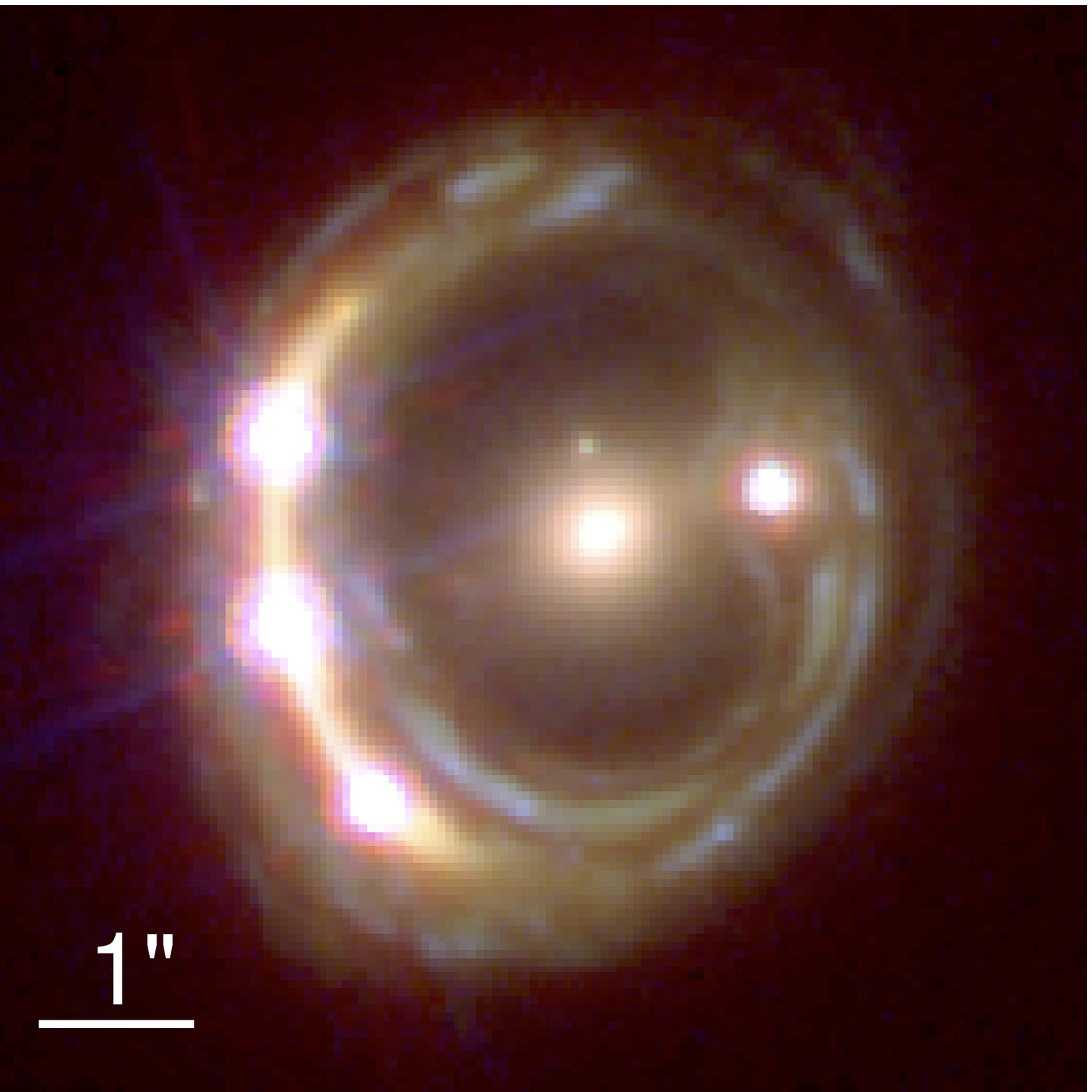}
\includegraphics[width=0.40\columnwidth]{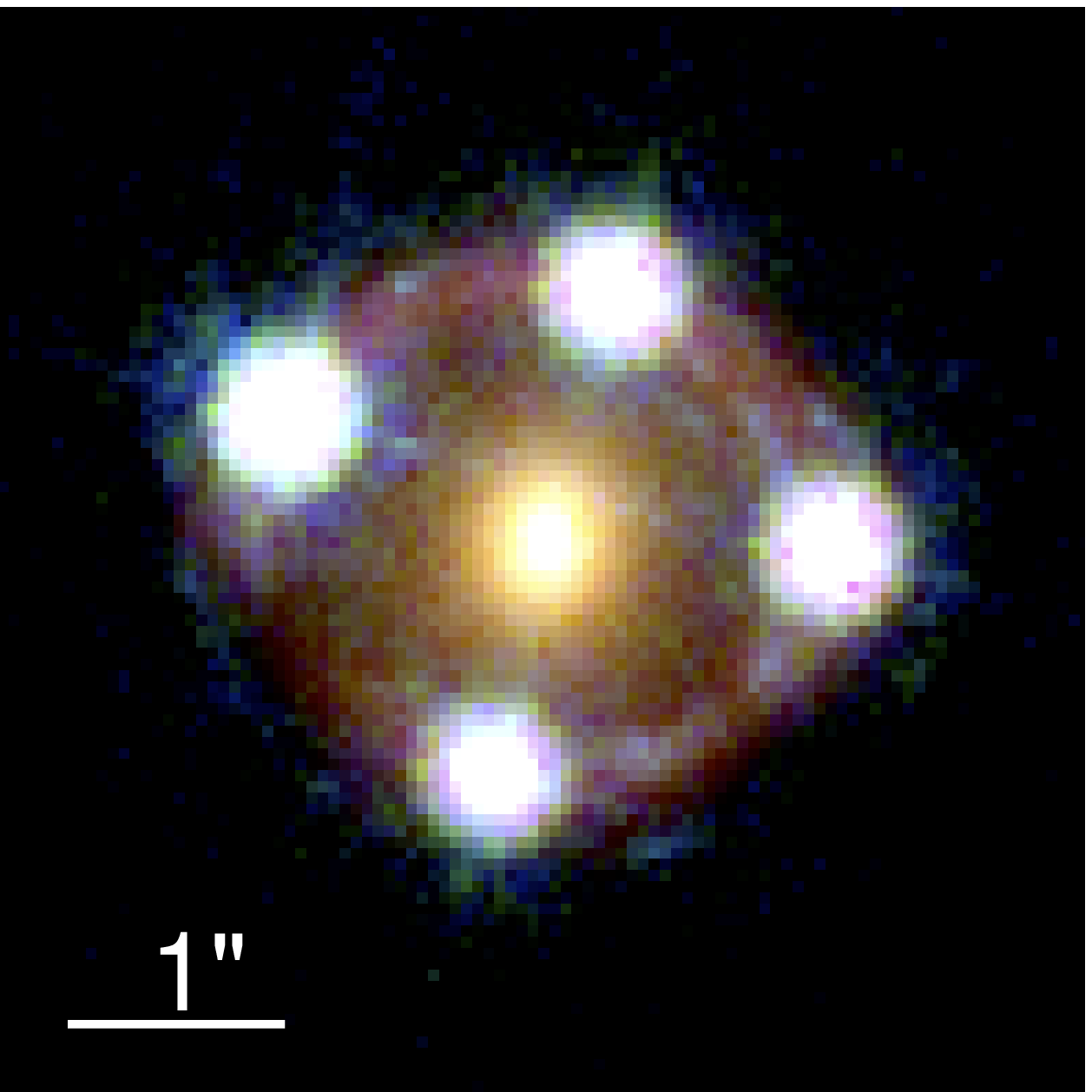}
\includegraphics[width=0.40\columnwidth]{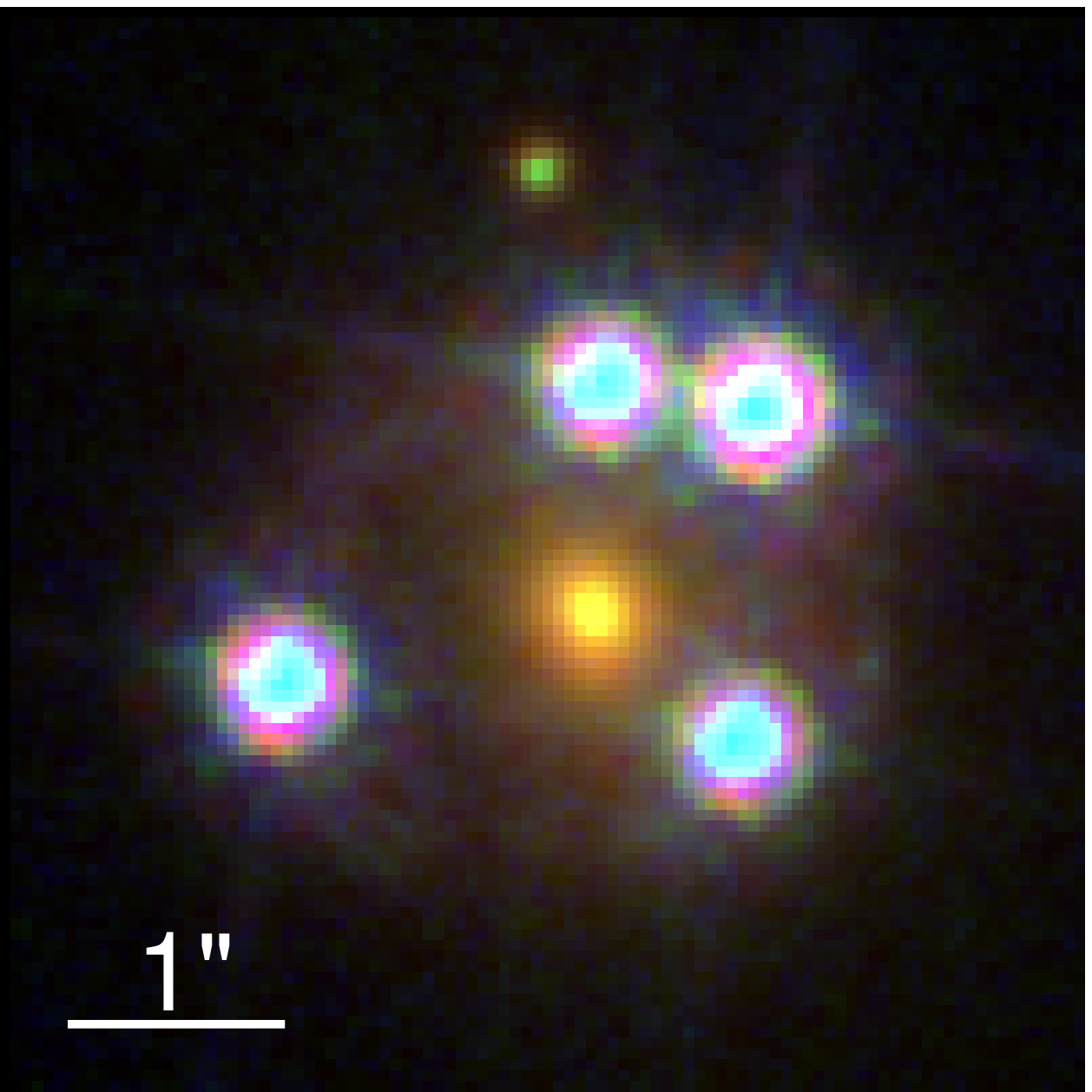}
\includegraphics[width=0.40\columnwidth]{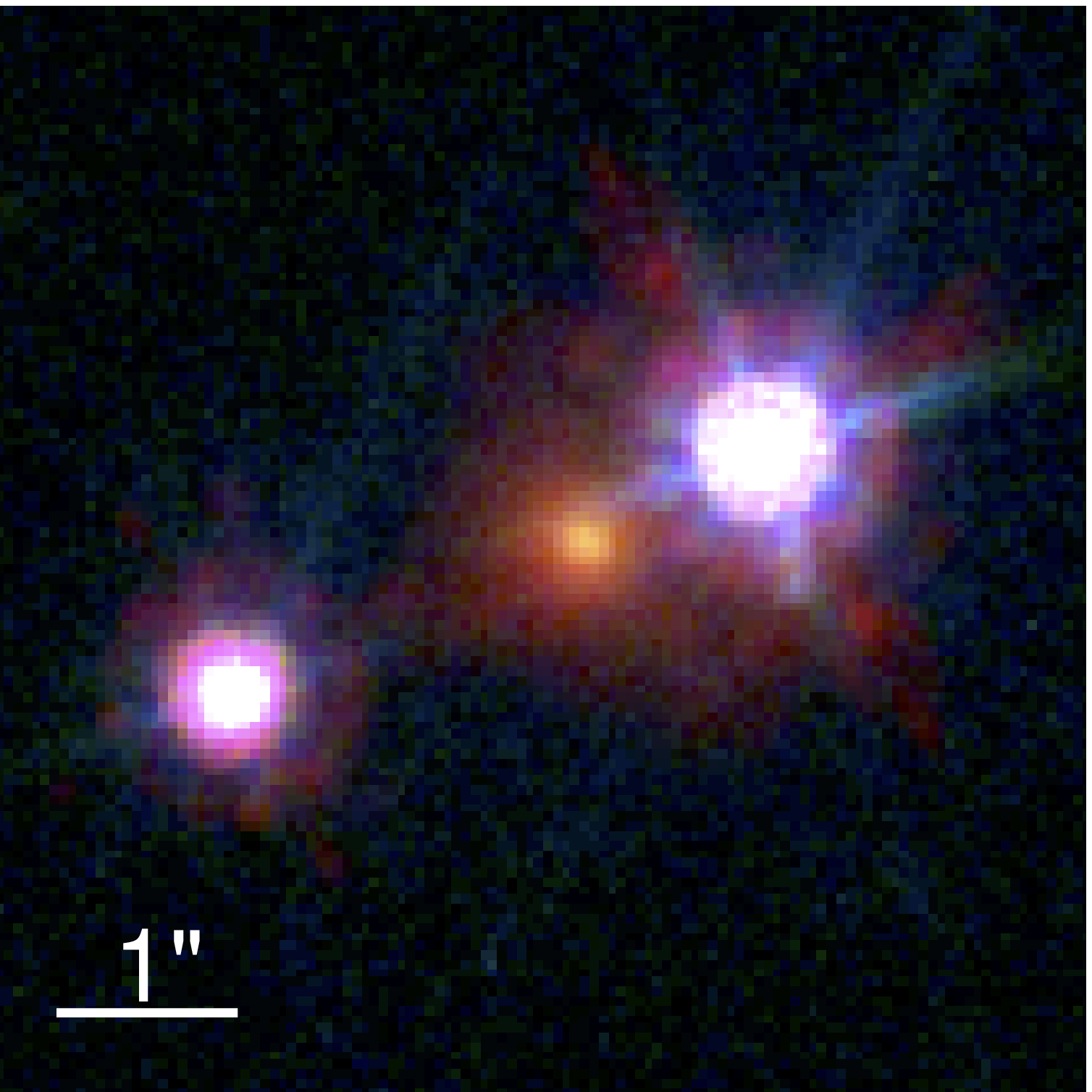}
\caption{H0LiCOW lens sample, consisting of four quadruply lensed
  quasar systems in various configurations and one doubly lensed
  quasar system.  The lens name is indicated above each panel.  The
  color images are composed using 2 (for \blens) or 3 (for other
  lenses) \hst\ imaging bands in the optical and near-infrared.  North
  is up and east is left.
}
\label{fig:lens_sample}
\end{figure*}

In \fref{fig:lens_sample}, we show the images of the five lenses in
our sample.  The left four lenses are quadruply lensed quasar systems
(``quads'') and the right-most lens system is a doubly lensed quasar
system (``double'').  As described below, the four quads span the three generic multiple
image configurations we have in galaxy-scale strong lenses: symmetric,
fold (with 2 merging images) and cusp (with 3 merging images).
Therefore, our sample will allow us to explore to some extent the
optimal image configuration for cosmographic studies.

Our sample of lenses was chosen based on three criteria: (1) availability
of accurate and precise time delays, (2) existing measurements of
spectroscopic redshifts for both the lens and the background source,
and (3) the lens system is not located near a galaxy cluster (to avoid
potentially large systematic effects due to mass along the line of sight).  We prefer quads to doubles
since quads provide more observational
constraints on the mass model (e.g., more time delays, image
positions).  The four quads in our sample were the only known quad
lenses that passed the above three criteria at the time of our sample
selection.  There were a few doubles that pass these criteria, and we
chose \hedouble\ as the first double in this pilot program given its
relative simplicity for mass modeling with only one strong-lens galaxy
(in contrast to other systems that have multiple massive lens
galaxies).  We describe in more detail each of the lenses below.

{\it \blens}.  The lens system was discovered in the Cosmic Lens
All-Sky Survey \citep[CLASS;][]{MyersEtal95, BrowneEtal03,
  MyersEtal03}.  The radio-loud AGN is lensed into four images that
are relatively dim in the optical wavelength, thus showing clearly the
extended Einstein ring of the AGN host galaxy in the \hst\ imaging
(\fref{fig:lens_sample}).  Two of the four multiple images are close
together, making this a standard ``fold'' configuration.  The system
contains two lens galaxies that appear to be interacting and resulting
in dust extinction in the system \citep[e.g.,][]{SurpiBlandford03,
  KoopmansEtal03, SuyuEtal09}.  The lens and source redshifts are,
respectively, $\zs=1.394$ \citep{FassnachtEtal96} and $\zd=0.6304$
\citep{MyersEtal95}.  This system was the first quad lens with all
three time delays measured with uncertainties of only a few percent
\citep{FassnachtEtal99, FassnachtEtal02}.

{\it \rxjlens}.  \citet{SluseEtal03} discovered \rxjlens\ serendipitously
during polarimetric imaging of a sample of radio
quasars.  This system shows a spectacular Einstein ring, with multiple
arclets that are the lensed images of the AGN host galaxy containing a
bulge and a disk with spiral arms and star formation clumps.  Three of
the four quasar images are close to each other, forming the typical
``cusp'' configuration.
The lens redshift is at $\zd=0.295$ \citep{SluseEtal03, SluseEtal07}, and the
source redshift is at $\zs=0.654$ \citep{SluseEtal07}\footnote{The
  source redshift of $\zs=0.654$ is based on the narrow emission lines, which is
  considered more accurate than the H$\alpha$ and MgII lines \citep{HewettWild10} that
  yield $\zs=0.657$ \citep{SluseEtal07}.  We note that a $0.003$
  change in $\zs$ corresponds to a $<0.4\%$ change in $\tdist$
for \rxjlens, and even less change in $\tdist$ for the other
higher-redshift lens systems.}.

{\it \hequad}.  This lens system was found by \citet{WisotzkiEtal02},
originally selected in the Hamburg/ESO survey \citep{WisotzkiEtal00}
as a highly probable quasar candidate.  The background quasar is
lensed into four multiple images that are nearly symmetrically
positioned in the ``cross'' configuration.  The background source is
at redshift $\zs=1.693$ \citep{SluseEtal12}\footnote{based on Mg II
  emission line, which results in a slightly higher redshift value than
  the previous
  measurement of $\zs=1.689$ \citep{WisotzkiEtal02} from C IV line that is
  known to be prone to systematic blueshifts in many quasars}
 and the foreground strong
lens is at redshift $\zd=0.4546$ \citep{MorganEtal05, EigenbrodEtal06}.
The \hst\ image
reveals an elliptical ring that connects the four images of the AGN.
This ring is produced by the extended lensed images of
the AGN galaxy.

{\it \wfilens}.  \citet{MorganEtal04} discovered this quad lens system
as part of an optical imaging survey using the MPG/ESO 2.2m telescope
at La Silla, Chile that is operated by the European Southern
Observatory (ESO).  The lens system exhibits a typical fold configuration,
since it contains two merging quasar images.  The quasar is at
redshift $\zs=1.662$
\citep{SluseEtal12}, which is consistent with the first measurement by
\citet{MorganEtal04}.  The quasar images are
substantially brighter than the background quasar host galaxy and the
foreground lens galaxy.   \citet{MorganEtal04} identified the
foreground lens galaxy, whose redshift was measured to be $\zd=0.661$
\citep{EigenbrodEtal06}, consistent with an earlier measurmeent by
\citet{OfekEtal06}.  The high-resolution \hst\ imaging shows several
galaxies in the vicinity of the lens system.  Since these galaxies
would likely influence the lens potential, their redshifts will be
obtained with our ancillary data (\sref{sec:followup:field}) in order
to incorporate them into the lens mass model.

{\it \hedouble}.  This system was also discovered in the early phase
of the Hamburg/ESO survey by \citet{WisotzkiEtal93}.  The two lensed
quasar images are separated by $\sim3''$ and is unusual in having the
brighter image as the one closer to the foreground lens galaxy, which
was first identified by \citet{CourbinEtal98} and \citet{RemyEtal98}.  The
source is at $\zs=2.316$ \citep{SmetteEtal95}, and the lens is at a
relatively high redshift of $\zd=0.729$ \citep{LidmanEtal00}.  The
\hst\ image shows multiple luminous structures/galaxies around the
lens system.


\section{Observational Follow-Up}
\label{sec:followup}

In collaboration with the COSMOGRAIL team, we carry out an
observational campaign in order to obtain each of the four ingredients
for distance measurements of the H0LiCOW lenses.  We describe the
monitoring in \sref{sec:followup:delays} to get the time delays, deep
\hst\ imaging
to constrain the lens galaxy mass
distribution in \sref{sec:followup:HST}, wide-field
spectroscopy and imaging to study the lens environment in
\sref{sec:followup:field} and spectroscopy of the foreground lens
galaxy to measure the stellar velocity dispersion in
\sref{sec:followup:vdisp}.

\subsection{Time delays}
\label{sec:followup:delays}

Of the five H0LiCOW lenses, \blens\ has been monitored previously by
\citet{FassnachtEtal99, FassnachtEtal02} using the Very Large Array,
whereas the other four lenses are currently being monitored by the
COSMOGRAIL and \citet{KochanekEtal06} collaborations using a network
of 1-2m optical telescopes, particularly the Euler telescope in Chile.

Using three seasons of monitoring of \blens, especially the third
season which showed significant variability that repeated in all four
quasar images, \citet{FassnachtEtal02} measured all three relative
time delays between the four quasar images with uncertainties of a few
percent.  The image fluxes were measured every 3--4 days during the
monitoring.  The time delays span $\sim30-80$ days, relative to the first
image that varies.

The monitoring of \rxjlens, \hequad, \wfilens\ and \hedouble\
by the COSMOGRAIL
and \citet{KochanekEtal06} teams started in 2003, with a
photometric point every 2--4 days.  The MCS deconvolution method
\citep{MagainEtal98, CantaleEtal16} is used to extract the photometry of the quasar
images for building the light curves. \citet{TewesEtal13a} set up an
automated pipeline to reduce the images, build the light curves, and
measure the time delays using a state-of-the-art curve-shifting
algorithm that simultaneously models both intrinsic variability of the
AGNs and microlensing variations.  With this pipeline,
\citet{BonvinEtal16} recovered the time delays with a precision of
$\sim3\%$ and negligible bias for simulated light curves mimicking
COSMOGRAIL monitoring in the blind strong lens time delay challenge
\citep{LiaoEtal15}, demonstrating the robustness of their curve-shifting
algorithms.

The monitoring and analysis yield time delays in \rxjlens\ with a
$1.5\%$ uncertainty on the longest delay \citep{TewesEtal13b}.  The
light curve has been separately modeled by A.~Hojjati and E.~Linder
using the Gaussian process technique \citep{HojjatiEtal13},
who have obtained delays that are consistent
with the measurements of \citet{TewesEtal13b} (Eric Linder, private
communications).
The monitoring and
analysis of \hequad\ is described in H0LiCOW Paper V, with a relative
uncertainty of
$6.5\%$ on the longest delay
(between images A and D).  The measurement
precision in the delays has improved by a factor of 2 compared to the
previous measurements by \citet{CourbinEtal11} with the five
additional years of monitoring and improvements in the curve-shifting
algorithms.   For \wfilens\ and \hedouble, we
expect to improve on the previous delay measurements by
\citet{VuissozEtal08} and \citet{PoindexterEtal07}, respectively, with the new
curve-shifting techniques, and estimate relative
uncertainties of $\sim4\%$ and $\sim2\%$, respectively, from the
monitoring campaign.

\subsection{\hst\ observations}
\label{sec:followup:HST}

Deep \hst\ Advanced Camera for Surveys (ACS) observations of \blens\
were obtained in Program 10158 (PI: C.~D.~Fassnacht) in two filters,
F606W and F814W.  \citet{SuyuEtal09} have described these observations
in detail.  Furthermore, \citet{SuyuEtal09} analyzed these data and
used a pixelated lens potential reconstruction technique to model the
lens mass distribution, which were subsequently used for cosmographic
analysis in \citet{SuyuEtal10}.

Archival \hst\ ACS observations of \rxjlens\ (Program 9744; PI: C.~S.~Kochanek)
are available in two filters, F555W and F814W.  Details of the
observations are described in, e.g., \citet{ClaeskensEtal06}.  These
have been used to model the lens mass distribution for cosmography,
accounting for uncertainties due to assumptions on the lens mass
profile \citep{SuyuEtal13, SuyuEtal14}.  Recently,
\citet{BirrerEtal16} have also used these observations to model
independently the lens mass distribution of \rxjlens\ for cosmography,
obtaining results that are consistent with \citet{SuyuEtal13}.

We have obtained new deep \hst\ Wide Field Camera 3 (WFC3) observations in
Program 12889 (PI: S.~H.~Suyu) of the remaining three lenses (\hequad,
\wfilens\ and \hedouble) in the infrared (IR) channel.  The goal of
these observations is to detect the Einstein rings of the AGN host
galaxies at high signal-to-noise ratios, in order to constrain the
foreground lens mass distribution (previous \hst\ observations
had insufficient signal-to-noise ratios of the rings for our analysis).
We use the F160W filter to
optimize the contrast between the AGN host galaxy and the AGN, since
the host galaxy is brighter in the infrared compared to the optical,
especially for \hedouble\ where the quasar is at a high redshift.

We employ four-point dither patterns that trace out parallelograms
with the lengths of the sides being non-integral numbers of pixels.
For each lens, we
use multiple parallelograms that are offset by non-integral pixels.
Specifically, we use 2, 5 and 3 parallelograms for \hequad, \wfilens,
\hedouble, respectively, depending on the total exposure time needed
to image the Einstein ring.  We
further ensure that the dithering points
do not overlap to avoid IR persistence effects.  This dithering
strategy allows us to recover effectively an angular resolution of
$\sim0\farcs08$ from the native $0\farcs13$ pixel scale.

Since the AGN host galaxy is substantially fainter than the AGN, we
further adopt an exposure sequence of short-long-long at each of the
dithering point\footnote{\label{footnote:exposeq} For \hequad\ one
  long exposure was lost due to a satellite passing  over the target.
  For one of the parallelogram dither pattern
  for \wfilens, we use an exposure sequence of short-long (rather than
  short-long-long) at each dither position to optimize target
  exposure time
  given overhead associated with observations.}.  The first short
exposure allows us to characterize the AGN, whereas the long exposures
would get the AGN host with possibly the pixels near the bright AGN
saturated.  We note that there are multiple non-destructive reads during each exposure with the MULTIACCUM mode of the WFC3/IR
detector, so we can have a count-rate estimate on the AGN pixels even
in the long
exposures if several non-destructive reads are available before
saturation.  The short exposures are taken to
ensure that there are sufficient reads to characterize
accurately the pixel count rates near the AGN positions, in case the long
exposures are indeed saturated with insufficient non-destructive
reads.  In essence, the
combination of the short and long exposures
allows us to reconstruct in full the brightness distribution of {\it
  both} the lensed AGN and the lensed host galaxy.  We summarize our
observations in \tref{tab:hstobs}.

\begin{table}
\caption{New \hst\ WFC3/IR Observations of \hequad, \wfilens\ and \hedouble}
\label{tab:hstobs}
\begin{center}
\begin{tabular}{l l l l}
\hline
Lens & Date & Number/type & Time (s) per \\
  &  & of exposures & exposure \\

\hline
\hequad & 2012-10-28 & 8 short exp.  & \phantom{0}44 \\
       &  & 15 long exp.$^{\ref{footnote:exposeq}}$ & 599 \\
\hline
\wfilens & 2013-05-03  & 20 short exp.  & \phantom{0}74  \\
       & to 2013-05-04  & 4 long exp.$^{\ref{footnote:exposeq}}$ & 599 \\
       &                & 32  long exp.$^{\ref{footnote:exposeq}}$ & 699 \\
\hline
\hedouble & 2013-03-18 & 12 short exp.  & \phantom{0}26 \\
       &                & 24 long exp.  & 599 \\
\hline

\end{tabular}
\end{center}
\begin{flushleft}
  Notes.  At each dither position, an exposure sequence of
  short-long-long exposure times is adopted in order to sample the
  large dynamical range of the AGN and its much fainter host galaxy$^{\ref{footnote:exposeq}}$.
\end{flushleft}
\end{table}

We reduce the images using {\sc DrizzlePac}\footnote{{\sc DrizzlePac}
  is a product of the Space Telescope Science Institute, which is
  operated by AURA for NASA.}. The images are drizzled to a final
pixel scale of $0\farcs08$, without masking the bright AGN pixels as
they are well characterized by the short exposures.  The uncertainty
on the flux in each pixel is estimated from the science image and the
drizzled exposure time map
by adding in quadrature the Poisson noise from the source and
the background noise due to the sky and detector readout.

\begin{figure*}
\large{\hequad\hspace{0.33\columnwidth} \wfilens\hspace{0.33\columnwidth} \hedouble} \\
\includegraphics[width=0.6\columnwidth]{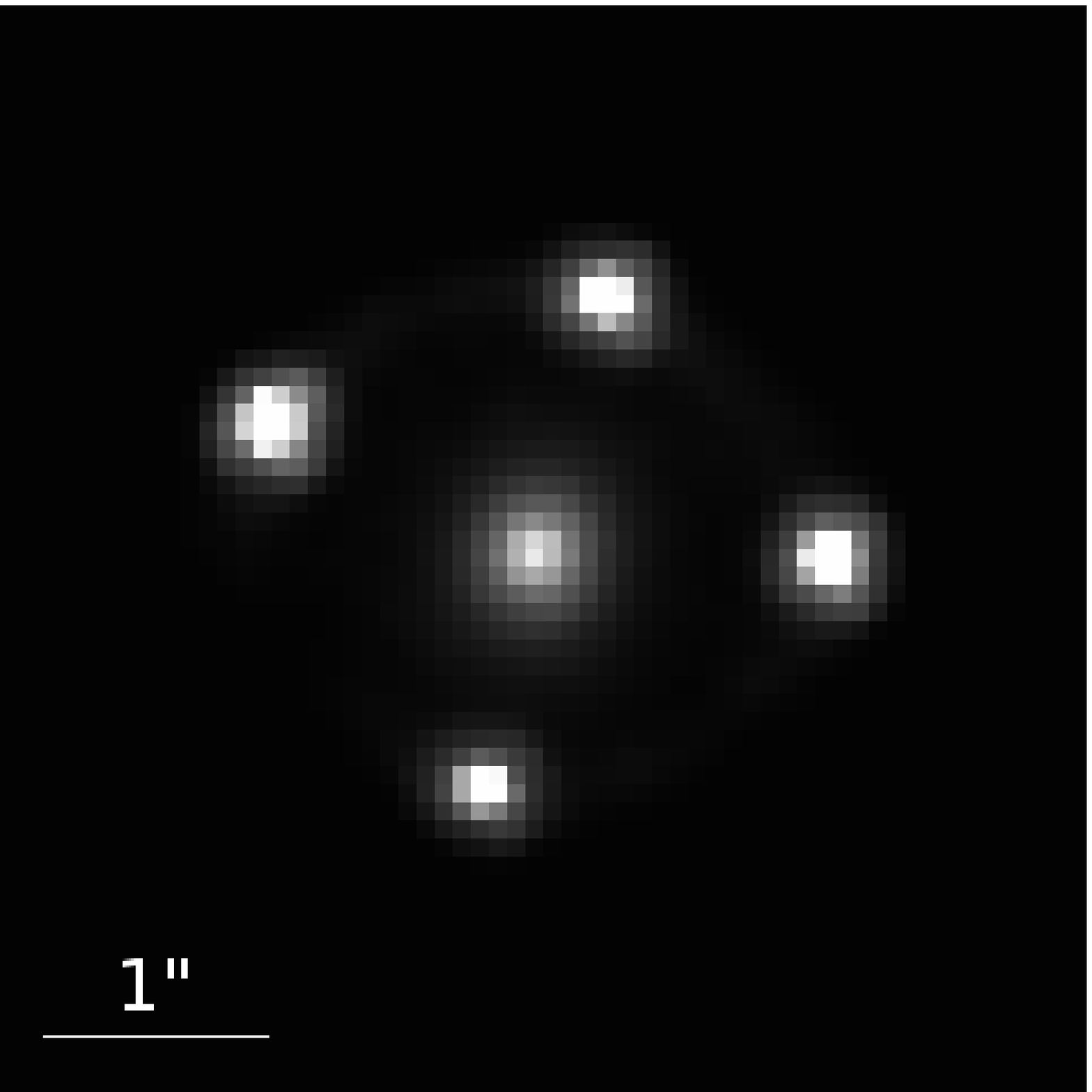}
\includegraphics[width=0.6\columnwidth]{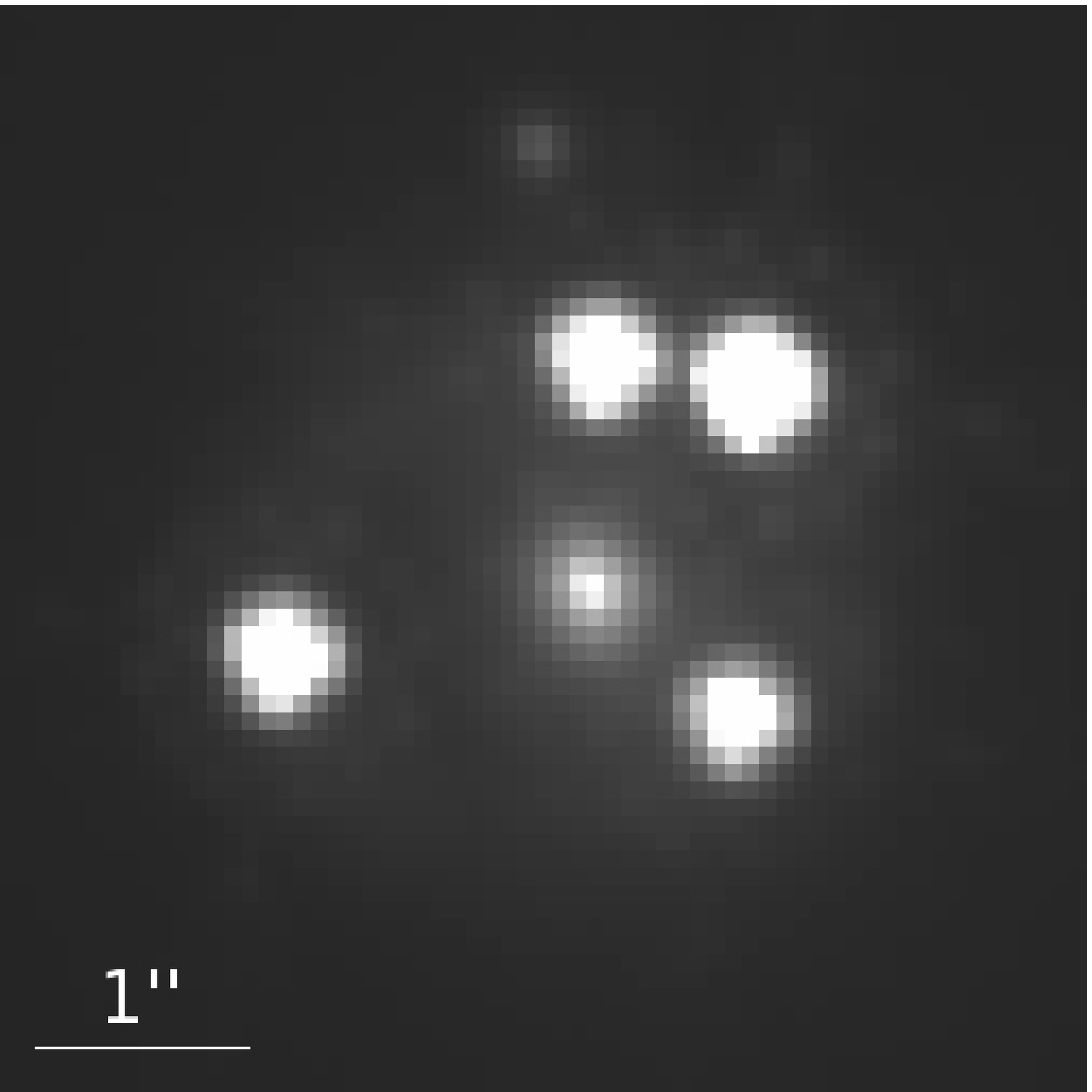}
\includegraphics[width=0.6\columnwidth]{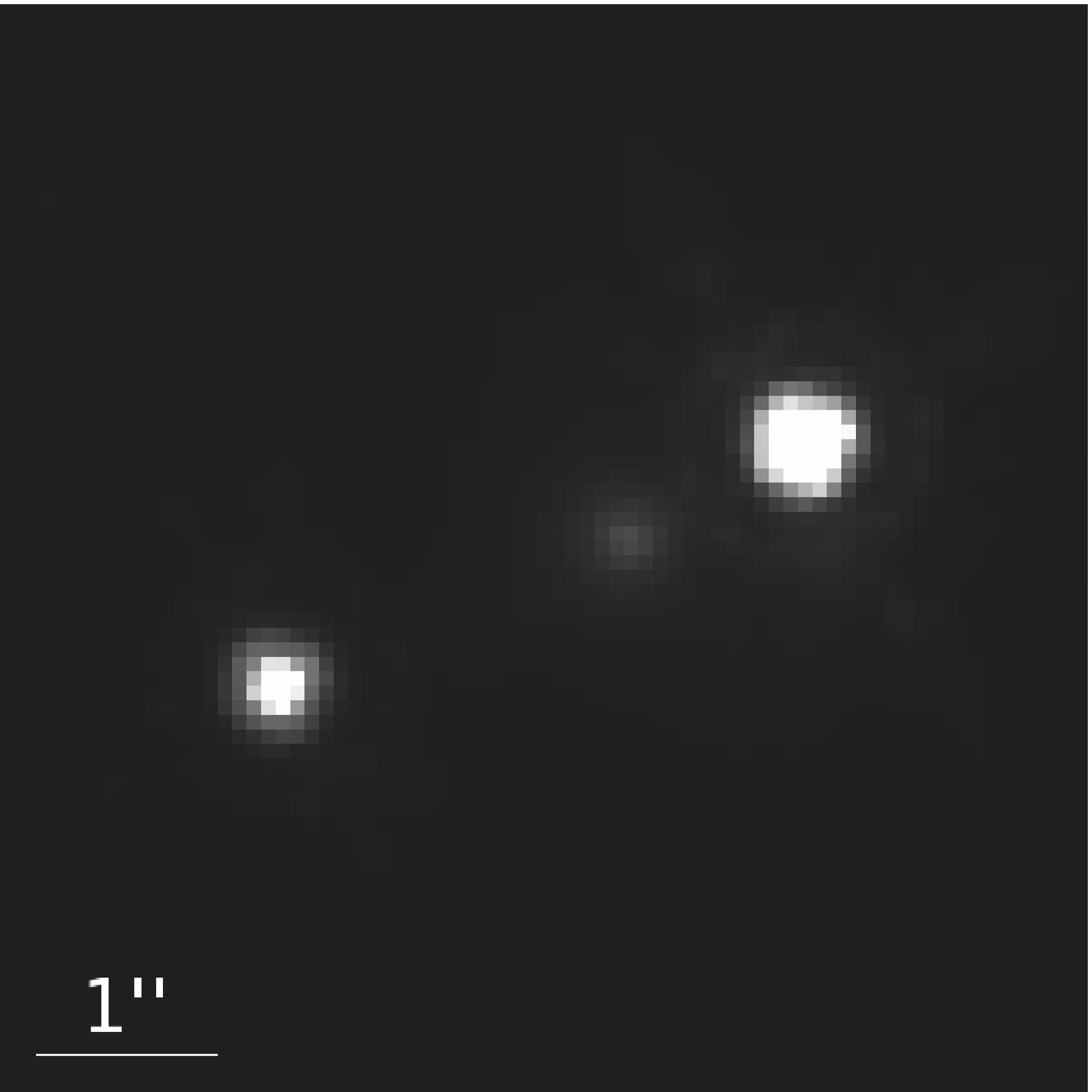}
\\
\includegraphics[width=0.6\columnwidth]{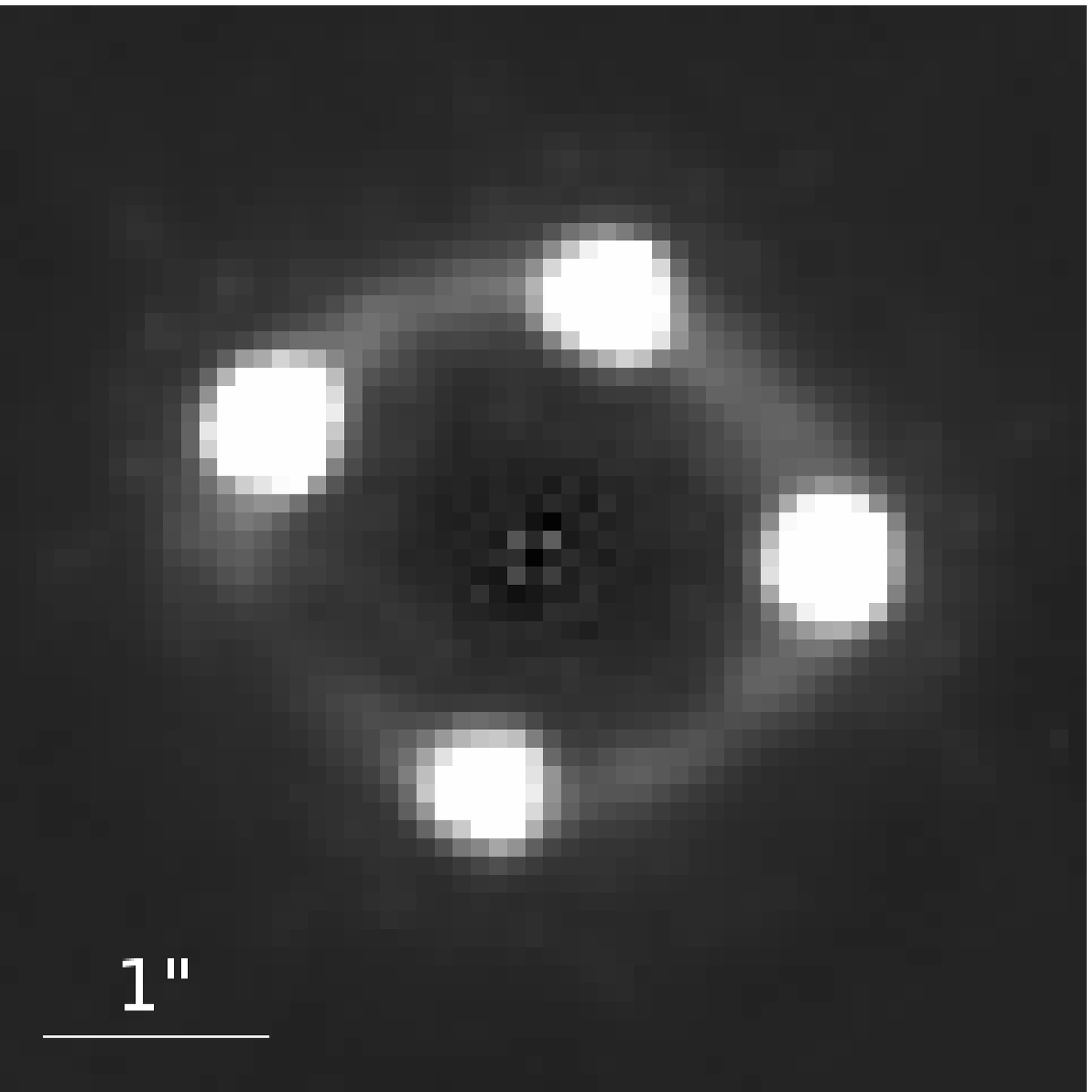}
\includegraphics[width=0.6\columnwidth]{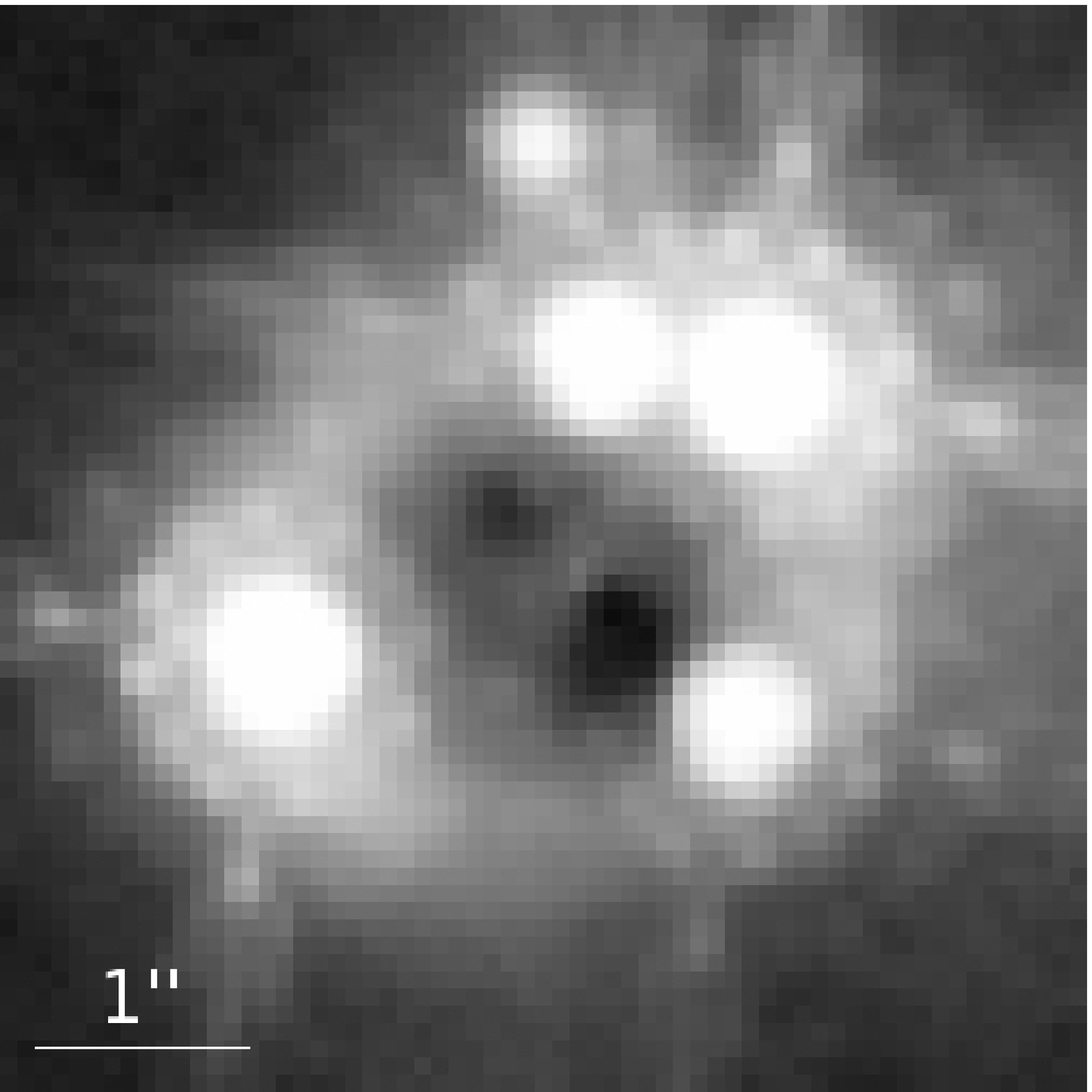}
\includegraphics[width=0.6\columnwidth]{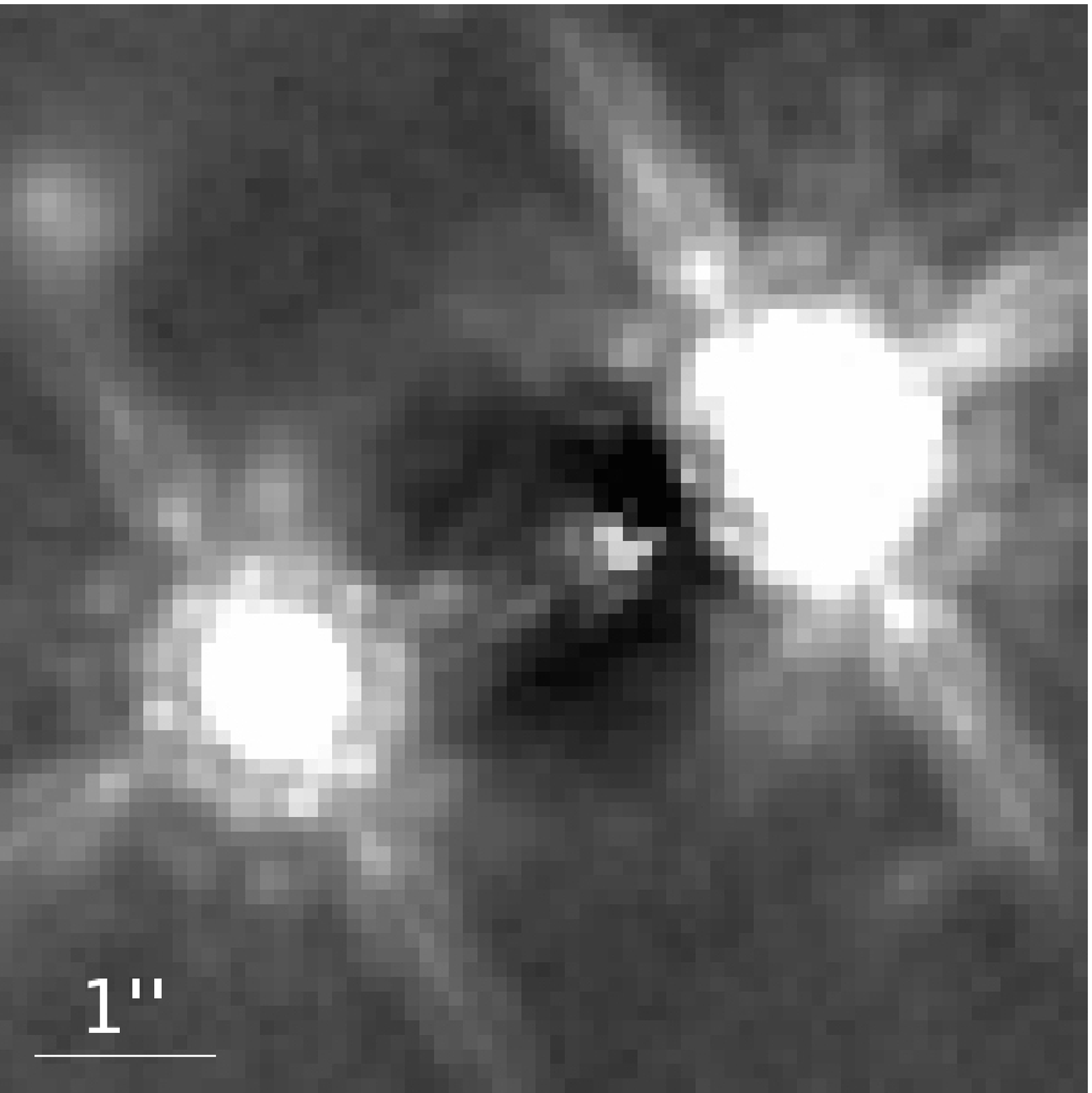}
  \caption{\hst\ WFC3 F160W observation of \hequad, \wfilens\ and
    \hedouble\ from left to right in the top panels.  In the bottom
    panels, the lens-galaxy light has been subtracted,
    revealing the Einstein ring of the AGN host galaxy that is needed
    for accurate and precise lens mass modeling.  The full modeling of
    \hequad\ is detailed in H0LiCOW Paper IV.  The lens subtraction
    for \wfilens\ and \hedouble\ in the bottom-middle and bottom-right
    panels, respectively, is based on an initial PSF model without
    PSF reconstruction (which we defer to future work), hence the
    visible residuals. In each of the
    panels, north is up, and east is left.
}
\label{fig:hstobs}
\end{figure*}

In \fref{fig:hstobs}, we show the reduced \hst\ WFC3 observations of
\hequad, \wfilens\ and \hequad\ in the top panels from left to right.
In the bottom, we show the images with the lens light
subtracted with \GLEE\footnote{A lens modeling software package developed
by A.~Halkola and S.~H.~Suyu \citep{SuyuHalkola10, SuyuEtal12a}},
revealing the Einstein ring of the AGN host galaxy.
In H0LiCOW Paper IV, we detail the modeling of \hequad\ using
multi-lens-plane ray tracing \citep[e.g.,][]{BlandfordNarayan86,
  SchneiderEtal92,BlandfordKochanek04}
 and point-spread-function reconstruction
techniques developed by Suyu et al.~(in preparation).  The subtraction
of lens light in \wfilens\ and \hedouble\ (bottom-middle and
bottom-right panels of \fref{fig:hstobs}, respectively) is based on an
initial point spread function built from stars in the field without
any lens mass modeling or iterative PSF reconstruction, hence the
lens-subtraction residuals.  Furthermore, the lens galaxy of
\hedouble\ is on a diffraction spike of the brighter AGN image -- an
accurate PSF model would be crucial for distinguishing the lens
galaxy, the two AGN images and the lensed host galaxy of the
AGN.
The full modeling and
analysis of \wfilens\ and \hedouble\ will appear in future
publications.

\subsection{Wide-field spectroscopy and imaging of lens environment}
\label{sec:followup:field}

We obtain wide-field spectroscopy to pinpoint the redshifts of
the bright galaxies in the fields of the H0LiCOW lenses, particularly
the ones close to the strong lens.  Redshifts of nearby galaxies,
especially those within a few arcseconds from the strong lens, are
crucial since the external convergence approximation is often
insufficient for these galaxies \citep[e.g.,][]{McCullyEtal14} and
they need to be incorporated directly into the strong lens modeling.
We use the multi-object spectrographs on the Very Large Telescope, the
Gemini Telescope and the W.~M.~Keck Telescope to target our lens
fields, as summarized in \tref{tab:followup:spec}.  The spectroscopic
redshifts and galaxy group identifications are detailed in
\citet{FassnachtEtal06},
H0LiCOW Paper II and forthcoming publications.

\begin{table}
\caption{Wide-field spectroscopy of H0LiCOW lenses as part of the
  H0LiCOW program}
\label{tab:followup:spec}
\begin{center}
\begin{tabular}{l l l l}
\hline
Lens & Facility/instrument & Proposal PI \\
\hline
\blens &  W.~M.~Keck / LRIS & C.~.D.~Fassnacht \\
       &  W.~M.~Keck / ESI & C.~.D.~Fassnacht \\
\rxjlens & W.~M.~Keck / LRIS & C.~.D.~Fassnacht \\
\hequad & W.~M.~Keck / LRIS & C.~D.~Fassnacht \\
  & VLT / FORS2 & D.~Sluse \\
  & Gemini / GMOS & T.~Treu \\
\wfilens &  VLT / FORS2 & D.~Sluse \\
  & Gemini / GMOS & T.~Treu \\
\hedouble & VLT / FORS2 & D.~Sluse \\
  & Gemini / GMOS & T.~Treu \\
\hline
\end{tabular}
\end{center}
\begin{flushleft}
Notes.  Abbreviations are LRIS
\citep[Low Resolution Imaging Spectrometer;][]{OkeEtal95,
  RockosiEtal10}, ESI \citep[Echellete Spectrograph and
Imager;][]{SheinisEtal02},
VLT (Very Large Telescope),
FORS2 \citep[FOcal Reducer and low dispersion Spectrograph;][]{AppenzellerEtal98}, and GMOS \citep[Gemini
Multi-Object Spectrographs;][]{HookEtal04}.  Details of the
observations for \blens\ are in \citet{FassnachtEtal06}, and for the
other four lenses are in
H0LiCOW Paper II and forthcoming publications.
Additional integral field spectroscopy of the central 30\arcmin\ around
\wfilens\ has been recently obtained with the Multi Unit Spectroscopic Explorer
\citep[MUSE;][]{BaconEtal12} on the VLT.
\end{flushleft}
\end{table}

To further characterize the lens environment and determine
$\kext$, we obtain
wide-field multiband imaging using the Canada France Hawaii Telescope,
Subaru Telescope, the Very Large Telescope, Gemini Telescope and
Spitzer Space Telescope.
\tref{tab:followup:img} summarizes the follow-up imaging that allow us
to compute the photometric redshifts of structures along the line of
sight as well as to estimate their stellar masses.  Details of the
observations and inference on $\kext$ are described in
H0LiCOW Paper III and forthcoming publications.

\begin{table*}
\caption{Wide-field imaging obtained as part of the H0LiCOW program}
\label{tab:followup:img}
\begin{center}
\begin{tabular}{l l l l}
\hline
Lens & Facility/instrument & Wavelength bands & Proposal PI \\
\hline
\blens & CFHT / MegaCam & u & S.~H.~Suyu \\
  & Subaru / Suprime-Cam & g, r, i & C.~D.~Fassnacht \\
  & Subaru / MOIRCS & J, H, Ks & C.~D.~Fassnacht \\
  & Gemini / NIRI & J, Ks & C.~D.~Fassnacht \\
  & Spitzer / IRAC & 3.6\,$\mu$m, 4.5\,$\mu$m & C.~E.~Rusu \\
\rxjlens & CFHT / MegaCam & u & S.~H.~Suyu \\
  & Subaru / Suprime-Cam & g, r, i & C.~D.~Fassnacht \\
  & Subaru / MOIRCS & J, H, Ks & C.~D.~Fassnacht \\
  & Gemini / NIRI & J, Ks & C.~D.~Fassnacht \\
\hequad & CFHT / MegaCam & u & S.~H.~Suyu \\
  & Subaru / Suprime-Cam & g, r, i & C.~D.~Fassnacht \\
  & Subaru / MOIRCS & H & C.~D.~Fassnacht \\
  & Gemini / NIRI & J, Ks & C.~D.~Fassnacht \\
\wfilens & CTIO Blanco / DECam & u & C.~E.~Rusu \\
  & VLT / HAWK-I & J, H, K & C.~D.~Fassnacht \\
\hedouble & CFHT / MegaCam & u & S.~H.~Suyu \\
  & Subaru / Suprime-Cam & g, r, i & C.~D.~Fassnacht \\
  & Subaru / MOIRCS & J, H, Ks & C.~D.~Fassnacht \\
  & Gemini / NIRI & J, Ks & C.~D.~Fassnacht \\
\hline
\end{tabular}
\end{center}
\begin{flushleft}
  Notes.
  Abbreviations and references for the instruments are
  CFHT (Canada-France-Hawaii Telescope) MegaCam \citep{BouladeEtal03},
  Suprime-Cam \citep{MiyazakiEtal02},
  MOIRCS \citep[Multi-Object InfraRed
  Camera and Spectrograph;][]{SuzukiEtal08, IchikawaEtal06}, NIRI \citep[Near
  InfraRed Imager and Spectrometer;][]{HodappEtal03}, IRAC
  \citep[Infrared Array Camera;][]{FazioEtal04},  CTIO (Cerro Tololo
  Inter-American
  Observatory) DECam \citep[Dark Energy Camera;][]{DiehlEtal12}, VLT (Very Large
  Telescope) HAWK-I \citep[High Acuity Wide field K-band
  Imager;][]{PirardEtal04, CasaliEtal06, Kissler-PatigEtal08}.  Details
  of the observations are in H0LiCOW Paper III and forthcoming
  publications.  \wfilens\ is in the footprint of the Dark Energy
  Survey with observations in g, r, i, z and Y bands, so we did not
  target \wfilens\ in these bands.  We observed only \blens\ with
  Spitzer since the other four lenses have archival Spitzer/IRAC
  observations (PI: C.~S.~Kochanek).
\end{flushleft}
\end{table*}

\citet[][and in preparation]{WilliamsEtal06} have independently
obtained I and either V or R images of all five H0LiCOW lenses using
the 4-m Cerro Tololo Inter-American Observatory (CTIO) Blanco
telescope for the sourthern fields and the 4-m Kitt Peak National
Observatory (KPNO) Mayall telescope for the nothern fields.  Using
these images to select spectroscopic targets, \citet{MomchevaEtal15}
have obtained spectroscopic observations of the five H0LiCOW lenses
using the 6.5m Magellan telescopes.  In H0LiCOW Paper II, we merge the
spectroscopic catalog from the multiple spectroscopic campaigns
on \hequad.

\subsection{Lens galaxy spectroscopy for lens velocity dispersion}
\label{sec:followup:vdisp}

For \blens\ and \rxjlens, we have obtained long-slit spectra of the
lens systems with the Low-Resolution Imaging Spectrometer
\citep[LRIS;][]{OkeEtal95} at the Keck Observatory for measuring the
lens stellar velocity dispersion \citep{SuyuEtal10, SuyuEtal13}.  For
\hequad, we observe the lens system with LRIS in multi-object mode to
obtain spectra of both the foreground lens galaxy for lens velocity
dispersion measurement (see H0LiCOW Paper IV) and also of nearby
galaxies (see H0LiCOW Paper II).  Both \wfilens\ and \hedouble\ have
bright AGNs relative to the lens galaxy, making the lens velocity
dispersion measurement challenging.  We have new observations of
\wfilens\ with
MUSE \citep{BaconEtal12} at the VLT, which we expect will 
allow us to reduce the
uncertainty on the current lens velocity dispersion by a factor of 2,
to $\sim 5-7\%$ precision.  The velocity dispersion is a key
ingredient to break the MSD/lensing degeneracies
\citep[e.g.,][]{SuyuEtal14}.
For \hedouble, we obtained one-sixth
of our proposed observations with XSHOOTER on the
VLT in priority B, which is not sufficient to measure the velocity dispersion.
We have time on OSIRIS \citep[OH-Suppressing Infra-Red Imaging
Spectrograph;][]{LarkinEtal06} on Keck to observe \hedouble,
\rxjlens, and \hequad\ with adaptive optics.  Because OSIRIS
is an integral field spectrograph, these observations have the goal
of obtaining two-dimensional kinematic
data of the foreground lens, which will then be used to further constrain the lens mass
models.
We summarize the spectroscopic
observations for lens velocity dispersion measurement in
\tref{tab:followup:vdisp}.

\begin{table}
\caption{Spectroscopy of foreground lens as part of the
  H0LiCOW program}
\label{tab:followup:vdisp}
\begin{center}
\begin{tabular}{l l l l}
\hline
Lens & Facility/instrument & Proposal PI \\
\hline
\blens & W.~M.~Keck / LRIS & C.~D.~Fassnacht \\
\rxjlens & W.~M.~Keck / LRIS & C.~D.~Fassnacht \\
        & W.~M.~Keck / OSIRIS & T.~Treu \\
\hequad & W.~M.~Keck / LRIS & C.~D.~Fassnacht \\
  & W.~M.~Keck / OSIRIS & T.~Treu \\
\wfilens &  VLT / MUSE & D.~Sluse \\
\hedouble & VLT / X-shooter & C.~Spiniello \\
  & W.~M.~Keck / OSIRIS & T.~Treu \\
\hline
\end{tabular}
\end{center}
\begin{flushleft}
Notes.
OSIRIS is the OH-Suppressing Infra-Red Imaging
Spectrograph \citep{LarkinEtal06}.
Details of the LRIS
observations for \blens\ are in \citet{SuyuEtal10}, for \rxjlens\ are
in \citet{SuyuEtal13}, and for \hequad\ are in H0LiCOW Paper IV; other
observations are in forthcoming publications.  Only one-sixth of the
\hedouble\ observations with X-shooter \citep{VernetEtal11} were
obtained, which were
insufficient for measuring the lens velocity dispersion.  
The 
observations with OSIRIS are pending.
\end{flushleft}
\end{table}


\section{Cosmography and astrophysics with \hequad: Key components}
\label{sec:papers}

We summarize the key ingredients and analysis of \hequad\ that are
described in upcoming publications of the H0LiCOW project (H0LiCOW
Papers II-V).  The titles of the papers begin with ``H0LiCOW'',
followed by the
specific titles written below.
\\
\indent II. Spectroscopic survey and galaxy-group identification of
the strong gravitational lens systems \hequad\ \citep{SluseEtal16}.
From our spectroscopic campaign of the lens environment, we present
the measured spectroscopic redshifts, focussing in particular on the massive and
nearby objects to the strong lens system that are necessary
ingredients for lens mass modeling and distance measurement.  By
combining with the spectroscopic catalog of independent efforts
\citep{MomchevaEtal15}, we identify potential galaxy groups towards
\hequad\ in order to control the systematic effect due to the
galaxies along the line of sight.  We use the
flexion-shift\footnote{The flexion-shift corresponds to the shift in
  the image positions due to the flexion (third order derivatives of
  the lens potential) of a line-of-sight perturber.
  \citet{McCullyEtal16} find through their study of simulated lens
  fields that perturbers with flexion-shifts larger than $\sim10^{-4}$
  arcseconds should be incorporated
  explicitly in the multi-plane lens mass model.  The threshold of
  $\sim10^{-4}$ arcseconds is conservative and is based on tests that
  only used image positions as constraints.  It may be that using the
  spatially extended images for modeling would push that threshold
  even lower.}  introduced
by \citet{McCullyEtal16} to determine which mass structures
(galaxies/groups) need to be 
incorporated explicitly in the lens mass model and which could be well
approximated by an external shear/convergence field. 
The flexion-shift analysis
presented in H0LiCOW Paper II shows that the most significant line-of-sight
perturber is the galaxy G1 that is closest to the lens system, which
justifies our inclusion of this particular galaxy in all of our strong
lensing models in H0LiCOW Paper IV.
Furthermore, the next four nearest perturbers from the lens system may also
produce higher order perturbations on the lens potential, and we
account for the effects of these four additional galaxies in one of
our systematic tests in H0LiCOW Paper IV.
\\
\indent III. Quantifying the effect of mass along the line of sight to
the gravitational lens \hequad\ through weighted galaxy counts
\citep{RusuEtal16b}.   Using the wide-field photometry and
spectroscopy in \sref{sec:followup:field}, we compute photometric
redshifts and stellar masses for objects in the field up to 120\arcsec\ from
the strong lens, and with i<24 mag.
We thoroughly test the weighted galaxy number
counts technique of \citet{GreeneEtal13}, and apply it to \hequad\
with the CFHTLenS survey \citep{HeymansEtal12} as the control field. By
comparing the weighted counts to simulated lines of sight that are ray
traced through numerical simulations \citep{HilbertEtal07,
  HilbertEtal09}, we infer the distribution for the external
convergence $\kext$ that excludes the strong lens redshift plane.
\\
\indent IV. Lens mass model of \hequad\ and blind measurement of its
time-delay distance for cosmology \citep{WongEtal16}.  Using the time
delays from H0LiCOW Paper V and our \hst/WFC3-IR imaging (F160W) and archival
\hst/ACS observations (F555W and F814W), we model the lens mass
distribution including explicitly the nearest, in projection from
\hequad, one (G1) or five (G1 plus the next four nearest/brightest) 
perturbers, with spectroscopic redshifts from H0LiCOW Paper II.    
We then incorporate
the velocity dispersion of the lens galaxy, and the external
convergence distribution from H0LiCOW Paper III to infer an effective
time-delay distance, which is blinded during the analysis stage.  We
unblind only after the completion of the analysis, and publish these
results without modifications.
\\
\indent V. New COSMOGRAIL time delays of \hequad: $H_0$ to 3.8\% from
strong lensing in flat-$\Lambda$CDM \citep{BonvinEtal16b}.
We present the 13-year monitoring of \hequad\ and measure the
time delays between the image pairs.  Using the resulting effective
time-delay distance of \hequad\ from the blind analysis in H0LiCOW
Paper IV, we create a Time Delay
Strong Lensing (TDSL) probe with \hequad, \rxjlens\ and \blens\ (we
note that the analysis of \rxjlens\ was also blinded in
\citet{SuyuEtal13}, whereas the analysis of \blens\ was not as it was
the first lens to be
analyzed using our modeling techniques). We infer
cosmological constraints from TDSL alone, and combine it with other
cosmological probes to constrain various cosmological models.
\\
\\
In addition to the above, there are more forthcoming publications.
The study of the AGN host galaxy properties
based on simulations are described in H0LiCOW Paper VI \citep{DingEtal16}. 
The newly developed multi-lens plane modeling, based on the multi-lens
plane equations \citep{SchneiderEtal92,BlandfordKochanek04}, and point-spread-function reconstruction will
be detailed by Suyu et al.~(in preparation).  The weak lensing
analysis of the field of \hequad\ will be presented by Tihhonova et
al.~(in preparation).   Following these publications, there will be the next
studies and analysis of the remaining sample (\wfilens\ and
\hedouble).


\section{H0LiCOW cosmographic forecast}
\label{sec:forecast}

We make predictions of the cosmographic constraints based on our
sample of H0LiCOW lenses.  We use the time-delay distance measurements
for \blens\ \citep[equation (35) of][]{SuyuEtal10}, \rxjlens\
\citep[equation (5) of][]{SuyuEtal14} and \hequad\ (equation (17) of
H0LiCOW Paper IV).  For the forecasted
time-delay distance measurements of the other two lenses, we adopt an
uncertainty with contributions from the time
delays, mass modeling and external convergence added in quadrature.
Specifically, we estimate time delay uncertainties of $4\%$ and
$2\%$, modeling uncertainties of $4\%$ and $8\%$, external convergence
uncertainties of $4\%$ and $4\%$,
yielding a total uncertainty of $7\%$ and
$9\%$ for \wfilens\ and \hedouble, respectively.  Furthermore, we
assume that the angular diameter distance to each lens
can be measured with an uncertainty of $15\%$ using our current data sets
\citep{JeeEtal15}.  More precise measurements of $\Dd$
($\sim5-10\%$ uncertainty) would
require additional kinematic data of the lens galaxy beyond what we
currently have, particularly
spatially resolved kinematics maps.  
For the forecasted $\tdist$ and $\Dd$ constraints,
we adopt a fiducial cosmological model with
$H_0=72\,\kmsMpc$, $\Om=1-\Ode=0.32$ and $w=-1$
to predict the distances with their estimated uncertainties mentioned
above, although we note that this assumption affects little the
fractional uncertainty, which is nearly scale-free.

\begin{figure*}
\flushleft \large{\hspace{0.13\columnwidth}Uniform Prior\hspace{0.23\columnwidth}
  WMAP9 Prior\hspace{0.22\columnwidth} Planck Prior
\hspace{0.25\columnwidth} Marginalized $H_0$} \\

\includegraphics[width=0.48\columnwidth]{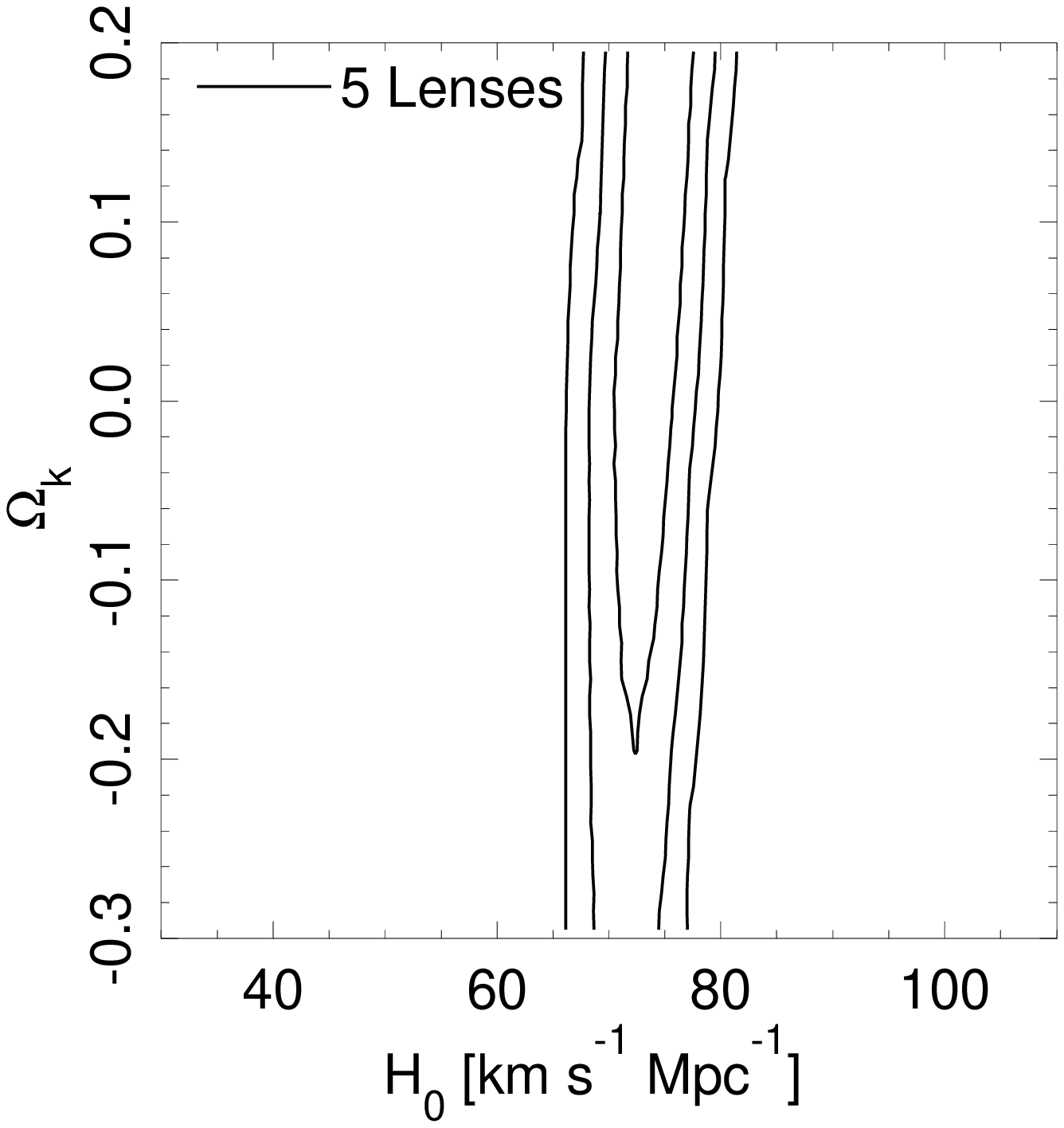}
\includegraphics[width=0.48\columnwidth]{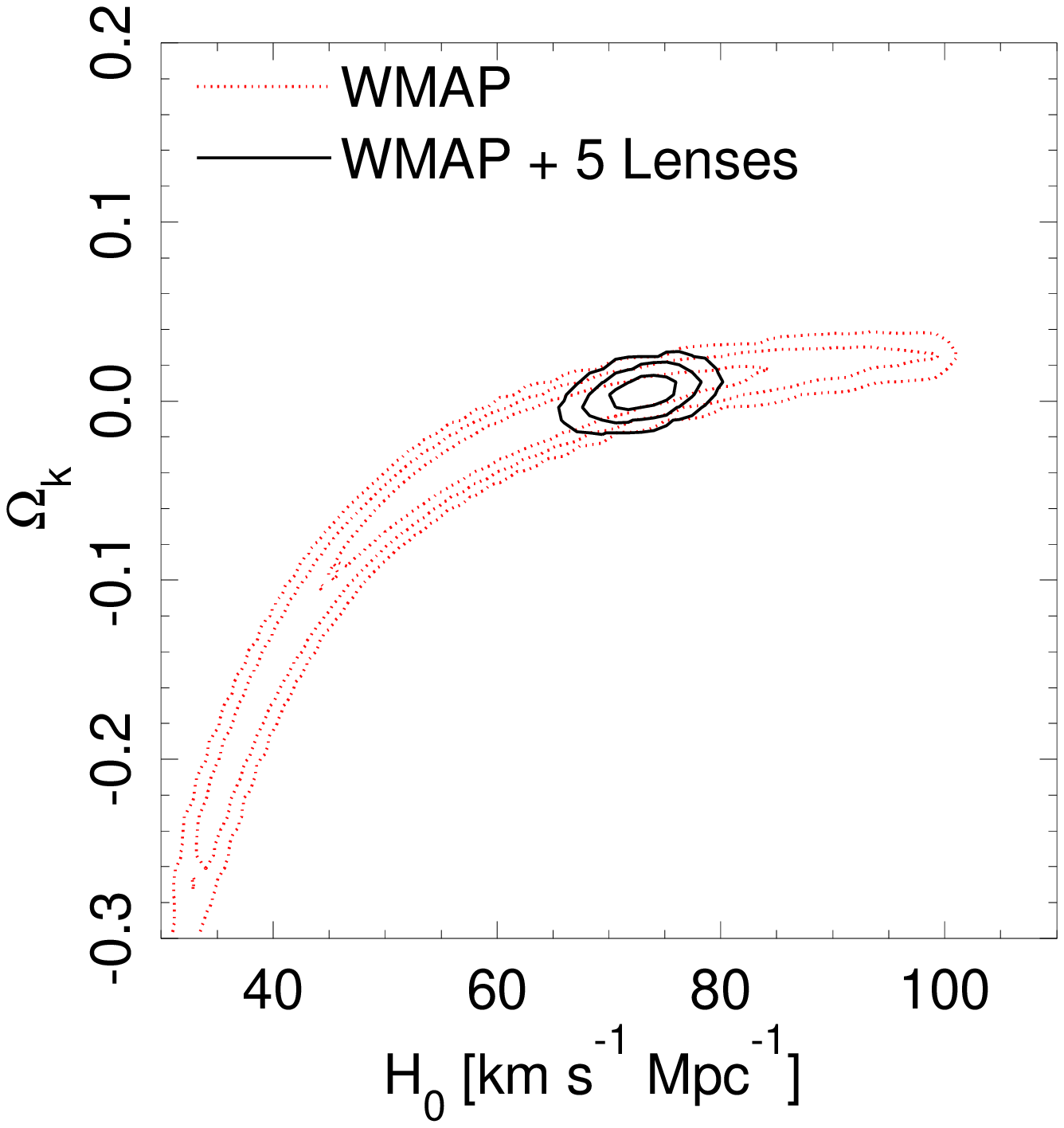}
\includegraphics[width=0.48\columnwidth]{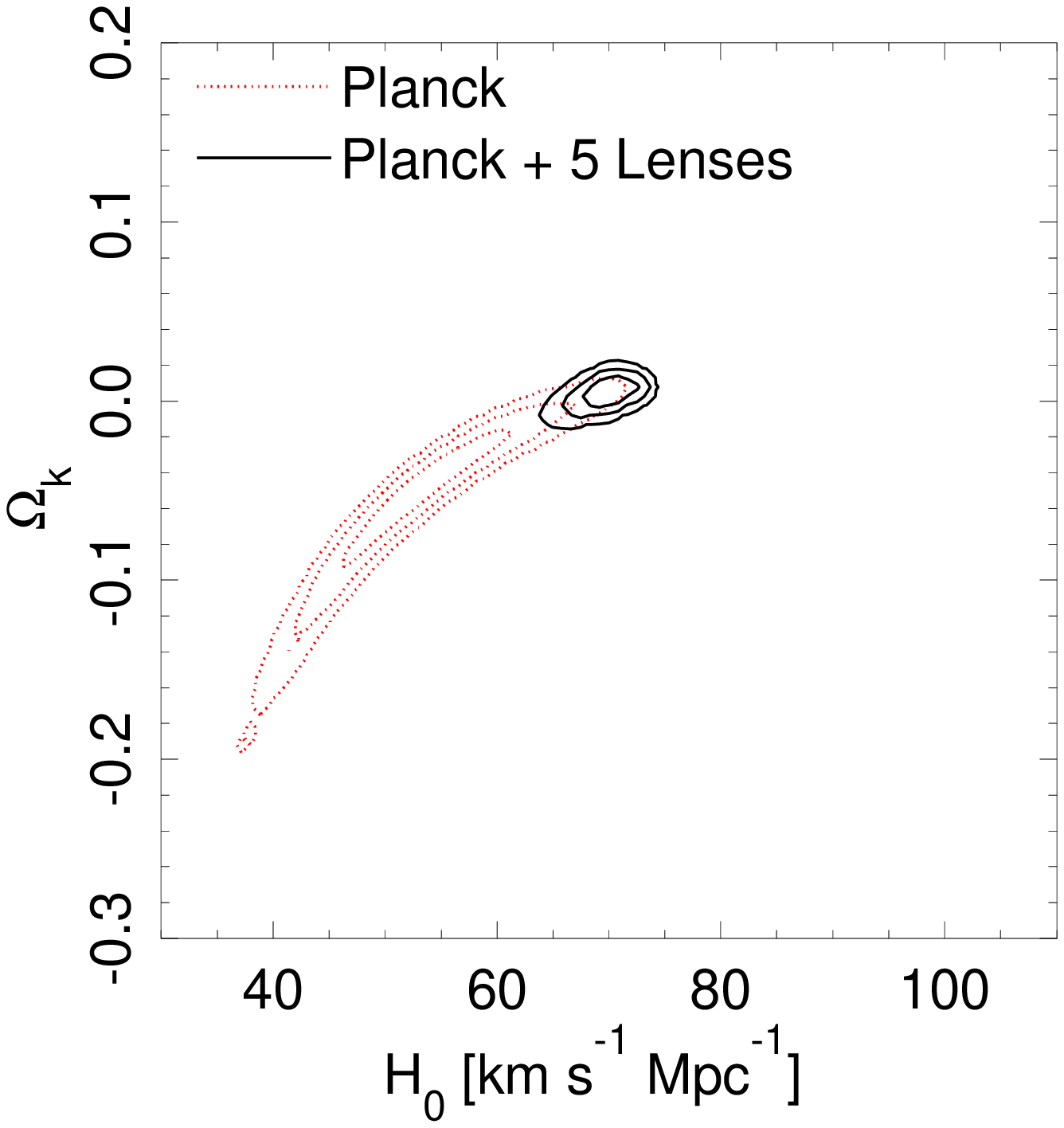}
\includegraphics[width=0.58\columnwidth]{fig3d.ps}

\includegraphics[width=0.48\columnwidth]{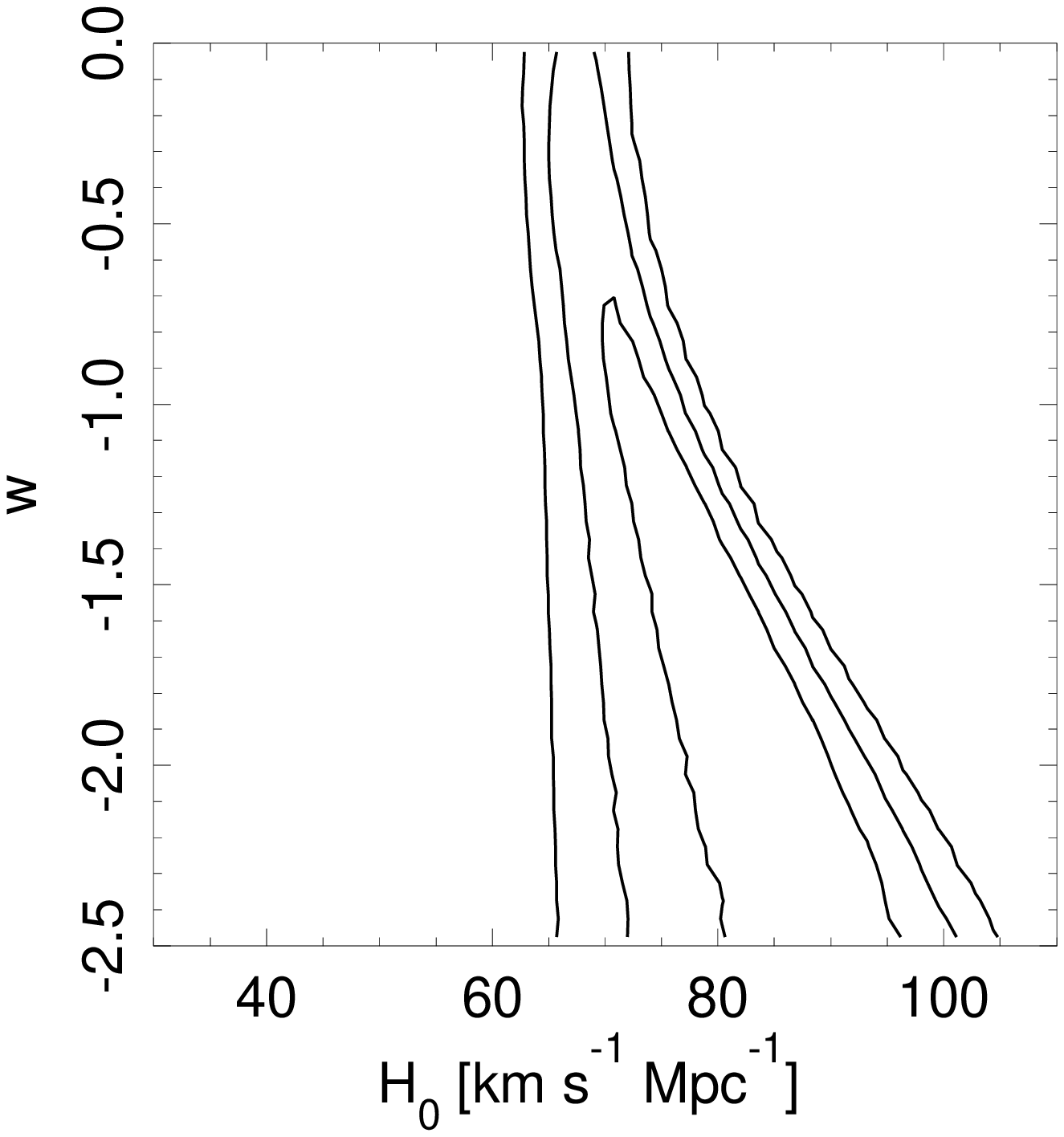}
\includegraphics[width=0.48\columnwidth]{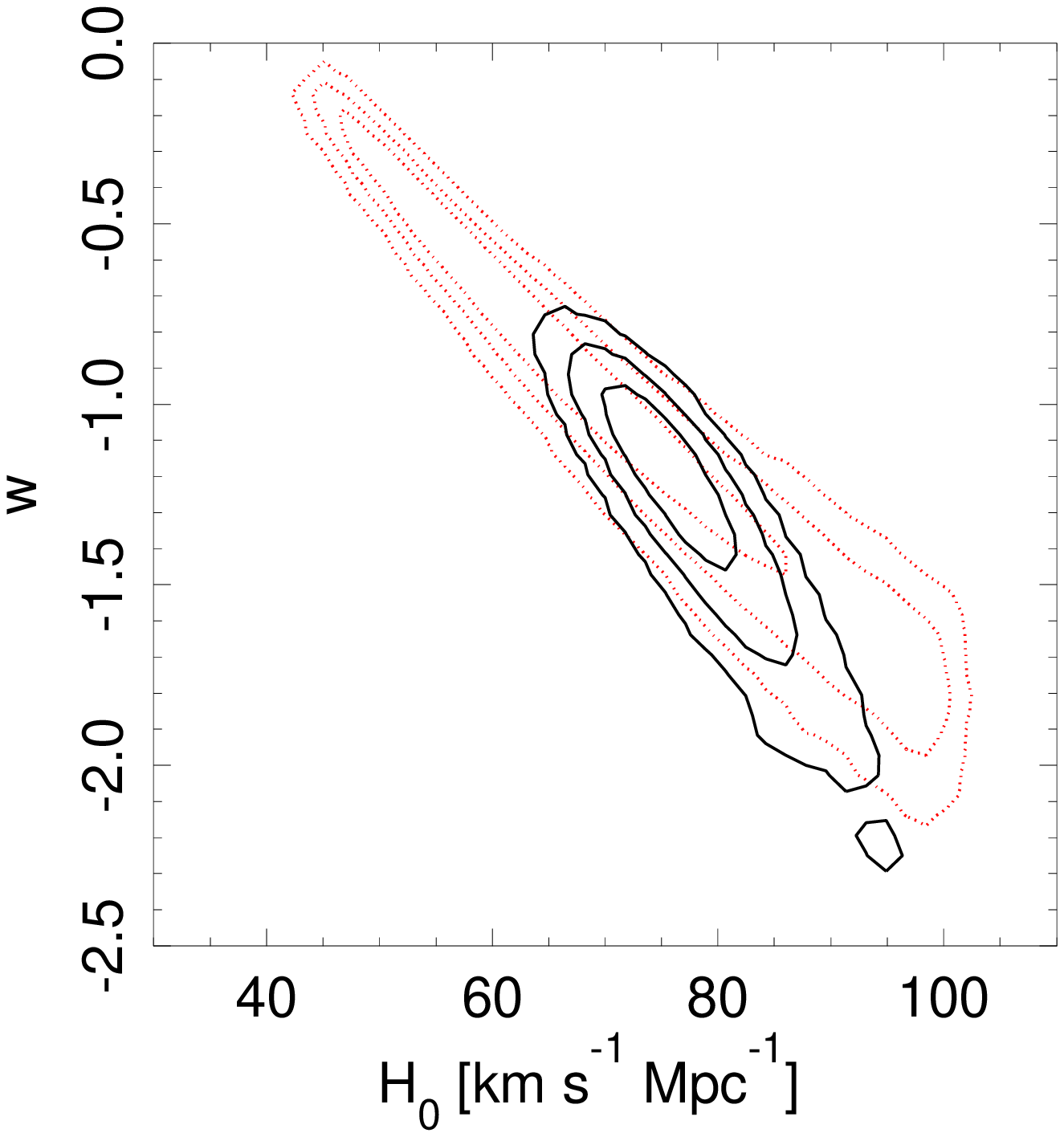}
\includegraphics[width=0.48\columnwidth]{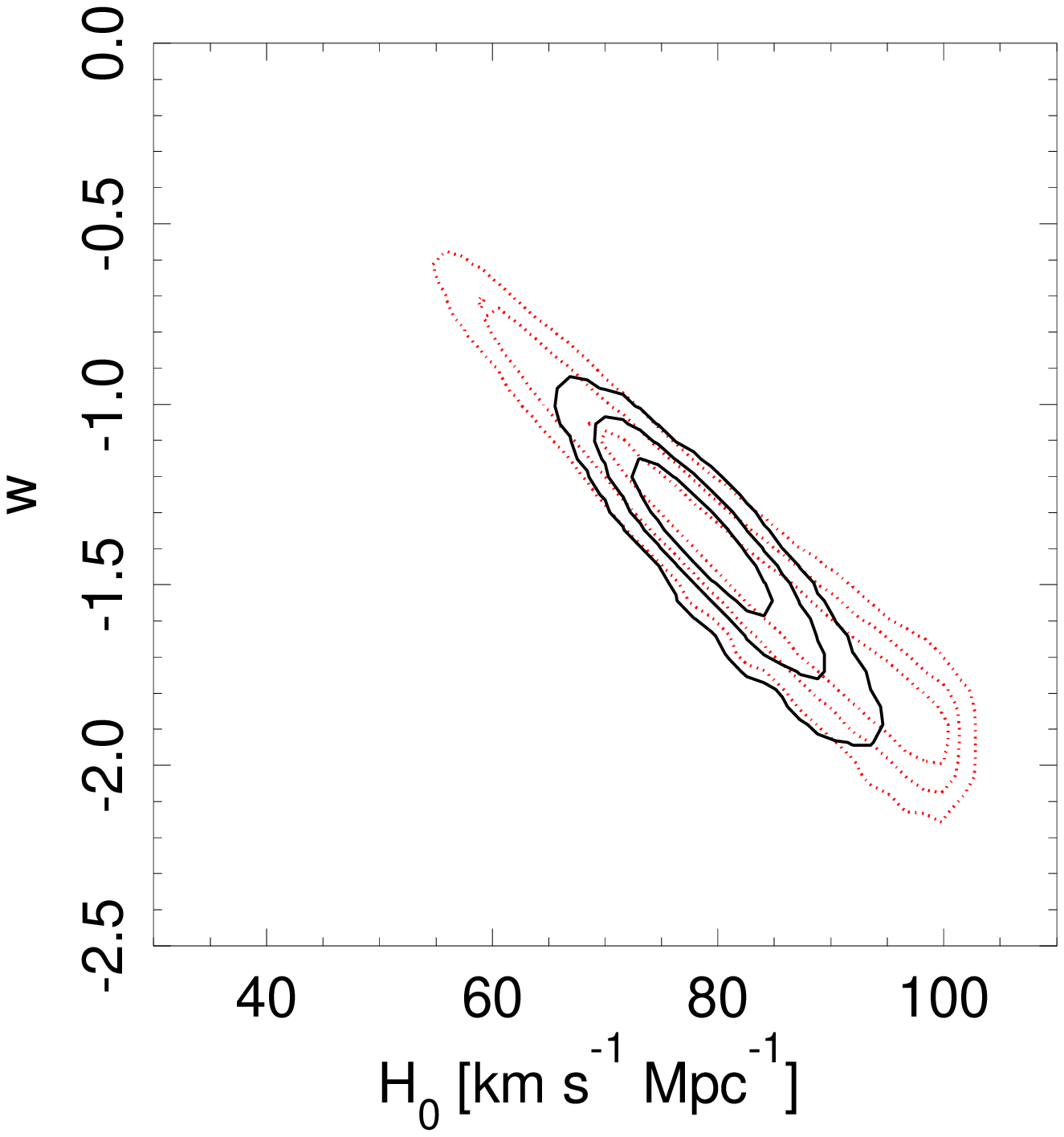}
\includegraphics[width=0.58\columnwidth]{fig3h.ps}

\includegraphics[width=0.48\columnwidth]{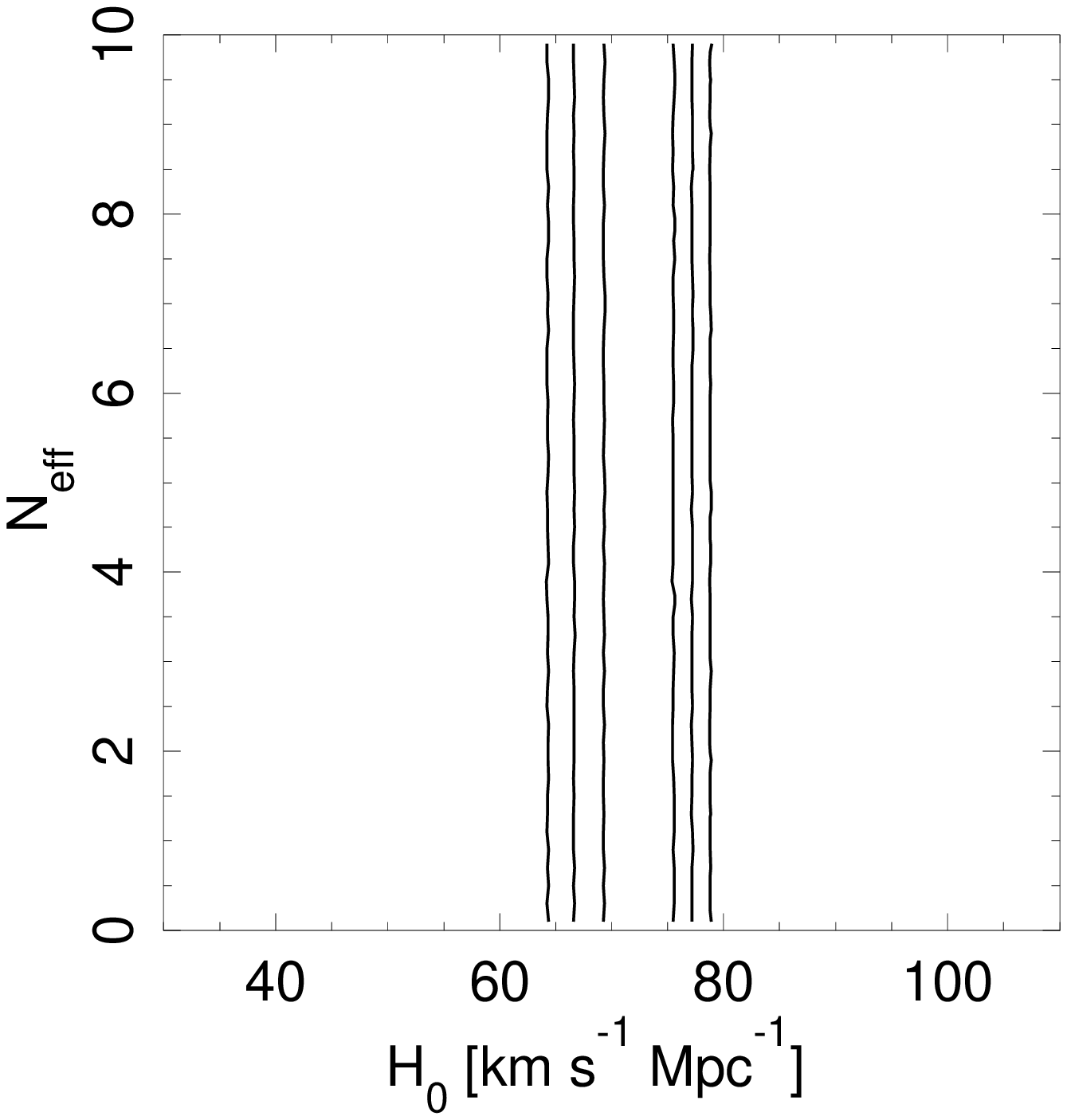}
\includegraphics[width=0.48\columnwidth]{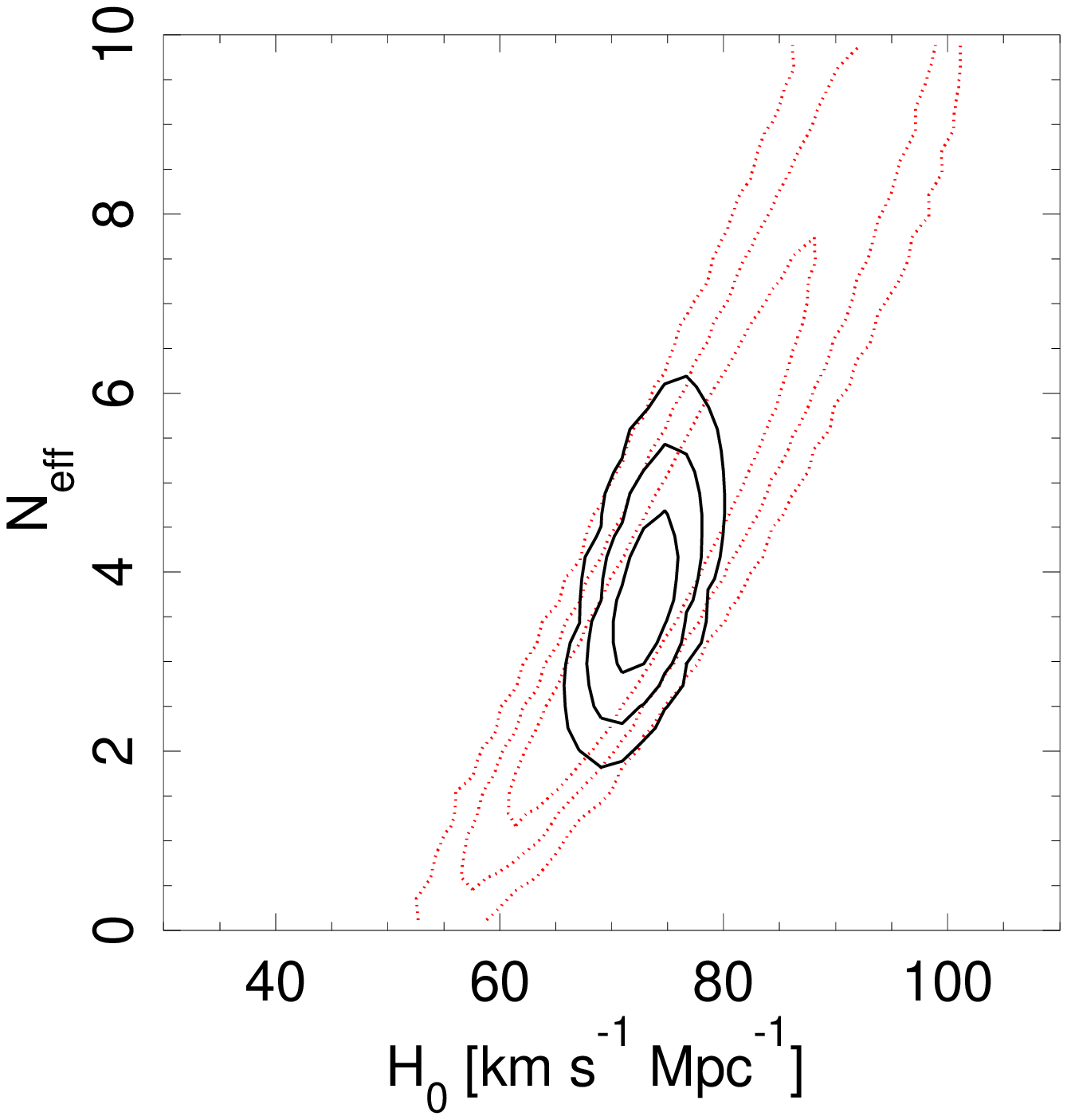}
\includegraphics[width=0.48\columnwidth]{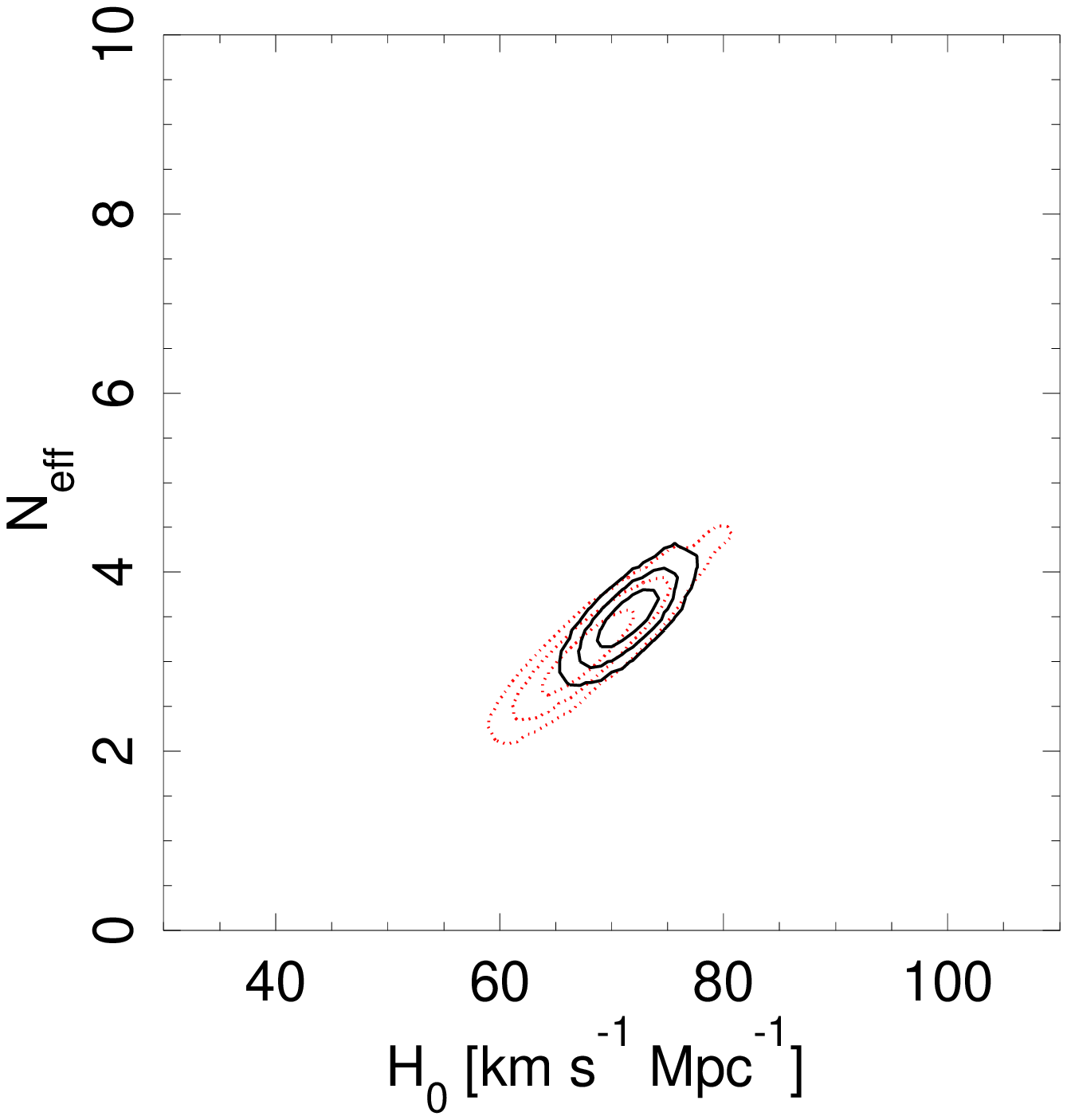}
\includegraphics[width=0.58\columnwidth]{fig3l.ps}

\caption{Forecasted cosmographic constraints from the H0LiCOW lens
  sample through measurements of $\tdist$ and $\Dd$.  Columns from left to right are, respectively, the
  constraints from the
  H0LiCOW lenses alone (with uniform prior on cosmological
  parameters), lenses in combination with WMAP 9-year
  results, lenses in combination with Planck 2015 results, and
  marginalized constraints on $H_0$ from the previous three columns.  The
  H0LiCOW lenses primarily constrain $H_0$, which in turn break CMB
  parameter degenercies to elucidate the spatial curvature of
  universe ($\Ok$, top row), dark energy equation of state ($w$, middle
  row) and effective number of relativistic species ($\Neff$, bottom
  row).  H0LiCOW lenses provide an independent, complementary and
  competitive probe of cosmology. }
\label{fig:cosmoforecast}
\end{figure*}

We show in \fref{fig:cosmoforecast} the cosmographic constraints of
our sample of lenses with uniform priors on the cosmological
parameters (left-column panels), in combination with WMAP 9-year results
\citep[][middle-left-column panels]{HinshawEtal12}, and in combination with
Planck 2015 results \citep[][middle-right-column
panels]{Planck2015P13}\footnote{we use the Planck chains designated by
``plikHM\_TT\_lowTEB'' that uses the baseline high-L Planck power
spectra and low-L temperature and LFI polarization} for three
different background cosmologies: (1) open $\Lambda$CDM with variable
spatial curvature $\Ok$ (top row), (2) spatially flat
$w$CDM with $w$ as the time-independent dark energy equation of state
(middle row), and (3) flat $\Lambda$CDM with varying effective
number of relativistic species $\Neff$ (bottom row).  In the
right-column panels, we show the one-dimensional marginalized
constraints of $H_0$ of our sample of lenses alone or in combination
with the CMB data sets (i.e., marginalized $H_0$ distributions of the
panels to the left), as indicated in the legend.  
We list in
\tref{tab:Ucosmo} the prior ranges for the uniform background
cosmologies.  The WMAP 9-year and Planck chains have a prior with
$H_0<100\,\kmsMpc$ imposed. 
The cosmographic constraints
of our lenses shown in \fref{fig:cosmoforecast} (from the forecasted
measurements of $\tdist$ and $\Dd$) mostly stem from the $\tdist$
measurements as a results of the substantially smaller uncertainties
of $\tdist$ than that of $\Dd$.  In fact, the cosmographic constraints
from $\tdist$ alone would increase the $H_0$ 1$\sigma$ uncertainties shown in
\fref{fig:cosmoforecast} by at most $0.8\,\kmsMpc$ (depending on the
background cosmology).  The additional cosmographic
information from $\Dd$ would become more significant when the $\Dd$
uncertainties are reduced to $\sim5-10\%$ \citep{JeeEtal16}.  

\begin{table}
\caption{Prior for ``uniform'' cosmological models}
\label{tab:Ucosmo}
\begin{center}
\begin{tabular}{l l}
\hline
Cosmology & Prior ranges \\
\hline
open $\Lambda$CDM  & $H_0 \in [0,120]\,\kmsMpc$  \\
    & $\Om \in [0,0.5]$    \\
    & $\OL \in [0.5,1]$ \\
    & $\Ok = 1-\Om-\OL$ \\
\hline
flat $w$CDM & $H_0 \in [0,120]\,\kmsMpc$   \\
    & $\Om \in [0,1]$ \\
    & $\Ode = 1 - \Om$ \\
    & $w \in [-2.5,0]$ \\
\hline
flat $\Neff\Lambda$CDM & $H_0 \in [0,120]\,\kmsMpc$  \\
    & $\Om \in [0,1]$ \\
    & $\OL = 1 - \Om $ \\
    & $\Neff \in [0,10]$ \\
\hline
\end{tabular}
\end{center}
\begin{flushleft}
\end{flushleft}
\end{table}

As seen in the left column, the time-delay lenses primarily constrain
$H_0$, and depend weakly (if at all) on other parameters.  Nonetheless, the
time-delay distances $\tdist$ and the lenses' angular
diameter distances $\Dd$ provide some information on $w$, as the
constraint contours are tilted rather than being vertical.  With more
lenses or smaller uncertainties on $\Dd$ measurements, the
constraints on cosmology become more prominent 
\citep{JeeEtal16}.  However, the H0LiCOW lenses provide strong cosmographic
constraints when combined with the CMB measurements since they help to
break parameter degeneracies in the CMB.  Thus, we should be able to
place substantially better constraints on, for example, the spatial
curvature, $w$ and $\Neff$ (middle two columns), compared to
constraints from CMB alone.  In particular, we expect better than
3.5\% precision on $H_0$ for the two cosmologies with $w=-1$ (open $\Lambda$CDM
and flat $\Neff\Lambda$CDM)\footnote{relative to $H_0=72\,\kmsMpc$};
when $w$ is allowed to vary, this constraint weakens to $\sim11\%$
without CMB priors and $\sim5\%$ with CMB priors in the $w$CDM
cosmology, as visible in the right-most panel in the middle row.  By combining our five
H0LiCOW lenses with Planck, we expect
to achieve the following precisions:  $\Ok$ to 0.004 in open $\Lambda$CDM, $w$ to
0.14 in flat wCDM, and $\Neff$ to 0.2 in flat $\Neff\Lambda$CDM (all
1$\sigma$ uncertainties).  These precisions are a factor of $\sim15$, $\sim2$ and
$\sim1.5$, respectively tighter than Planck on its own.  Our H0LiCOW
sample provides not only an independent check of systematics, but also
a great complement to other cosmological probes for pinning down
cosmological parameters.

\section{Summary and outlook}
\label{sec:summary}

We present the H0LiCOW program that aims to measure $H_0$ to $<3.5\%$
in precision and accuracy (in most background cosmological models)
with a sample of five time-delay lenses, completely independent of
the cosmic distance ladder and other direct
measurements of $H_0$.  Our cosmographic information comes from
measuring the distances to the lens systems, specifically $\tdist$ and
$\Dd$.

To achieve our goal, we have obtained almost all the 
key ingredients for our lens sample\footnote{with spectroscopic observations of \hedouble\ pending for lens velocity dispersion measurement}: (1) the
time delays from the COSMOGRAIL and Very Large Array monitoring, (2)
high-resolution $\hst$ imaging for
modeling the lens mass distributions, (3) wide-field imaging and
spectroscopy to quantify the effects of the lens environment, and (4)
lens velocity dispersion measurements to augment our lensing mass
models.  Our new \hst\ observations reveal Einstein rings in the lens
systems that allow us to perform precision lens mass modeling.

The results of our recent blind analysis of \hequad\ will appear in
the companion H0LiCOW publications.  H0LiCOW Paper II
\citep{SluseEtal16} presents the spectroscopic campaign on the
\hequad\ field 
and identifies galaxy groups in the light cone containing the lens.  H0LiCOW Paper III
\citep{RusuEtal16b} combines the spectroscopy, the wide-field
imaging data, and the Millennium Simulation to derive the
external convergence of the line-of-sight mass distributions.  H0LiCOW
Paper
IV \citep{WongEtal16} models the lens mass distribution using
the \hst\ data, the time delays and the lens velocity dispersion to
infer the time-delay distance, that is blinded throughout the
analysis.  H0LiCOW Paper V \citep{BonvinEtal16b} presents the
COSMOGRAIL monitoring of \hequad\ and 
investigates the cosmological implications based on the three lenses
(\blens, \rxjlens, \hequad) that we have so far analyzed.

With our sample of five lenses, we expect to measure $H_0$ to $<3.5\%$
in precision and accuracy for the non-flat $\Lambda$CDM cosmology or flat
$\Neff\Lambda$CDM cosmology, with $w=-1$.  When $w$ is allowed to vary, the
constraint on $H_0$ degrades to $\sim11\%$ with time-delay data only,
and to $\sim5\%$ when augmented with CMB data.  Our independent
strong-lensing distances significantly improve cosmological
constraints from the Planck data: the precisions on $\Ok$, $w$ and
$\Neff$ improve by a factor of $\sim15$, $\sim2$ and $1.5$
respectively when we combine our lenses with Planck.  Time-delay
lenses are therefore highly complementary to other cosmological probes.

Our data set provides an excellent opportunity to study, in addition
to cosmography, galaxy formation and evolution.  For example, we can
study the distribution of dark matter in the lens galaxies by
combining lensing and kinematics data, and also infer the stellar mass
of the lens galaxies \citep[e.g.,][]{TreuKoopmans04, BarnabeEtal11,
  SuyuEtal12a, SonnenfeldEtal12, SonnenfeldEtal15}.  By separately
determining the stellar mass based on either (1) stellar population
synthesis using multiband photometry \citep[e.g.,][]{AugerEtal09,
  TreuEtal10, OguriEtal14}, or (2) identification/characterization of
spectral features \citep[e.g.,][]{vanDokkumConroy10,
  ConroyVanDokkum12, BarnabeEtal13, SpinielloEtal12, SpinielloEtal14,
  SpinielloEtal15}, and comparing this stellar mass to that obtained from
lensing and dynamics, we can study properties of the stellar
population and infer the stellar initial-mass-function (IMF) slope
\citep[e.g.,][]{GrilloEtal09, AugerEtal10, TreuEtal10,
  SpinielloEtal11, BarnabeEtal13, SpinielloEtal15}.
There are about a dozen early-type lens galaxies that have been
studied in detail 
for constraining the stellar IMF slope individually
\citep[e.g.,][]{SonnenfeldEtal12, BarnabeEtal13, SpinielloEtal15,
  NewmanEtal16}, and these galaxies are all at redshifts below 0.35.
Four of our H0LiCOW lens galaxies are at redshifts between 0.45 and
0.73, which would allow us to explore the stellar IMF with comparable
precisions per lens galaxy as previous studies, but at substantially
higher redshifts.  Given the current tension in the IMF measurement
between nearby ($\zd<0.06$) lens galaxies and $\zd\sim0.2-0.3$ lens
galaxies \citep[e.g.,][]{SmithLucey13, NewmanEtal16}, our H0LiCOW lenses
would help assess whether the tensions are just limited to those
particular objects or if they reflect a more general problem in our
understanding of stellar populations.
In addition, our lenses are natural telescopes that magnify the
background sources, allowing us to study the host galaxies of the AGNs
in detail and probe the origin of the co-evolution between
supermassive black holes and their host galaxies
\citep[][]{PengEtal06, RusuEtal16, DingEtal16}.

Our H0LiCOW program aims to establish gravitational lens time delays as an
independent and competitive probe of cosmology, and paves the way for
determining $H_0$ to 1\% in the future.  Given the hundreds, if not
thousands, of time-delay lens systems that are expected to be
discovered in ongoing and future surveys such as the Sloan Digital Sky
Survey \citep[e.g.,][]{OguriEtal06, InadaEtal12, MoreEtal16}, the Dark
Energy Survey \citep[e.g.,][]{AgnelloEtal15}, the Hyper Suprime-Cam
Survey \citep[e.g.,][]{ChanEtal16}, the Kilo-Degree Survey
\citep[e.g.,][]{NapolitanoEtal15}, Euclid and the Large Synoptic
Survey Telescope \citep{OguriMarshall10}, and continuous advances in
high-resolution imaging and spectroscopy in the current and next
generation of telescopes for observational follow-up
\citep{MengEtal15, Linder15}, the H0LiCOW program will provide the basis
for extracting cosmological information from the wealth of strong
lensing data sets.  In particular, we expect the combination of
facilities at different wavelengths such as the \hst\ in the
optical/near-infrared, James Webb Space Telescope in the infrared,
large and extremely large telescopes with adaptive optics,
the Atacama Large Millimeter/submilliter Array in the sub mm waveband,
and the Square Kilometer Array in the radio,
will be of great synergistic value for studying these fruitful lenses.

\section*{Acknowledgments}

We thank M.~Barnab{\`e}, J.~Chan, Y.~Hezaveh, E.~Komatsu, E.~Linder, J.~McKean,
D.~Paraficz, P.~Schneider, and S.~Vegetti for helpful discussions, and
the anonymous referee for detailed comments that improved the
presentation of this work.  
H0LiCOW and COSMOGRAIL are made possible thanks to the continuous work
of all observers and technical staff obtaining the monitoring
observations, in particular at the Swiss Euler telescope at La Silla
Observatory. Euler is supported by the Swiss National Science
Foundation.
S.H.S.~is supported by the Max Planck Society through the Max
Planck Research Group.  This work is supported in part by the Ministry
of Science and Technology in Taiwan via grant
MOST-103-2112-M-001-003-MY3.
V.B., F.C.~and G.M.~acknowledge the support of the Swiss National
Science Foundation (SNSF).
C.D.F.~and C.E.R.~were funded through the NSF grant AST-1312329,
``Collaborative Research: Accurate cosmology with strong gravitational
lens time delays,''
and C.D.F.~and N.R.~were funded by the HST grant GO-12889.
D.S.~acknowledges funding support from a {\it {Back to Belgium}} grant from the Belgian Federal Science Policy (BELSPO).
T.T.~thanks the Packard Foundation for generous support through a Packard Research Fellowship, the NSF for funding through NSF grant AST-1450141, ``Collaborative Research: Accurate cosmology with strong gravitational lens time delays".
K.C.W.~is supported by an EACOA Fellowship awarded by the East Asia Core
Observatories Association, which consists of the Academia Sinica
Institute of Astronomy and Astrophysics, the National Astronomical
Observatory of Japan, the National Astronomical Observatories of the
Chinese Academy of Sciences, and the Korea Astronomy and Space Science
Institute.
X.D.~is supported by the China Scholarship Council.
S.H.~acknowledges support by the DFG cluster of excellence \lq{}Origin and Structure of the Universe\rq{} (\href{http://www.universe-cluster.de}{\texttt{www.universe-cluster.de}}).
P.J.M.~acknowledges support from the U.S.\ Department of Energy under
contract number DE-AC02-76SF00515.
M.T.~acknowledges support by a fellowship of the Alexander von Humboldt
Foundation and the DFG grant Hi 1495/2-1.
L.V.E.K.~is supported in part through an NWO-VICI career grant (project number
639.043.308).
%
Based on observations made with the NASA/ESA Hubble Space Telescope,
obtained at the Space Telescope Science Institute, which is operated
by the Association of Universities for Research in Astronomy, Inc.,
under NASA contract NAS 5-26555. These observations are associated
with programs \#12889, \#10158, \#7422 and \#9744.  Support for program
\#12889 was provided by NASA through a grant from the Space Telescope
Science Institute, which is operated by the Association of
Universities for Research in Astronomy, Inc., under NASA contract NAS
5-26555.
Some of the data presented herein were obtained at the W.M. Keck
Observatory, which is operated as a scientific partnership among the
California Institute of Technology, the University of California and
the National Aeronautics and Space Administration. The Observatory was
made possible by the generous financial support of the W.M. Keck
Foundation.
The authors wish to recognize and acknowledge the very significant
cultural role and reverence that the summit of Mauna Kea has always
had within the indigenous Hawaiian community.  We are most fortunate
to have the opportunity to conduct observations from this mountain.
Access to the CFHT was made possible by the Institute of Astronomy and
Astrophysics, Academia Sinica, National Tsing Hua University, and
National Science Council, Taiwan.
Based in part on data collected at Subaru Telescope, which is
operated by the National Astronomical Observatory of Japan.
Based on observations obtained at the Gemini Observatory, which is
operated by the Association of Universities for Research in Astronomy,
Inc., under a cooperative agreement with the NSF on behalf of the
Gemini partnership: the National Science Foundation (United States),
the National Research Council (Canada), CONICYT (Chile), Ministerio de
Ciencia, Tecnolog\'{i}a e Innovaci\'{o}n Productiva (Argentina), and
Minist\'{e}rio da Ci\^{e}ncia, Tecnologia e Inova\c{c}\~{a}o
(Brazil).  These observations are associated with programs
GN-2012B-Q-11, GN-2013A-Q-72, 
GS-2013A-Q-2 and GS-2013B-Q-28. 
Based on observations made with ESO Telescopes at the La Silla Paranal
Observatory under programme ID
090.A-0531(A), 
P91.B-0346(B), 
091.A-0642(A), 092.A-0515(A), 092.A-0515(B), 60.A-9306(A),
and 097.A-0454(A). 
This work is based in part on observations and archival data obtained
with the Spitzer
Space Telescope, which is operated by the Jet Propulsion Laboratory,
California Institute of Technology under a contract with NASA.
    This project used data obtained with the Dark Energy Camera (DECam), which was constructed by the Dark Energy Survey (DES) collaboration.
Funding for the DES Projects has been provided by
the U.S. Department of Energy,
the U.S. National Science Foundation,
the Ministry of Science and Education of Spain,
the Science and Technology Facilities Council of the United Kingdom,
the Higher Education Funding Council for England,
the National Center for Supercomputing Applications at the University of Illinois at Urbana-Champaign,
the Kavli Institute of Cosmological Physics at the University of Chicago,
the Center for Cosmology and Astro-Particle Physics at the Ohio State University,
the Mitchell Institute for Fundamental Physics and Astronomy at Texas A\&M University,
Financiadora de Estudos e Projetos, Funda{\c c}{\~a}o Carlos Chagas Filho de Amparo {\`a} Pesquisa do Estado do Rio de Janeiro,
Conselho Nacional de Desenvolvimento Cient{\'i}fico e Tecnol{\'o}gico and the Minist{\'e}rio da Ci{\^e}ncia, Tecnologia e Inovac{\~a}o,
the Deutsche Forschungsgemeinschaft,
and the Collaborating Institutions in the Dark Energy Survey.
The Collaborating Institutions are
Argonne National Laboratory,
the University of California at Santa Cruz,
the University of Cambridge,
Centro de Investigaciones En{\'e}rgeticas, Medioambientales y Tecnol{\'o}gicas-Madrid,
the University of Chicago,
University College London,
the DES-Brazil Consortium,
the University of Edinburgh,
the Eidgen{\"o}ssische Technische Hoch\-schule (ETH) Z{\"u}rich,
Fermi National Accelerator Laboratory,
the University of Illinois at Urbana-Champaign,
the Institut de Ci{\`e}ncies de l'Espai (IEEC/CSIC),
the Institut de F{\'i}sica d'Altes Energies,
Lawrence Berkeley National Laboratory,
the Ludwig-Maximilians Universit{\"a}t M{\"u}nchen and the associated Excellence Cluster Universe,
the University of Michigan,
{the} National Optical Astronomy Observatory,
the University of Nottingham,
the Ohio State University,
the University of Pennsylvania,
the University of Portsmouth,
SLAC National Accelerator Laboratory,
Stanford University,
the University of Sussex,
and Texas A\&M University.
%


\bibliography{ms}

\begin{thebibliography}{}
\makeatletter
\relax
\def\mn@urlcharsother{\let\do\@makeother \do\$\do\&\do\#\do\^\do\_\do\%\do\~}
\def\mn@doi{\begingroup\mn@urlcharsother \@ifnextchar [ {\mn@doi@}
  {\mn@doi@[]}}
\def\mn@doi@[#1]#2{\def\@tempa{#1}\ifx\@tempa\@empty \href
  {http://dx.doi.org/#2} {doi:#2}\else \href {http://dx.doi.org/#2} {#1}\fi
  \endgroup}
\def\mn@eprint#1#2{\mn@eprint@#1:#2::\@nil}
\def\mn@eprint@arXiv#1{\href {http://arxiv.org/abs/#1} {{\tt arXiv:#1}}}
\def\mn@eprint@dblp#1{\href {http://dblp.uni-trier.de/rec/bibtex/#1.xml}
  {dblp:#1}}
\def\mn@eprint@#1:#2:#3:#4\@nil{\def\@tempa {#1}\def\@tempb {#2}\def\@tempc
  {#3}\ifx \@tempc \@empty \let \@tempc \@tempb \let \@tempb \@tempa \fi \ifx
  \@tempb \@empty \def\@tempb {arXiv}\fi \@ifundefined
  {mn@eprint@\@tempb}{\@tempb:\@tempc}{\expandafter \expandafter \csname
  mn@eprint@\@tempb\endcsname \expandafter{\@tempc}}}

\bibitem[\protect\citeauthoryear{{Agnello} et~al.,}{{Agnello}
  et~al.}{2015}]{AgnelloEtal15}
{Agnello} A.,  et~al., 2015, \mn@doi [\mnras] {10.1093/mnras/stv2171}, \href
  {http://adsabs.harvard.edu/abs/2015MNRAS.454.1260A} {454, 1260}

\bibitem[\protect\citeauthoryear{{Anderson} et~al.,}{{Anderson}
  et~al.}{2014}]{AndersonEtal14}
{Anderson} L.,  et~al., 2014, \mn@doi [\mnras] {10.1093/mnras/stu523}, \href
  {http://adsabs.harvard.edu/abs/2014MNRAS.441...24A} {441, 24}

\bibitem[\protect\citeauthoryear{{Appenzeller} et~al.,}{{Appenzeller}
  et~al.}{1998}]{AppenzellerEtal98}
{Appenzeller} I.,  et~al., 1998, The Messenger, \href
  {http://esoads.eso.org/abs/1998Msngr..94....1A} {94, 1}

\bibitem[\protect\citeauthoryear{{Aubourg} et~al.,}{{Aubourg}
  et~al.}{2015}]{AubourgEtal15}
{Aubourg} {\'E}.,  et~al., 2015, \mn@doi [\prd] {10.1103/PhysRevD.92.123516},
  \href {http://adsabs.harvard.edu/abs/2015PhRvD..92l3516A} {92, 123516}

\bibitem[\protect\citeauthoryear{{Auger}, {Treu}, {Bolton}, {Gavazzi},
  {Koopmans}, {Marshall}, {Bundy}  \& {Moustakas}}{{Auger}
  et~al.}{2009}]{AugerEtal09}
{Auger} M.~W.,  {Treu} T.,  {Bolton} A.~S.,  {Gavazzi} R.,  {Koopmans}
  L.~V.~E.,  {Marshall} P.~J.,  {Bundy} K.,   {Moustakas} L.~A.,  2009, \mn@doi
  [\apj] {10.1088/0004-637X/705/2/1099}, \href
  {http://adsabs.harvard.edu/abs/2009ApJ...705.1099A} {705, 1099}

\bibitem[\protect\citeauthoryear{{Auger}, {Treu}, {Bolton}, {Gavazzi},
  {Koopmans}, {Marshall}, {Moustakas}  \& {Burles}}{{Auger}
  et~al.}{2010}]{AugerEtal10}
{Auger} M.~W.,  {Treu} T.,  {Bolton} A.~S.,  {Gavazzi} R.,  {Koopmans}
  L.~V.~E.,  {Marshall} P.~J.,  {Moustakas} L.~A.,   {Burles} S.,  2010,
  \mn@doi [\apj] {10.1088/0004-637X/724/1/511}, \href
  {http://adsabs.harvard.edu/abs/2010ApJ...724..511A} {724, 511}

\bibitem[\protect\citeauthoryear{{Bacon} et~al.,}{{Bacon}
  et~al.}{2012}]{BaconEtal12}
{Bacon} R.,  et~al., 2012, The Messenger, \href
  {http://adsabs.harvard.edu/abs/2012Msngr.147....4B} {147, 4}

\bibitem[\protect\citeauthoryear{{Barnab{\`e}}, {Czoske}, {Koopmans}, {Treu},
  {Bolton}  \& {Gavazzi}}{{Barnab{\`e}} et~al.}{2009}]{BarnabeEtal09}
{Barnab{\`e}} M.,  {Czoske} O.,  {Koopmans} L.~V.~E.,  {Treu} T.,  {Bolton}
  A.~S.,   {Gavazzi} R.,  2009, \mn@doi [\mnras]
  {10.1111/j.1365-2966.2009.14941.x}, \href
  {http://adsabs.harvard.edu/abs/2009MNRAS.399...21B} {399, 21}

\bibitem[\protect\citeauthoryear{{Barnab{\`e}}, {Czoske}, {Koopmans}, {Treu}
  \& {Bolton}}{{Barnab{\`e}} et~al.}{2011}]{BarnabeEtal11}
{Barnab{\`e}} M.,  {Czoske} O.,  {Koopmans} L.~V.~E.,  {Treu} T.,   {Bolton}
  A.~S.,  2011, \mn@doi [\mnras] {10.1111/j.1365-2966.2011.18842.x}, \href
  {http://adsabs.harvard.edu/abs/2011MNRAS.415.2215B} {415, 2215}

\bibitem[\protect\citeauthoryear{{Barnab{\`e}}, {Spiniello}, {Koopmans},
  {Trager}, {Czoske}  \& {Treu}}{{Barnab{\`e}} et~al.}{2013}]{BarnabeEtal13}
{Barnab{\`e}} M.,  {Spiniello} C.,  {Koopmans} L.~V.~E.,  {Trager} S.~C.,
  {Czoske} O.,   {Treu} T.,  2013, \mn@doi [\mnras] {10.1093/mnras/stt1727},
  \href {http://adsabs.harvard.edu/abs/2013MNRAS.436..253B} {436, 253}

\bibitem[\protect\citeauthoryear{{Beaton} et~al.,}{{Beaton}
  et~al.}{2016}]{BeatonEtal16}
{Beaton} R.~L.,  et~al., 2016, ArXiv e-prints (1604.01788), \href
  {http://adsabs.harvard.edu/abs/2016arXiv160401788B} {}

\bibitem[\protect\citeauthoryear{{Betoule} et~al.,}{{Betoule}
  et~al.}{2014}]{BetouleEtal14}
{Betoule} M.,  et~al., 2014, \mn@doi [\aap] {10.1051/0004-6361/201423413},
  \href {http://adsabs.harvard.edu/abs/2014A%26A...568A..22B} {568, A22}

\bibitem[\protect\citeauthoryear{{Birrer}, {Amara}  \& {Refregier}}{{Birrer}
  et~al.}{2015a}]{BirrerEtal16}
{Birrer} S.,  {Amara} A.,   {Refregier} A.,  2015a, ArXiv e-prints
  (1511.03662), \href {http://adsabs.harvard.edu/abs/2015arXiv151103662B} {}

\bibitem[\protect\citeauthoryear{{Birrer}, {Amara}  \& {Refregier}}{{Birrer}
  et~al.}{2015b}]{BirrerEtal15}
{Birrer} S.,  {Amara} A.,   {Refregier} A.,  2015b, \mn@doi [\apj]
  {10.1088/0004-637X/813/2/102}, \href
  {http://adsabs.harvard.edu/abs/2015ApJ...813..102B} {813, 102}

\bibitem[\protect\citeauthoryear{{Blake} et~al.,}{{Blake}
  et~al.}{2011}]{BlakeEtal11}
{Blake} C.,  et~al., 2011, \mn@doi [\mnras] {10.1111/j.1365-2966.2011.19077.x},
  \href {http://adsabs.harvard.edu/abs/2011MNRAS.415.2892B} {415, 2892}

\bibitem[\protect\citeauthoryear{{Blandford} \& {Kochanek}}{{Blandford} \&
  {Kochanek}}{2004}]{BlandfordKochanek04}
{Blandford} R.~D.,  {Kochanek} C.~S.,  2004, {Gravitational Lenses}.
World Scientific Publishing Company, p.~103

\bibitem[\protect\citeauthoryear{{Blandford} \& {Narayan}}{{Blandford} \&
  {Narayan}}{1986}]{BlandfordNarayan86}
{Blandford} R.,  {Narayan} R.,  1986, \mn@doi [\apj] {10.1086/164709}, \href
  {http://adsabs.harvard.edu/cgi-bin/nph-bib_query?bibcode=1986ApJ...3
  10..568B&db_key=AST} {310, 568}

\bibitem[\protect\citeauthoryear{{Bonvin}, {Tewes}, {Courbin}, {Kuntzer},
  {Sluse}  \& {Meylan}}{{Bonvin} et~al.}{2016}]{BonvinEtal16}
{Bonvin} V.,  {Tewes} M.,  {Courbin} F.,  {Kuntzer} T.,  {Sluse} D.,   {Meylan}
  G.,  2016, \mn@doi [Astronomy and Astrophysics]
  {10.1051/0004-6361/201526704}, \href
  {http://adsabs.harvard.edu/abs/2016A%26A...585A..88B} {585, A88}

\bibitem[\protect\citeauthoryear{{Bonvin} et~al.,}{{Bonvin}
  et~al.}{2017}]{BonvinEtal16b}
{Bonvin} V.,  et~al., 2017, MNRAS in press, ArXiv e-prints (1607.01790), \href
  {http://adsabs.harvard.edu/abs/2016arXiv160701790B} {}

\bibitem[\protect\citeauthoryear{{Boulade} et~al.,}{{Boulade}
  et~al.}{2003}]{BouladeEtal03}
{Boulade} O.,  et~al., 2003, in {Iye} M.,  {Moorwood} A.~F.~M.,  eds,
  \procspie Vol. 4841, Instrument Design and Performance for Optical/Infrared
  Ground-based Telescopes. pp 72--81, \mn@doi{10.1117/12.459890}

\bibitem[\protect\citeauthoryear{{Browne} et~al.,}{{Browne}
  et~al.}{2003}]{BrowneEtal03}
{Browne} I.~W.~A.,  et~al., 2003, \mn@doi [\mnras]
  {10.1046/j.1365-8711.2003.06257.x}, \href
  {http://adsabs.harvard.edu/abs/2003MNRAS.341...13B} {341, 13}

\bibitem[\protect\citeauthoryear{{Burud} et~al.,}{{Burud}
  et~al.}{2002}]{BurudEtal02}
{Burud} I.,  et~al., 2002, \mn@doi [\aap] {10.1051/0004-6361:20011731}, \href
  {http://adsabs.harvard.edu/abs/2002A%26A...383...71B} {383, 71}

\bibitem[\protect\citeauthoryear{{Cantale}, {Courbin}, {Tewes}, {Jablonka.}  \&
  {Meylan}}{{Cantale} et~al.}{2016}]{CantaleEtal16}
{Cantale} N.,  {Courbin} F.,  {Tewes} M.,  {Jablonka.} P.,   {Meylan} G.,
  2016, preprint, \href {http://adsabs.harvard.edu/abs/2016arXiv160202167C} {}
  (\mn@eprint {arXiv} {1602.02167})

\bibitem[\protect\citeauthoryear{{Casali} et~al.,}{{Casali}
  et~al.}{2006}]{CasaliEtal06}
{Casali} M.,  et~al., 2006, in Society of Photo-Optical Instrumentation
  Engineers (SPIE) Conference Series. p. 62690W, \mn@doi{10.1117/12.670150}

\bibitem[\protect\citeauthoryear{{Chan} et~al.,}{{Chan}
  et~al.}{2016}]{ChanEtal16}
{Chan} J.~H.~H.,  et~al., 2016, preprint, \href
  {http://adsabs.harvard.edu/abs/2016arXiv160408215C} {} (\mn@eprint {arXiv}
  {1604.08215})

\bibitem[\protect\citeauthoryear{{Chen} et~al.,}{{Chen}
  et~al.}{2016}]{ChenEtal16}
{Chen} G.~C.~F.,  et~al., 2016, ArXiv e-prints (1601.01321), \href
  {http://adsabs.harvard.edu/abs/2016arXiv160101321C} {}

\bibitem[\protect\citeauthoryear{{Claeskens}, {Sluse}, {Riaud}  \&
  {Surdej}}{{Claeskens} et~al.}{2006}]{ClaeskensEtal06}
{Claeskens} J.-F.,  {Sluse} D.,  {Riaud} P.,   {Surdej} J.,  2006, \mn@doi
  [\aap] {10.1051/0004-6361:20054352}, \href
  {http://adsabs.harvard.edu/abs/2006A%26A...451..865C} {451, 865}

\bibitem[\protect\citeauthoryear{{Collett} \& {Cunnington}}{{Collett} \&
  {Cunnington}}{2016}]{CollettCunnington16}
{Collett} T.~E.,  {Cunnington} S.~D.,  2016, preprint, \href
  {http://adsabs.harvard.edu/abs/2016arXiv160508341C} {} (\mn@eprint {arXiv}
  {1605.08341})

\bibitem[\protect\citeauthoryear{{Collett} et~al.,}{{Collett}
  et~al.}{2013}]{CollettEtal13}
{Collett} T.~E.,  et~al., 2013, \mn@doi [\mnras] {10.1093/mnras/stt504}, \href
  {http://adsabs.harvard.edu/abs/2013MNRAS.432..679C} {432, 679}

\bibitem[\protect\citeauthoryear{{Conley} et~al.,}{{Conley}
  et~al.}{2006}]{ConleyEtal06}
{Conley} A.,  et~al., 2006, \mn@doi [\apj] {10.1086/503533}, \href
  {http://adsabs.harvard.edu/abs/2006ApJ...644....1C} {644, 1}

\bibitem[\protect\citeauthoryear{{Conley} et~al.,}{{Conley}
  et~al.}{2011}]{ConleyEtal11}
{Conley} A.,  et~al., 2011, \mn@doi [\apjs] {10.1088/0067-0049/192/1/1}, \href
  {http://adsabs.harvard.edu/abs/2011ApJS..192....1C} {192, 1}

\bibitem[\protect\citeauthoryear{{Conroy} \& {van Dokkum}}{{Conroy} \& {van
  Dokkum}}{2012}]{ConroyVanDokkum12}
{Conroy} C.,  {van Dokkum} P.~G.,  2012, \mn@doi [\apj]
  {10.1088/0004-637X/760/1/71}, \href
  {http://adsabs.harvard.edu/abs/2012ApJ...760...71C} {760, 71}

\bibitem[\protect\citeauthoryear{{Courbin}, {Lidman}  \& {Magain}}{{Courbin}
  et~al.}{1998}]{CourbinEtal98}
{Courbin} F.,  {Lidman} C.,   {Magain} P.,  1998, \aap, \href
  {http://adsabs.harvard.edu/abs/1998A%26A...330...57C} {330, 57}

\bibitem[\protect\citeauthoryear{{Courbin}, {Eigenbrod}, {Vuissoz}, {Meylan}
  \& {Magain}}{{Courbin} et~al.}{2005}]{CourbinEtal05}
{Courbin} F.,  {Eigenbrod} A.,  {Vuissoz} C.,  {Meylan} G.,   {Magain} P.,
  2005, in {Mellier} Y.,  {Meylan} G.,  eds,  IAU Symposium Vol. 225,
  Gravitational Lensing Impact on Cosmology. pp 297--303,
  \mn@doi{10.1017/S1743921305002097}

\bibitem[\protect\citeauthoryear{{Courbin} et~al.,}{{Courbin}
  et~al.}{2011}]{CourbinEtal11}
{Courbin} F.,  et~al., 2011, \mn@doi [\aap] {10.1051/0004-6361/201015709},
  \href {http://adsabs.harvard.edu/abs/2011A%26A...536A..53C} {536, A53}

\bibitem[\protect\citeauthoryear{{Diehl} \& {Dark Energy Survey
  Collaboration}}{{Diehl} \& {Dark Energy Survey
  Collaboration}}{2012}]{DiehlEtal12}
{Diehl} T.,  {Dark Energy Survey Collaboration} 2012, \mn@doi [Physics
  Procedia] {10.1016/j.phpro.2012.02.472}, \href
  {http://esoads.eso.org/abs/2012PhPro..37.1332D} {37, 1332}

\bibitem[\protect\citeauthoryear{{Ding} et~al.,}{{Ding}
  et~al.}{2017}]{DingEtal16}
{Ding} X.,  et~al., 2017, MNRAS in press, ArXiv e-prints (1610.08504), \href
  {http://adsabs.harvard.edu/abs/2016arXiv161008504D} {}

\bibitem[\protect\citeauthoryear{{Dobler}, {Fassnacht}, {Treu}, {Marshall},
  {Liao}, {Hojjati}, {Linder}  \& {Rumbaugh}}{{Dobler}
  et~al.}{2015}]{DoblerEtal15}
{Dobler} G.,  {Fassnacht} C.~D.,  {Treu} T.,  {Marshall} P.,  {Liao} K.,
  {Hojjati} A.,  {Linder} E.,   {Rumbaugh} N.,  2015, \mn@doi [\apj]
  {10.1088/0004-637X/799/2/168}, \href
  {http://adsabs.harvard.edu/abs/2015ApJ...799..168D} {799, 168}

\bibitem[\protect\citeauthoryear{{Dye}, {Evans}, {Belokurov}, {Warren}  \&
  {Hewett}}{{Dye} et~al.}{2008}]{DyeEtal08}
{Dye} S.,  {Evans} N.~W.,  {Belokurov} V.,  {Warren} S.~J.,   {Hewett} P.,
  2008, \mn@doi [\mnras] {10.1111/j.1365-2966.2008.13401.x}, \href
  {http://adsabs.harvard.edu/abs/2008MNRAS.388..384D} {388, 384}

\bibitem[\protect\citeauthoryear{{Eigenbrod}, {Courbin}, {Meylan}, {Vuissoz}
  \& {Magain}}{{Eigenbrod} et~al.}{2006}]{EigenbrodEtal06}
{Eigenbrod} A.,  {Courbin} F.,  {Meylan} G.,  {Vuissoz} C.,   {Magain} P.,
  2006, \mn@doi [\aap] {10.1051/0004-6361:20054454}, \href
  {http://adsabs.harvard.edu/abs/2006A%26A...451..759E} {451, 759}

\bibitem[\protect\citeauthoryear{{Eisenstein} et~al.,}{{Eisenstein}
  et~al.}{2005}]{EisensteinEtal05}
{Eisenstein} D.~J.,  et~al., 2005, \mn@doi [\apj] {10.1086/466512}, \href
  {http://adsabs.harvard.edu/abs/2005ApJ...633..560E} {633, 560}

\bibitem[\protect\citeauthoryear{{Falco}, {Gorenstein}  \& {Shapiro}}{{Falco}
  et~al.}{1985}]{FalcoEtal85}
{Falco} E.~E.,  {Gorenstein} M.~V.,   {Shapiro} I.~I.,  1985, \mn@doi [\apjl]
  {10.1086/184422}, \href {http://adsabs.harvard.edu/abs/1985ApJ...289L...1F}
  {289, L1}

\bibitem[\protect\citeauthoryear{{Fassnacht}, {Womble}, {Neugebauer}, {Browne},
  {Readhead}, {Matthews}  \& {Pearson}}{{Fassnacht}
  et~al.}{1996}]{FassnachtEtal96}
{Fassnacht} C.~D.,  {Womble} D.~S.,  {Neugebauer} G.,  {Browne} I.~W.~A.,
  {Readhead} A.~C.~S.,  {Matthews} K.,   {Pearson} T.~J.,  1996, \mn@doi
  [\apjl] {10.1086/309984}, \href
  {http://adsabs.harvard.edu/abs/1996ApJ...460L.103F} {460, L103}

\bibitem[\protect\citeauthoryear{{Fassnacht}, {Pearson}, {Readhead}, {Browne},
  {Koopmans}, {Myers}  \& {Wilkinson}}{{Fassnacht}
  et~al.}{1999}]{FassnachtEtal99}
{Fassnacht} C.~D.,  {Pearson} T.~J.,  {Readhead} A.~C.~S.,  {Browne} I.~W.~A.,
  {Koopmans} L.~V.~E.,  {Myers} S.~T.,   {Wilkinson} P.~N.,  1999, \mn@doi
  [\apj] {10.1086/308118}, \href
  {http://adsabs.harvard.edu/cgi-bin/nph-bib_query?bibcode=1999ApJ...5
  27..498F&db_key=AST} {527, 498}

\bibitem[\protect\citeauthoryear{{Fassnacht}, {Xanthopoulos}, {Koopmans}  \&
  {Rusin}}{{Fassnacht} et~al.}{2002}]{FassnachtEtal02}
{Fassnacht} C.~D.,  {Xanthopoulos} E.,  {Koopmans} L.~V.~E.,   {Rusin} D.,
  2002, \mn@doi [\apj] {10.1086/344368}, \href
  {http://adsabs.harvard.edu/cgi-bin/nph-bib_query?bibcode=2002ApJ...5
  81..823F&db_key=AST} {581, 823}

\bibitem[\protect\citeauthoryear{{Fassnacht}, {Gal}, {Lubin}, {McKean},
  {Squires}  \& {Readhead}}{{Fassnacht} et~al.}{2006}]{FassnachtEtal06}
{Fassnacht} C.~D.,  {Gal} R.~R.,  {Lubin} L.~M.,  {McKean} J.~P.,  {Squires}
  G.~K.,   {Readhead} A.~C.~S.,  2006, \mn@doi [\apj] {10.1086/500927}, \href
  {http://adsabs.harvard.edu/abs/2006ApJ...642...30F} {642, 30}

\bibitem[\protect\citeauthoryear{{Fassnacht}, {Koopmans}  \&
  {Wong}}{{Fassnacht} et~al.}{2011}]{FassnachtEtal11}
{Fassnacht} C.~D.,  {Koopmans} L.~V.~E.,   {Wong} K.~C.,  2011, \mn@doi
  [\mnras] {10.1111/j.1365-2966.2010.17591.x}, \href
  {http://adsabs.harvard.edu/abs/2011MNRAS.410.2167F} {410, 2167}

\bibitem[\protect\citeauthoryear{{Fazio} et~al.,}{{Fazio}
  et~al.}{2004}]{FazioEtal04}
{Fazio} G.~G.,  et~al., 2004, \mn@doi [\apjs] {10.1086/422843}, \href
  {http://esoads.eso.org/abs/2004ApJS..154...10F} {154, 10}

\bibitem[\protect\citeauthoryear{{Freedman}, {Madore}, {Scowcroft}, {Burns},
  {Monson}, {Persson}, {Seibert}  \& {Rigby}}{{Freedman}
  et~al.}{2012}]{FreedmanEtal12}
{Freedman} W.~L.,  {Madore} B.~F.,  {Scowcroft} V.,  {Burns} C.,  {Monson} A.,
  {Persson} S.~E.,  {Seibert} M.,   {Rigby} J.,  2012, \mn@doi [\apj]
  {10.1088/0004-637X/758/1/24}, \href
  {http://adsabs.harvard.edu/abs/2012ApJ...758...24F} {758, 24}

\bibitem[\protect\citeauthoryear{{Futamase} \& {Hamana}}{{Futamase} \&
  {Hamana}}{1999}]{FutamaseHamana99}
{Futamase} T.,  {Hamana} T.,  1999, \mn@doi [Progress of Theoretical Physics]
  {10.1143/PTP.102.1037}, \href
  {http://adsabs.harvard.edu/abs/1999PThPh.102.1037F} {102, 1037}

\bibitem[\protect\citeauthoryear{{Futamase} \& {Yoshida}}{{Futamase} \&
  {Yoshida}}{2001}]{FutamaseYoshida01}
{Futamase} T.,  {Yoshida} S.,  2001, \mn@doi [Progress of Theoretical Physics]
  {10.1143/PTP.105.887}, \href
  {http://adsabs.harvard.edu/abs/2001PThPh.105..887F} {105, 887}

\bibitem[\protect\citeauthoryear{{Gao} et~al.,}{{Gao} et~al.}{2016}]{GaoEtal16}
{Gao} F.,  et~al., 2016, \mn@doi [\apj] {10.3847/0004-637X/817/2/128}, \href
  {http://adsabs.harvard.edu/abs/2016ApJ...817..128G} {817, 128}

\bibitem[\protect\citeauthoryear{{Goobar} et~al.,}{{Goobar}
  et~al.}{2016}]{GoobarEtal16}
{Goobar} A.,  et~al., 2016, preprint, \href
  {http://adsabs.harvard.edu/abs/2016arXiv161100014G} {} (\mn@eprint {arXiv}
  {1611.00014})

\bibitem[\protect\citeauthoryear{{Greene} et~al.,}{{Greene}
  et~al.}{2013}]{GreeneEtal13}
{Greene} Z.~S.,  et~al., 2013, \mn@doi [\apj] {10.1088/0004-637X/768/1/39},
  \href {http://adsabs.harvard.edu/abs/2013ApJ...768...39G} {768, 39}

\bibitem[\protect\citeauthoryear{{Grillo}, {Lombardi}  \& {Bertin}}{{Grillo}
  et~al.}{2008}]{GrilloEtal08}
{Grillo} C.,  {Lombardi} M.,   {Bertin} G.,  2008, \mn@doi [\aap]
  {10.1051/0004-6361:20077534}, \href
  {http://adsabs.harvard.edu/abs/2008A%26A...477..397G} {477, 397}

\bibitem[\protect\citeauthoryear{{Grillo}, {Gobat}, {Lombardi}  \&
  {Rosati}}{{Grillo} et~al.}{2009}]{GrilloEtal09}
{Grillo} C.,  {Gobat} R.,  {Lombardi} M.,   {Rosati} P.,  2009, \mn@doi [\aap]
  {10.1051/0004-6361/200811604}, \href
  {http://adsabs.harvard.edu/abs/2009A%26A...501..461G} {501, 461}

\bibitem[\protect\citeauthoryear{{Grillo} et~al.,}{{Grillo}
  et~al.}{2016}]{GrilloEtal16}
{Grillo} C.,  et~al., 2016, \mn@doi [\apj] {10.3847/0004-637X/822/2/78}, \href
  {http://adsabs.harvard.edu/abs/2016ApJ...822...78G} {822, 78}

\bibitem[\protect\citeauthoryear{Harva \& Raychaudhury}{Harva \&
  Raychaudhury}{2008}]{HarvaEtal08}
Harva M.,  Raychaudhury S.,  2008, Neurocomputing, 72, 32

\bibitem[\protect\citeauthoryear{{Heavens}, {Jimenez}  \& {Verde}}{{Heavens}
  et~al.}{2014}]{HeavensEtal14}
{Heavens} A.,  {Jimenez} R.,   {Verde} L.,  2014, \mn@doi [Physical Review
  Letters] {10.1103/PhysRevLett.113.241302}, \href
  {http://adsabs.harvard.edu/abs/2014PhRvL.113x1302H} {113, 241302}

\bibitem[\protect\citeauthoryear{{Hewett} \& {Wild}}{{Hewett} \&
  {Wild}}{2010}]{HewettWild10}
{Hewett} P.~C.,  {Wild} V.,  2010, \mn@doi [\mnras]
  {10.1111/j.1365-2966.2010.16648.x}, \href
  {http://adsabs.harvard.edu/abs/2010MNRAS.405.2302H} {405, 2302}

\bibitem[\protect\citeauthoryear{{Heymans} et~al.,}{{Heymans}
  et~al.}{2012}]{HeymansEtal12}
{Heymans} C.,  et~al., 2012, \mn@doi [\mnras]
  {10.1111/j.1365-2966.2012.21952.x}, \href
  {http://adsabs.harvard.edu/abs/2012MNRAS.427..146H} {427, 146}

\bibitem[\protect\citeauthoryear{{Hilbert}, {White}, {Hartlap}  \&
  {Schneider}}{{Hilbert} et~al.}{2007}]{HilbertEtal07}
{Hilbert} S.,  {White} S.~D.~M.,  {Hartlap} J.,   {Schneider} P.,  2007,
  \mn@doi [\mnras] {10.1111/j.1365-2966.2007.12391.x}, \href
  {http://adsabs.harvard.edu/abs/2007MNRAS.382..121H} {382, 121}

\bibitem[\protect\citeauthoryear{{Hilbert}, {Hartlap}, {White}  \&
  {Schneider}}{{Hilbert} et~al.}{2009}]{HilbertEtal09}
{Hilbert} S.,  {Hartlap} J.,  {White} S.~D.~M.,   {Schneider} P.,  2009,
  \mn@doi [\aap] {10.1051/0004-6361/200811054}, \href
  {http://adsabs.harvard.edu/abs/2009A%26A...499...31H} {499, 31}

\bibitem[\protect\citeauthoryear{{Hinshaw} et~al.,}{{Hinshaw}
  et~al.}{2013}]{HinshawEtal12}
{Hinshaw} G.,  et~al., 2013, \mn@doi [\apjs] {10.1088/0067-0049/208/2/19},
  \href {http://adsabs.harvard.edu/abs/2013ApJS..208...19H} {208, 19}

\bibitem[\protect\citeauthoryear{{Hirv}, {Olspert}  \& {Pelt}}{{Hirv}
  et~al.}{2011}]{HirvEtal11}
{Hirv} A.,  {Olspert} N.,   {Pelt} J.,  2011, Baltic Astronomy, \href
  {http://adsabs.harvard.edu/abs/2011BaltA..20..125H} {20, 125}

\bibitem[\protect\citeauthoryear{{Hjorth} et~al.,}{{Hjorth}
  et~al.}{2002}]{HjorthEtal02}
{Hjorth} J.,  et~al., 2002, \mn@doi [\apjl] {10.1086/341603}, \href
  {http://adsabs.harvard.edu/abs/2002ApJ...572L..11H} {572, L11}

\bibitem[\protect\citeauthoryear{{Hodapp} et~al.,}{{Hodapp}
  et~al.}{2003}]{HodappEtal03}
{Hodapp} K.~W.,  et~al., 2003, \mn@doi [\pasp] {10.1086/379669}, \href
  {http://esoads.eso.org/abs/2003PASP..115.1388H} {115, 1388}

\bibitem[\protect\citeauthoryear{{Hojjati}, {Kim}  \& {Linder}}{{Hojjati}
  et~al.}{2013}]{HojjatiEtal13}
{Hojjati} A.,  {Kim} A.~G.,   {Linder} E.~V.,  2013, \mn@doi [\prd]
  {10.1103/PhysRevD.87.123512}, \href
  {http://adsabs.harvard.edu/abs/2013PhRvD..87l3512H} {87, 123512}

\bibitem[\protect\citeauthoryear{{Holder} \& {Schechter}}{{Holder} \&
  {Schechter}}{2003}]{HolderSchechter03}
{Holder} G.~P.,  {Schechter} P.~L.,  2003, \mn@doi [\apj] {10.1086/374688},
  \href {http://adsabs.harvard.edu/abs/2003ApJ...589..688H} {589, 688}

\bibitem[\protect\citeauthoryear{{Hook}, {J{\o}rgensen}, {Allington-Smith},
  {Davies}, {Metcalfe}, {Murowinski}  \& {Crampton}}{{Hook}
  et~al.}{2004}]{HookEtal04}
{Hook} I.~M.,  {J{\o}rgensen} I.,  {Allington-Smith} J.~R.,  {Davies} R.~L.,
  {Metcalfe} N.,  {Murowinski} R.~G.,   {Crampton} D.,  2004, \mn@doi [\pasp]
  {10.1086/383624}, \href {http://adsabs.harvard.edu/abs/2004PASP..116..425H}
  {116, 425}

\bibitem[\protect\citeauthoryear{{Hu}}{{Hu}}{2005}]{Hu05}
{Hu} W.,  2005, in {Wolff} S.~C.,  {Lauer} T.~R.,  eds,  Astronomical Society
  of the Pacific Conference Series Vol. 339, Observing Dark Energy. p.~215
  (\mn@eprint {} {astro-ph/0407158})

\bibitem[\protect\citeauthoryear{{Ichikawa} et~al.,}{{Ichikawa}
  et~al.}{2006}]{IchikawaEtal06}
{Ichikawa} T.,  et~al., 2006, in Society of Photo-Optical Instrumentation
  Engineers (SPIE) Conference Series. p. 626916, \mn@doi{10.1117/12.670078}

\bibitem[\protect\citeauthoryear{{Inada} et~al.,}{{Inada}
  et~al.}{2012}]{InadaEtal12}
{Inada} N.,  et~al., 2012, \mn@doi [\aj] {10.1088/0004-6256/143/5/119}, \href
  {http://adsabs.harvard.edu/abs/2012AJ....143..119I} {143, 119}

\bibitem[\protect\citeauthoryear{{Jee}, {Komatsu}  \& {Suyu}}{{Jee}
  et~al.}{2015}]{JeeEtal15}
{Jee} I.,  {Komatsu} E.,   {Suyu} S.~H.,  2015, \mn@doi [\jcap]
  {10.1088/1475-7516/2015/11/033}, \href
  {http://adsabs.harvard.edu/abs/2015JCAP...11..033J} {11, 033}

\bibitem[\protect\citeauthoryear{{Jee}, {Komatsu}, {Suyu}  \& {Huterer}}{{Jee}
  et~al.}{2016}]{JeeEtal16}
{Jee} I.,  {Komatsu} E.,  {Suyu} S.~H.,   {Huterer} D.,  2016, \jcap, in press
  (ArXiv e-prints 1509.03310), \href
  {http://adsabs.harvard.edu/abs/2015arXiv150903310J} {}

\bibitem[\protect\citeauthoryear{{Kawamata}, {Oguri}, {Ishigaki}, {Shimasaku}
  \& {Ouchi}}{{Kawamata} et~al.}{2016}]{KawamataEtal16}
{Kawamata} R.,  {Oguri} M.,  {Ishigaki} M.,  {Shimasaku} K.,   {Ouchi} M.,
  2016, \mn@doi [\apj] {10.3847/0004-637X/819/2/114}, \href
  {http://adsabs.harvard.edu/abs/2016ApJ...819..114K} {819, 114}

\bibitem[\protect\citeauthoryear{{Kazin} et~al.,}{{Kazin}
  et~al.}{2014}]{KazinEtal14}
{Kazin} E.~A.,  et~al., 2014, \mn@doi [\mnras] {10.1093/mnras/stu778}, \href
  {http://adsabs.harvard.edu/abs/2014MNRAS.441.3524K} {441, 3524}

\bibitem[\protect\citeauthoryear{{Kelly} et~al.,}{{Kelly}
  et~al.}{2015}]{KellyEtal15}
{Kelly} P.~L.,  et~al., 2015, \mn@doi [Science] {10.1126/science.aaa3350},
  \href {http://adsabs.harvard.edu/abs/2015Sci...347.1123K} {347, 1123}

\bibitem[\protect\citeauthoryear{{Kelly} et~al.,}{{Kelly}
  et~al.}{2016}]{KellyEtal16}
{Kelly} P.~L.,  et~al., 2016, \mn@doi [\apjl] {10.3847/2041-8205/819/1/L8},
  \href {http://adsabs.harvard.edu/abs/2016ApJ...819L...8K} {819, L8}

\bibitem[\protect\citeauthoryear{{Kissler-Patig} et~al.,}{{Kissler-Patig}
  et~al.}{2008}]{Kissler-PatigEtal08}
{Kissler-Patig} M.,  et~al., 2008, \mn@doi [\aap]
  {10.1051/0004-6361:200809910}, \href
  {http://esoads.eso.org/abs/2008A%26A...491..941K} {491, 941}

\bibitem[\protect\citeauthoryear{{Kochanek}}{{Kochanek}}{2002}]{Kochanek02}
{Kochanek} C.~S.,  2002, \mn@doi [\apj] {10.1086/342476}, \href
  {http://adsabs.harvard.edu/abs/2002ApJ...578...25K} {578, 25}

\bibitem[\protect\citeauthoryear{{Kochanek}, {Keeton}  \& {McLeod}}{{Kochanek}
  et~al.}{2001}]{KochanekEtal01}
{Kochanek} C.~S.,  {Keeton} C.~R.,   {McLeod} B.~A.,  2001, \mn@doi [\apj]
  {10.1086/318350}, \href {http://adsabs.harvard.edu/abs/2001ApJ...547...50K}
  {547, 50}

\bibitem[\protect\citeauthoryear{{Kochanek}, {Morgan}, {Falco}, {McLeod},
  {Winn}, {Dembicky}  \& {Ketzeback}}{{Kochanek} et~al.}{2006}]{KochanekEtal06}
{Kochanek} C.~S.,  {Morgan} N.~D.,  {Falco} E.~E.,  {McLeod} B.~A.,  {Winn}
  J.~N.,  {Dembicky} J.,   {Ketzeback} B.,  2006, \mn@doi [\apj]
  {10.1086/499766}, \href {http://adsabs.harvard.edu/abs/2006ApJ...640...47K}
  {640, 47}

\bibitem[\protect\citeauthoryear{{Komatsu} et~al.,}{{Komatsu}
  et~al.}{2011}]{KomatsuEtal11}
{Komatsu} E.,  et~al., 2011, \mn@doi [\apjs] {10.1088/0067-0049/192/2/18},
  \href {http://adsabs.harvard.edu/abs/2011ApJS..192...18K} {192, 18}

\bibitem[\protect\citeauthoryear{{Koopmans}}{{Koopmans}}{2005}]{Koopmans05}
{Koopmans} L.~V.~E.,  2005, \mn@doi [\mnras]
  {10.1111/j.1365-2966.2005.09523.x}, \href
  {http://adsabs.harvard.edu/cgi-bin/nph-bib_query?bibcode=2005MNRAS.363.1136K&db_key=AST}
  {363, 1136}

\bibitem[\protect\citeauthoryear{{Koopmans}, {Treu}, {Fassnacht}, {Blandford}
  \& {Surpi}}{{Koopmans} et~al.}{2003}]{KoopmansEtal03}
{Koopmans} L.~V.~E.,  {Treu} T.,  {Fassnacht} C.~D.,  {Blandford} R.~D.,
  {Surpi} G.,  2003, \mn@doi [\apj] {10.1086/379226}, \href
  {http://adsabs.harvard.edu/cgi-bin/nph-bib_query?bibcode=2003ApJ...5
  99...70K&db_key=AST} {599, 70}

\bibitem[\protect\citeauthoryear{{Kuo} et~al.,}{{Kuo} et~al.}{2015}]{KuoEtal15}
{Kuo} C.~Y.,  et~al., 2015, \mn@doi [\apj] {10.1088/0004-637X/800/1/26}, \href
  {http://adsabs.harvard.edu/abs/2015ApJ...800...26K} {800, 26}

\bibitem[\protect\citeauthoryear{{Larkin} et~al.,}{{Larkin}
  et~al.}{2006}]{LarkinEtal06}
{Larkin} J.,  et~al., 2006, in Society of Photo-Optical Instrumentation
  Engineers (SPIE) Conference Series. p. 62691A, \mn@doi{10.1117/12.672061}

\bibitem[\protect\citeauthoryear{{Liao} et~al.,}{{Liao}
  et~al.}{2015}]{LiaoEtal15}
{Liao} K.,  et~al., 2015, \mn@doi [\apj] {10.1088/0004-637X/800/1/11}, \href
  {http://adsabs.harvard.edu/abs/2015ApJ...800...11L} {800, 11}

\bibitem[\protect\citeauthoryear{{Lidman}, {Courbin}, {Kneib}, {Golse},
  {Castander}  \& {Soucail}}{{Lidman} et~al.}{2000}]{LidmanEtal00}
{Lidman} C.,  {Courbin} F.,  {Kneib} J.-P.,  {Golse} G.,  {Castander} F.,
  {Soucail} G.,  2000, \aap, \href
  {http://adsabs.harvard.edu/abs/2000A%26A...364L..62L} {364, L62}

\bibitem[\protect\citeauthoryear{{Linder}}{{Linder}}{2015}]{Linder15}
{Linder} E.~V.,  2015, \mn@doi [\prd] {10.1103/PhysRevD.91.083511}, \href
  {http://adsabs.harvard.edu/abs/2015PhRvD..91h3511L} {91, 083511}

\bibitem[\protect\citeauthoryear{{Magain}, {Courbin}  \& {Sohy}}{{Magain}
  et~al.}{1998}]{MagainEtal98}
{Magain} P.,  {Courbin} F.,   {Sohy} S.,  1998, \mn@doi [\apj]
  {10.1086/305187}, \href {http://adsabs.harvard.edu/abs/1998ApJ...494..472M}
  {494, 472}

\bibitem[\protect\citeauthoryear{{McCully}, {Keeton}, {Wong}  \&
  {Zabludoff}}{{McCully} et~al.}{2014}]{McCullyEtal14}
{McCully} C.,  {Keeton} C.~R.,  {Wong} K.~C.,   {Zabludoff} A.~I.,  2014,
  \mn@doi [\mnras] {10.1093/mnras/stu1316}, \href
  {http://adsabs.harvard.edu/abs/2014MNRAS.443.3631M} {443, 3631}

\bibitem[\protect\citeauthoryear{{McCully}, {Keeton}, {Wong}  \&
  {Zabludoff}}{{McCully} et~al.}{2016}]{McCullyEtal16}
{McCully} C.,  {Keeton} C.~R.,  {Wong} K.~C.,   {Zabludoff} A.~I.,  2016,
  preprint, \href {http://adsabs.harvard.edu/abs/2016arXiv160105417M} {}
  (\mn@eprint {arXiv} {1601.05417})

\bibitem[\protect\citeauthoryear{{Meng}, {Treu}, {Agnello}, {Auger}, {Liao}  \&
  {Marshall}}{{Meng} et~al.}{2015}]{MengEtal15}
{Meng} X.-L.,  {Treu} T.,  {Agnello} A.,  {Auger} M.~W.,  {Liao} K.,
  {Marshall} P.~J.,  2015, \mn@doi [\jcap] {10.1088/1475-7516/2015/09/059},
  \href {http://adsabs.harvard.edu/abs/2015JCAP...09..059M} {9, 059}

\bibitem[\protect\citeauthoryear{{Miyazaki} et~al.,}{{Miyazaki}
  et~al.}{2002}]{MiyazakiEtal02}
{Miyazaki} S.,  et~al., 2002, \mn@doi [\pasj] {10.1093/pasj/54.6.833}, \href
  {http://esoads.eso.org/abs/2002PASJ...54..833M} {54, 833}

\bibitem[\protect\citeauthoryear{{Momcheva}, {Williams}, {Keeton}  \&
  {Zabludoff}}{{Momcheva} et~al.}{2006}]{MomchevaEtal06}
{Momcheva} I.,  {Williams} K.,  {Keeton} C.,   {Zabludoff} A.,  2006, \mn@doi
  [\apj] {10.1086/500382}, \href
  {http://adsabs.harvard.edu/abs/2006ApJ...641..169M} {641, 169}

\bibitem[\protect\citeauthoryear{{Momcheva}, {Williams}, {Cool}, {Keeton}  \&
  {Zabludoff}}{{Momcheva} et~al.}{2015}]{MomchevaEtal15}
{Momcheva} I.~G.,  {Williams} K.~A.,  {Cool} R.~J.,  {Keeton} C.~R.,
  {Zabludoff} A.~I.,  2015, \mn@doi [\apjs] {10.1088/0067-0049/219/2/29}, \href
  {http://adsabs.harvard.edu/abs/2015ApJS..219...29M} {219, 29}

\bibitem[\protect\citeauthoryear{{More}, {Suyu}, {Oguri}, {More}  \&
  {Lee}}{{More} et~al.}{2016a}]{MoreEtal16b}
{More} A.,  {Suyu} S.~H.,  {Oguri} M.,  {More} S.,   {Lee} C.-H.,  2016a,
  preprint, \href {http://adsabs.harvard.edu/abs/2016arXiv161104866M} {}
  (\mn@eprint {arXiv} {1611.04866})

\bibitem[\protect\citeauthoryear{{More} et~al.,}{{More}
  et~al.}{2016b}]{MoreEtal16}
{More} A.,  et~al., 2016b, \mn@doi [\mnras] {10.1093/mnras/stv2813}, \href
  {http://adsabs.harvard.edu/abs/2016MNRAS.456.1595M} {456, 1595}

\bibitem[\protect\citeauthoryear{{Morgan}, {Caldwell}, {Schechter}, {Dressler},
  {Egami}  \& {Rix}}{{Morgan} et~al.}{2004}]{MorganEtal04}
{Morgan} N.~D.,  {Caldwell} J.~A.~R.,  {Schechter} P.~L.,  {Dressler} A.,
  {Egami} E.,   {Rix} H.-W.,  2004, \mn@doi [\aj] {10.1086/383295}, \href
  {http://adsabs.harvard.edu/abs/2004AJ....127.2617M} {127, 2617}

\bibitem[\protect\citeauthoryear{{Morgan}, {Kochanek}, {Pevunova}  \&
  {Schechter}}{{Morgan} et~al.}{2005}]{MorganEtal05}
{Morgan} N.~D.,  {Kochanek} C.~S.,  {Pevunova} O.,   {Schechter} P.~L.,  2005,
  \mn@doi [\aj] {10.1086/430145}, \href
  {http://adsabs.harvard.edu/abs/2005AJ....129.2531M} {129, 2531}

\bibitem[\protect\citeauthoryear{{Morgan}, {Eyler}, {Kochanek}, {Morgan},
  {Falco}, {Vuissoz}, {Courbin}  \& {Meylan}}{{Morgan}
  et~al.}{2008}]{MorganEtal08}
{Morgan} C.~W.,  {Eyler} M.~E.,  {Kochanek} C.~S.,  {Morgan} N.~D.,  {Falco}
  E.~E.,  {Vuissoz} C.,  {Courbin} F.,   {Meylan} G.,  2008, \mn@doi [\apj]
  {10.1086/527371}, \href {http://adsabs.harvard.edu/abs/2008ApJ...676...80M}
  {676, 80}

\bibitem[\protect\citeauthoryear{{Myers} et~al.,}{{Myers}
  et~al.}{1995}]{MyersEtal95}
{Myers} S.~T.,  et~al., 1995, \mn@doi [\apjl] {10.1086/309556}, \href
  {http://adsabs.harvard.edu/abs/1995ApJ...447L...5M} {447, L5}

\bibitem[\protect\citeauthoryear{{Myers} et~al.,}{{Myers}
  et~al.}{2003}]{MyersEtal03}
{Myers} S.~T.,  et~al., 2003, \mn@doi [\mnras]
  {10.1046/j.1365-8711.2003.06256.x}, \href
  {http://adsabs.harvard.edu/abs/2003MNRAS.341....1M} {341, 1}

\bibitem[\protect\citeauthoryear{{Napolitano} et~al.,}{{Napolitano}
  et~al.}{2015}]{NapolitanoEtal15}
{Napolitano} N.~R.,  et~al., 2015, ArXiv e-prints (1507.00733), \href
  {http://adsabs.harvard.edu/abs/2015arXiv150700733N} {}

\bibitem[\protect\citeauthoryear{{Newman}, {Smith}, {Conroy}, {Villaume}  \&
  {van Dokkum}}{{Newman} et~al.}{2016}]{NewmanEtal16}
{Newman} A.~B.,  {Smith} R.~J.,  {Conroy} C.,  {Villaume} A.,   {van Dokkum}
  P.,  2016, ArXiv e-prints (1612.00065), \href
  {http://adsabs.harvard.edu/abs/2016arXiv161200065N} {}

\bibitem[\protect\citeauthoryear{{Ofek}, {Maoz}, {Rix}, {Kochanek}  \&
  {Falco}}{{Ofek} et~al.}{2006}]{OfekEtal06}
{Ofek} E.~O.,  {Maoz} D.,  {Rix} H.-W.,  {Kochanek} C.~S.,   {Falco} E.~E.,
  2006, \mn@doi [\apj] {10.1086/500403}, \href
  {http://adsabs.harvard.edu/abs/2006ApJ...641...70O} {641, 70}

\bibitem[\protect\citeauthoryear{{Oguri} \& {Marshall}}{{Oguri} \&
  {Marshall}}{2010}]{OguriMarshall10}
{Oguri} M.,  {Marshall} P.~J.,  2010, \mn@doi [\mnras]
  {10.1111/j.1365-2966.2010.16639.x}, \href
  {http://adsabs.harvard.edu/abs/2010MNRAS.405.2579O} {405, 2579}

\bibitem[\protect\citeauthoryear{{Oguri} et~al.,}{{Oguri}
  et~al.}{2006}]{OguriEtal06}
{Oguri} M.,  et~al., 2006, \mn@doi [\aj] {10.1086/506019}, \href
  {http://adsabs.harvard.edu/abs/2006AJ....132..999O} {132, 999}

\bibitem[\protect\citeauthoryear{{Oguri}, {Rusu}  \& {Falco}}{{Oguri}
  et~al.}{2014}]{OguriEtal14}
{Oguri} M.,  {Rusu} C.~E.,   {Falco} E.~E.,  2014, \mn@doi [\mnras]
  {10.1093/mnras/stu106}, \href
  {http://adsabs.harvard.edu/abs/2014MNRAS.439.2494O} {439, 2494}

\bibitem[\protect\citeauthoryear{{Oke} et~al.,}{{Oke} et~al.}{1995}]{OkeEtal95}
{Oke} J.~B.,  et~al., 1995, \pasp, \href
  {http://adsabs.harvard.edu/abs/1995PASP..107..375O} {107, 375}

\bibitem[\protect\citeauthoryear{{Paraficz} \& {Hjorth}}{{Paraficz} \&
  {Hjorth}}{2009}]{ParaficzHjorth09}
{Paraficz} D.,  {Hjorth} J.,  2009, \mn@doi [\aap]
  {10.1051/0004-6361/200913307}, \href
  {http://adsabs.harvard.edu/abs/2009A%26A...507L..49P} {507, L49}

\bibitem[\protect\citeauthoryear{{Pelt}, {Kayser}, {Refsdal}  \&
  {Schramm}}{{Pelt} et~al.}{1996}]{PeltEtal96}
{Pelt} J.,  {Kayser} R.,  {Refsdal} S.,   {Schramm} T.,  1996, \aap, \href
  {http://adsabs.harvard.edu/abs/1996A%26A...305...97P} {305, 97}

\bibitem[\protect\citeauthoryear{{Peng}, {Impey}, {Rix}, {Kochanek}, {Keeton},
  {Falco}, {Leh{\'a}r}  \& {McLeod}}{{Peng} et~al.}{2006}]{PengEtal06}
{Peng} C.~Y.,  {Impey} C.~D.,  {Rix} H.-W.,  {Kochanek} C.~S.,  {Keeton} C.~R.,
   {Falco} E.~E.,  {Leh{\'a}r} J.,   {McLeod} B.~A.,  2006, \mn@doi [\apj]
  {10.1086/506266}, \href {http://adsabs.harvard.edu/abs/2006ApJ...649..616P}
  {649, 616}

\bibitem[\protect\citeauthoryear{{Percival} et~al.,}{{Percival}
  et~al.}{2010}]{PercivalEtal10}
{Percival} W.~J.,  et~al., 2010, \mn@doi [\mnras]
  {10.1111/j.1365-2966.2009.15812.x}, \href
  {http://adsabs.harvard.edu/abs/2010MNRAS.401.2148P} {401, 2148}

\bibitem[\protect\citeauthoryear{{Perlmutter} et~al.,}{{Perlmutter}
  et~al.}{1999}]{PerlmutterEtal99}
{Perlmutter} S.,  et~al., 1999, \mn@doi [\apj] {10.1086/307221}, \href
  {http://adsabs.harvard.edu/abs/1999ApJ...517..565P} {517, 565}

\bibitem[\protect\citeauthoryear{{Pirard} et~al.,}{{Pirard}
  et~al.}{2004}]{PirardEtal04}
{Pirard} J.-F.,  et~al., 2004, in {Moorwood} A.~F.~M.,  {Iye} M.,  eds,
  \procspie Vol. 5492, Ground-based Instrumentation for Astronomy. pp
  1763--1772, \mn@doi{10.1117/12.578293}

\bibitem[\protect\citeauthoryear{{Planck Collaboration} et~al.,}{{Planck
  Collaboration} et~al.}{2013}]{Planck2013P16}
{Planck Collaboration} et~al., 2013, ArXiv e-prints (1303.5076), \href
  {http://adsabs.harvard.edu/abs/2013arXiv1303.5076P} {}

\bibitem[\protect\citeauthoryear{{Planck Collaboration} et~al.,}{{Planck
  Collaboration} et~al.}{2015}]{Planck2015P13}
{Planck Collaboration} et~al., 2015, preprint, \href
  {http://adsabs.harvard.edu/abs/2015arXiv150201589P} {} (\mn@eprint {arXiv}
  {1502.01589})

\bibitem[\protect\citeauthoryear{{Poindexter}, {Morgan}, {Kochanek}  \&
  {Falco}}{{Poindexter} et~al.}{2007}]{PoindexterEtal07}
{Poindexter} S.,  {Morgan} N.,  {Kochanek} C.~S.,   {Falco} E.~E.,  2007,
  \mn@doi [\apj] {10.1086/512773}, \href
  {http://adsabs.harvard.edu/abs/2007ApJ...660..146P} {660, 146}

\bibitem[\protect\citeauthoryear{{Press}, {Rybicki}  \& {Hewitt}}{{Press}
  et~al.}{1992}]{PressEtal92b}
{Press} W.~H.,  {Rybicki} G.~B.,   {Hewitt} J.~N.,  1992, \mn@doi [\apj]
  {10.1086/170951}, \href {http://adsabs.harvard.edu/abs/1992ApJ...385..404P}
  {385, 404}

\bibitem[\protect\citeauthoryear{{Quimby} et~al.,}{{Quimby}
  et~al.}{2014}]{QuimbyEtal14}
{Quimby} R.~M.,  et~al., 2014, \mn@doi [Science] {10.1126/science.1250903},
  \href {http://adsabs.harvard.edu/abs/2014Sci...344..396Q} {344, 396}

\bibitem[\protect\citeauthoryear{{Refsdal}}{{Refsdal}}{1964}]{Refsdal64}
{Refsdal} S.,  1964, \mnras, \href
  {http://adsabs.harvard.edu/cgi-bin/nph-bib_query?bibcode=1964MNRAS.128..307R&db_key=AST}
  {128, 307}

\bibitem[\protect\citeauthoryear{{Reid}, {Braatz}, {Condon}, {Lo}, {Kuo},
  {Impellizzeri}  \& {Henkel}}{{Reid} et~al.}{2013}]{ReidEtal13}
{Reid} M.~J.,  {Braatz} J.~A.,  {Condon} J.~J.,  {Lo} K.~Y.,  {Kuo} C.~Y.,
  {Impellizzeri} C.~M.~V.,   {Henkel} C.,  2013, \mn@doi [\apj]
  {10.1088/0004-637X/767/2/154}, \href
  {http://adsabs.harvard.edu/abs/2013ApJ...767..154R} {767, 154}

\bibitem[\protect\citeauthoryear{{Remy}, {Claeskens}, {Surdej}, {Hjorth},
  {Refsdal}, {Wucknitz}, {S{\o}rensen}  \& {Grundahl}}{{Remy}
  et~al.}{1998}]{RemyEtal98}
{Remy} M.,  {Claeskens} J.-F.,  {Surdej} J.,  {Hjorth} J.,  {Refsdal} S.,
  {Wucknitz} O.,  {S{\o}rensen} A.~N.,   {Grundahl} F.,  1998, \mn@doi [\na]
  {10.1016/S1384-1076(98)00005-0}, \href
  {http://adsabs.harvard.edu/abs/1998NewA....3..379R} {3, 379}

\bibitem[\protect\citeauthoryear{{Riess} et~al.,}{{Riess}
  et~al.}{1998}]{RiessEtal98}
{Riess} A.~G.,  et~al., 1998, \mn@doi [\aj] {10.1086/300499}, \href
  {http://adsabs.harvard.edu/abs/1998AJ....116.1009R} {116, 1009}

\bibitem[\protect\citeauthoryear{{Riess} et~al.,}{{Riess}
  et~al.}{2016}]{RiessEtal16}
{Riess} A.~G.,  et~al., 2016, ArXiv e-prints (1604.01424), \href
  {http://adsabs.harvard.edu/abs/2016arXiv160401424R} {}

\bibitem[\protect\citeauthoryear{{Rockosi} et~al.,}{{Rockosi}
  et~al.}{2010}]{RockosiEtal10}
{Rockosi} C.,  et~al., 2010, in Ground-based and Airborne Instrumentation for
  Astronomy III. p. 77350R, \mn@doi{10.1117/12.856818}

\bibitem[\protect\citeauthoryear{{Romanowsky} \& {Kochanek}}{{Romanowsky} \&
  {Kochanek}}{1999}]{RomanowskyKochanek99}
{Romanowsky} A.~J.,  {Kochanek} C.~S.,  1999, \mn@doi [\apj] {10.1086/307077},
  \href {http://adsabs.harvard.edu/abs/1999ApJ...516...18R} {516, 18}

\bibitem[\protect\citeauthoryear{{Ross}, {Samushia}, {Howlett}, {Percival},
  {Burden}  \& {Manera}}{{Ross} et~al.}{2015}]{RossEtal15}
{Ross} A.~J.,  {Samushia} L.,  {Howlett} C.,  {Percival} W.~J.,  {Burden} A.,
  {Manera} M.,  2015, \mn@doi [\mnras] {10.1093/mnras/stv154}, \href
  {http://adsabs.harvard.edu/abs/2015MNRAS.449..835R} {449, 835}

\bibitem[\protect\citeauthoryear{{Rumbaugh}, {Fassnacht}, {McKean}, {Koopmans},
  {Auger}  \& {Suyu}}{{Rumbaugh} et~al.}{2015}]{RumbaughEtal15}
{Rumbaugh} N.,  {Fassnacht} C.~D.,  {McKean} J.~P.,  {Koopmans} L.~V.~E.,
  {Auger} M.~W.,   {Suyu} S.~H.,  2015, \mn@doi [\mnras]
  {10.1093/mnras/stv672}, \href
  {http://adsabs.harvard.edu/abs/2015MNRAS.450.1042R} {450, 1042}

\bibitem[\protect\citeauthoryear{{Rusu} et~al.,}{{Rusu}
  et~al.}{2016}]{RusuEtal16}
{Rusu} C.~E.,  et~al., 2016, \mn@doi [\mnras] {10.1093/mnras/stw092}, \href
  {http://adsabs.harvard.edu/abs/2016MNRAS.458....2R} {458, 2}

\bibitem[\protect\citeauthoryear{{Rusu} et~al.,}{{Rusu}
  et~al.}{2017}]{RusuEtal16b}
{Rusu} C.~E.,  et~al., 2017, ArXiv e-prints (1607.01047), \href
  {http://adsabs.harvard.edu/abs/2016arXiv160701047R} {}

\bibitem[\protect\citeauthoryear{{Schechter} et~al.,}{{Schechter}
  et~al.}{1997}]{SchechterEtal97}
{Schechter} P.~L.,  et~al., 1997, \mn@doi [\apjl] {10.1086/310478}, \href
  {http://adsabs.harvard.edu/abs/1997ApJ...475L..85S} {475, L85}

\bibitem[\protect\citeauthoryear{{Schneider} \& {Sluse}}{{Schneider} \&
  {Sluse}}{2013}]{SchneiderSluse13}
{Schneider} P.,  {Sluse} D.,  2013, \mn@doi [\aap]
  {10.1051/0004-6361/201321882}, \href
  {http://adsabs.harvard.edu/abs/2013A%26A...559A..37S} {559, A37}

\bibitem[\protect\citeauthoryear{{Schneider} \& {Sluse}}{{Schneider} \&
  {Sluse}}{2014}]{SchneiderSluse14}
{Schneider} P.,  {Sluse} D.,  2014, \mn@doi [\aap]
  {10.1051/0004-6361/201322106}, \href
  {http://adsabs.harvard.edu/abs/2014A%26A...564A.103S} {564, A103}

\bibitem[\protect\citeauthoryear{{Schneider}, {Ehlers}  \& {Falco}}{{Schneider}
  et~al.}{1992}]{SchneiderEtal92}
{Schneider} P.,  {Ehlers} J.,   {Falco} E.~E.,  1992, Gravitational Lenses,
  XIV, 560 pp.~112 figs..~Springer-Verlag Berlin Heidelberg New York., \href
  {http://adsabs.harvard.edu/abs/1992grle.book.....S} {}

\bibitem[\protect\citeauthoryear{{Sheinis}, {Bolte}, {Epps}, {Kibrick},
  {Miller}, {Radovan}, {Bigelow}  \& {Sutin}}{{Sheinis}
  et~al.}{2002}]{SheinisEtal02}
{Sheinis} A.~I.,  {Bolte} M.,  {Epps} H.~W.,  {Kibrick} R.~I.,  {Miller} J.~S.,
   {Radovan} M.~V.,  {Bigelow} B.~C.,   {Sutin} B.~M.,  2002, \pasp, \href
  {http://adsabs.harvard.edu/abs/2002PASP..114..851S} {114, 851}

\bibitem[\protect\citeauthoryear{{Sluse} \& {Tewes}}{{Sluse} \&
  {Tewes}}{2014}]{SluseTewes14}
{Sluse} D.,  {Tewes} M.,  2014, \mn@doi [\aap] {10.1051/0004-6361/201424776},
  \href {http://adsabs.harvard.edu/abs/2014A%26A...571A..60S} {571, A60}

\bibitem[\protect\citeauthoryear{{Sluse} et~al.,}{{Sluse}
  et~al.}{2003}]{SluseEtal03}
{Sluse} D.,  et~al., 2003, \mn@doi [\aap] {10.1051/0004-6361:20030904}, \href
  {http://adsabs.harvard.edu/abs/2003A%26A...406L..43S} {406, L43}

\bibitem[\protect\citeauthoryear{{Sluse}, {Claeskens}, {Hutsem{\'e}kers}  \&
  {Surdej}}{{Sluse} et~al.}{2007}]{SluseEtal07}
{Sluse} D.,  {Claeskens} J.-F.,  {Hutsem{\'e}kers} D.,   {Surdej} J.,  2007,
  \mn@doi [\aap] {10.1051/0004-6361:20066821}, \href
  {http://adsabs.harvard.edu/abs/2007A%26A...468..885S} {468, 885}

\bibitem[\protect\citeauthoryear{{Sluse}, {Hutsem{\'e}kers}, {Courbin},
  {Meylan}  \& {Wambsganss}}{{Sluse} et~al.}{2012}]{SluseEtal12}
{Sluse} D.,  {Hutsem{\'e}kers} D.,  {Courbin} F.,  {Meylan} G.,   {Wambsganss}
  J.,  2012, \mn@doi [\aap] {10.1051/0004-6361/201219125}, \href
  {http://adsabs.harvard.edu/abs/2012A%26A...544A..62S} {544, A62}

\bibitem[\protect\citeauthoryear{{Sluse} et~al.,}{{Sluse}
  et~al.}{2017}]{SluseEtal16}
{Sluse} D.,  et~al., 2017, ArXiv e-prints (1607.00382), \href
  {http://adsabs.harvard.edu/abs/2016arXiv160700382S} {}

\bibitem[\protect\citeauthoryear{{Smette}, {Robertson}, {Shaver}, {Reimers},
  {Wisotzki}  \& {Koehler}}{{Smette} et~al.}{1995}]{SmetteEtal95}
{Smette} A.,  {Robertson} J.~G.,  {Shaver} P.~A.,  {Reimers} D.,  {Wisotzki}
  L.,   {Koehler} T.,  1995, \aaps, \href
  {http://adsabs.harvard.edu/abs/1995A%26AS..113..199S} {113, 199}

\bibitem[\protect\citeauthoryear{{Smith} \& {Lucey}}{{Smith} \&
  {Lucey}}{2013}]{SmithLucey13}
{Smith} R.~J.,  {Lucey} J.~R.,  2013, \mn@doi [\mnras] {10.1093/mnras/stt1141},
  \href {http://adsabs.harvard.edu/abs/2013MNRAS.434.1964S} {434, 1964}

\bibitem[\protect\citeauthoryear{{Sonnenfeld}, {Treu}, {Gavazzi}, {Marshall},
  {Auger}, {Suyu}, {Koopmans}  \& {Bolton}}{{Sonnenfeld}
  et~al.}{2012}]{SonnenfeldEtal12}
{Sonnenfeld} A.,  {Treu} T.,  {Gavazzi} R.,  {Marshall} P.~J.,  {Auger} M.~W.,
  {Suyu} S.~H.,  {Koopmans} L.~V.~E.,   {Bolton} A.~S.,  2012, \mn@doi [\apj]
  {10.1088/0004-637X/752/2/163}, \href
  {http://adsabs.harvard.edu/abs/2012ApJ...752..163S} {752, 163}

\bibitem[\protect\citeauthoryear{{Sonnenfeld}, {Treu}, {Marshall}, {Suyu},
  {Gavazzi}, {Auger}  \& {Nipoti}}{{Sonnenfeld}
  et~al.}{2015}]{SonnenfeldEtal15}
{Sonnenfeld} A.,  {Treu} T.,  {Marshall} P.~J.,  {Suyu} S.~H.,  {Gavazzi} R.,
  {Auger} M.~W.,   {Nipoti} C.,  2015, \mn@doi [\apj]
  {10.1088/0004-637X/800/2/94}, \href
  {http://adsabs.harvard.edu/abs/2015ApJ...800...94S} {800, 94}

\bibitem[\protect\citeauthoryear{{Spiniello}, {Koopmans}, {Trager}, {Czoske}
  \& {Treu}}{{Spiniello} et~al.}{2011}]{SpinielloEtal11}
{Spiniello} C.,  {Koopmans} L.~V.~E.,  {Trager} S.~C.,  {Czoske} O.,   {Treu}
  T.,  2011, \mn@doi [\mnras] {10.1111/j.1365-2966.2011.19458.x}, \href
  {http://adsabs.harvard.edu/abs/2011MNRAS.417.3000S} {417, 3000}

\bibitem[\protect\citeauthoryear{{Spiniello}, {Trager}, {Koopmans}  \&
  {Chen}}{{Spiniello} et~al.}{2012}]{SpinielloEtal12}
{Spiniello} C.,  {Trager} S.~C.,  {Koopmans} L.~V.~E.,   {Chen} Y.~P.,  2012,
  \mn@doi [\apjl] {10.1088/2041-8205/753/2/L32}, \href
  {http://adsabs.harvard.edu/abs/2012ApJ...753L..32S} {753, L32}

\bibitem[\protect\citeauthoryear{{Spiniello}, {Trager}, {Koopmans}  \&
  {Conroy}}{{Spiniello} et~al.}{2014}]{SpinielloEtal14}
{Spiniello} C.,  {Trager} S.,  {Koopmans} L.~V.~E.,   {Conroy} C.,  2014,
  \mn@doi [\mnras] {10.1093/mnras/stt2282}, \href
  {http://adsabs.harvard.edu/abs/2014MNRAS.438.1483S} {438, 1483}

\bibitem[\protect\citeauthoryear{{Spiniello}, {Barnab{\`e}}, {Koopmans}  \&
  {Trager}}{{Spiniello} et~al.}{2015}]{SpinielloEtal15}
{Spiniello} C.,  {Barnab{\`e}} M.,  {Koopmans} L.~V.~E.,   {Trager} S.~C.,
  2015, \mn@doi [\mnras] {10.1093/mnrasl/slv079}, \href
  {http://adsabs.harvard.edu/abs/2015MNRAS.452L..21S} {452, L21}

\bibitem[\protect\citeauthoryear{{Surpi} \& {Blandford}}{{Surpi} \&
  {Blandford}}{2003}]{SurpiBlandford03}
{Surpi} G.,  {Blandford} R.~D.,  2003, \mn@doi [\apj] {10.1086/345592}, \href
  {http://adsabs.harvard.edu/abs/2003ApJ...584..100S} {584, 100}

\bibitem[\protect\citeauthoryear{{Suyu}}{{Suyu}}{2012}]{Suyu12}
{Suyu} S.~H.,  2012, ArXiv e-prints (1202.0287), \href
  {http://adsabs.harvard.edu/abs/2012arXiv1202.0287S} {}

\bibitem[\protect\citeauthoryear{{Suyu} \& {Halkola}}{{Suyu} \&
  {Halkola}}{2010}]{SuyuHalkola10}
{Suyu} S.~H.,  {Halkola} A.,  2010, \mn@doi [\aap]
  {10.1051/0004-6361/201015481}, \href
  {http://adsabs.harvard.edu/abs/2010A%26A...524A..94S} {524, A94}

\bibitem[\protect\citeauthoryear{{Suyu}, {Marshall}, {Blandford}, {Fassnacht},
  {Koopmans}, {McKean}  \& {Treu}}{{Suyu} et~al.}{2009}]{SuyuEtal09}
{Suyu} S.~H.,  {Marshall} P.~J.,  {Blandford} R.~D.,  {Fassnacht} C.~D.,
  {Koopmans} L.~V.~E.,  {McKean} J.~P.,   {Treu} T.,  2009, \mn@doi [\apj]
  {10.1088/0004-637X/691/1/277}, \href
  {http://adsabs.harvard.edu/abs/2009ApJ...691..277S} {691, 277}

\bibitem[\protect\citeauthoryear{{Suyu}, {Marshall}, {Auger}, {Hilbert},
  {Blandford}, {Koopmans}, {Fassnacht}  \& {Treu}}{{Suyu}
  et~al.}{2010}]{SuyuEtal10}
{Suyu} S.~H.,  {Marshall} P.~J.,  {Auger} M.~W.,  {Hilbert} S.,  {Blandford}
  R.~D.,  {Koopmans} L.~V.~E.,  {Fassnacht} C.~D.,   {Treu} T.,  2010, \mn@doi
  [\apj] {10.1088/0004-637X/711/1/201}, \href
  {http://adsabs.harvard.edu/abs/2010ApJ...711..201S} {711, 201}

\bibitem[\protect\citeauthoryear{{Suyu} et~al.,}{{Suyu}
  et~al.}{2012a}]{SuyuEtal12b}
{Suyu} S.~H.,  et~al., 2012a, ArXiv e-prints (1202.4459), \href
  {http://adsabs.harvard.edu/abs/2012arXiv1202.4459S} {}

\bibitem[\protect\citeauthoryear{{Suyu} et~al.,}{{Suyu}
  et~al.}{2012b}]{SuyuEtal12a}
{Suyu} S.~H.,  et~al., 2012b, \mn@doi [\apj] {10.1088/0004-637X/750/1/10},
  \href {http://adsabs.harvard.edu/abs/2012ApJ...750...10S} {750, 10}

\bibitem[\protect\citeauthoryear{{Suyu} et~al.,}{{Suyu}
  et~al.}{2013}]{SuyuEtal13}
{Suyu} S.~H.,  et~al., 2013, \mn@doi [\apj] {10.1088/0004-637X/766/2/70}, \href
  {http://adsabs.harvard.edu/abs/2013ApJ...766...70S} {766, 70}

\bibitem[\protect\citeauthoryear{{Suyu} et~al.,}{{Suyu}
  et~al.}{2014}]{SuyuEtal14}
{Suyu} S.~H.,  et~al., 2014, \mn@doi [\apjl] {10.1088/2041-8205/788/2/L35},
  \href {http://adsabs.harvard.edu/abs/2014ApJ...788L..35S} {788, L35}

\bibitem[\protect\citeauthoryear{{Suzuki} et~al.,}{{Suzuki}
  et~al.}{2008}]{SuzukiEtal08}
{Suzuki} R.,  et~al., 2008, \mn@doi [\pasj] {10.1093/pasj/60.6.1347}, \href
  {http://esoads.eso.org/abs/2008PASJ...60.1347S} {60, 1347}

\bibitem[\protect\citeauthoryear{{Suzuki} et~al.,}{{Suzuki}
  et~al.}{2012}]{SuzukiEtal12}
{Suzuki} N.,  et~al., 2012, \mn@doi [\apj] {10.1088/0004-637X/746/1/85}, \href
  {http://adsabs.harvard.edu/abs/2012ApJ...746...85S} {746, 85}

\bibitem[\protect\citeauthoryear{{Tewes}, {Courbin}  \& {Meylan}}{{Tewes}
  et~al.}{2013a}]{TewesEtal13a}
{Tewes} M.,  {Courbin} F.,   {Meylan} G.,  2013a, \mn@doi [\aap]
  {10.1051/0004-6361/201220123}, \href
  {http://adsabs.harvard.edu/abs/2013A%26A...553A.120T} {553, A120}

\bibitem[\protect\citeauthoryear{{Tewes} et~al.,}{{Tewes}
  et~al.}{2013b}]{TewesEtal13b}
{Tewes} M.,  et~al., 2013b, \mn@doi [\aap] {10.1051/0004-6361/201220352}, \href
  {http://adsabs.harvard.edu/abs/2013A%26A...556A..22T} {556, A22}

\bibitem[\protect\citeauthoryear{{Treu} \& {Koopmans}}{{Treu} \&
  {Koopmans}}{2002}]{TreuKoopmans02}
{Treu} T.,  {Koopmans} L.~V.~E.,  2002, \mn@doi [\mnras]
  {10.1046/j.1365-8711.2002.06107.x}, \href
  {http://adsabs.harvard.edu/abs/2002MNRAS.337L...6T} {337, L6}

\bibitem[\protect\citeauthoryear{{Treu} \& {Koopmans}}{{Treu} \&
  {Koopmans}}{2004}]{TreuKoopmans04}
{Treu} T.,  {Koopmans} L.~V.~E.,  2004, \mn@doi [\apj] {10.1086/422245}, \href
  {http://adsabs.harvard.edu/cgi-bin/nph-bib_query?bibcode=2004ApJ...611..739T&db_key=AST}
  {611, 739}

\bibitem[\protect\citeauthoryear{Treu \& Marshall}{Treu \&
  Marshall}{2016}]{TreuMarshall16}
Treu T.,  Marshall P.~J.,  2016, \mn@doi [The Astronomy and Astrophysics
  Review] {10.1007/s00159-016-0096-8}, 24, 11

\bibitem[\protect\citeauthoryear{{Treu}, {Gavazzi}, {Gorecki}, {Marshall},
  {Koopmans}, {Bolton}, {Moustakas}  \& {Burles}}{{Treu}
  et~al.}{2009}]{TreuEtal09}
{Treu} T.,  {Gavazzi} R.,  {Gorecki} A.,  {Marshall} P.~J.,  {Koopmans}
  L.~V.~E.,  {Bolton} A.~S.,  {Moustakas} L.~A.,   {Burles} S.,  2009, \mn@doi
  [\apj] {10.1088/0004-637X/690/1/670}, \href
  {http://adsabs.harvard.edu/abs/2009ApJ...690..670T} {690, 670}

\bibitem[\protect\citeauthoryear{{Treu}, {Auger}, {Koopmans}, {Gavazzi},
  {Marshall}  \& {Bolton}}{{Treu} et~al.}{2010}]{TreuEtal10}
{Treu} T.,  {Auger} M.~W.,  {Koopmans} L.~V.~E.,  {Gavazzi} R.,  {Marshall}
  P.~J.,   {Bolton} A.~S.,  2010, \mn@doi [\apj]
  {10.1088/0004-637X/709/2/1195}, \href
  {http://adsabs.harvard.edu/abs/2010ApJ...709.1195T} {709, 1195}

\bibitem[\protect\citeauthoryear{{Treu} et~al.,}{{Treu}
  et~al.}{2016}]{TreuEtal16}
{Treu} T.,  et~al., 2016, \mn@doi [\apj] {10.3847/0004-637X/817/1/60}, \href
  {http://adsabs.harvard.edu/abs/2016ApJ...817...60T} {817, 60}

\bibitem[\protect\citeauthoryear{{Unruh}, {Schneider}  \& {Sluse}}{{Unruh}
  et~al.}{2016}]{UnruhEtal16}
{Unruh} S.,  {Schneider} P.,   {Sluse} D.,  2016, preprint, \href
  {http://adsabs.harvard.edu/abs/2016arXiv160604321U} {} (\mn@eprint {arXiv}
  {1606.04321})

\bibitem[\protect\citeauthoryear{{Vanderriest}, {Schneider}, {Herpe},
  {Chevreton}, {Moles}  \& {Wlerick}}{{Vanderriest}
  et~al.}{1989}]{VanderriestEtal89}
{Vanderriest} C.,  {Schneider} J.,  {Herpe} G.,  {Chevreton} M.,  {Moles} M.,
  {Wlerick} G.,  1989, \aap, \href
  {http://adsabs.harvard.edu/abs/1989A%26A...215....1V} {215, 1}

\bibitem[\protect\citeauthoryear{{Vegetti} \& {Koopmans}}{{Vegetti} \&
  {Koopmans}}{2009}]{VegettiKoopmans09}
{Vegetti} S.,  {Koopmans} L.~V.~E.,  2009, \mn@doi [\mnras]
  {10.1111/j.1365-2966.2008.14005.x}, \href
  {http://adsabs.harvard.edu/abs/2009MNRAS.392..945V} {392, 945}

\bibitem[\protect\citeauthoryear{{Vernet} et~al.,}{{Vernet}
  et~al.}{2011}]{VernetEtal11}
{Vernet} J.,  et~al., 2011, \mn@doi [\aap] {10.1051/0004-6361/201117752}, \href
  {http://adsabs.harvard.edu/abs/2011A%26A...536A.105V} {536, A105}

\bibitem[\protect\citeauthoryear{{Vuissoz} et~al.,}{{Vuissoz}
  et~al.}{2007}]{VuissozEtal07}
{Vuissoz} C.,  et~al., 2007, \mn@doi [\aap] {10.1051/0004-6361:20065823}, \href
  {http://adsabs.harvard.edu/abs/2007A%26A...464..845V} {464, 845}

\bibitem[\protect\citeauthoryear{{Vuissoz} et~al.,}{{Vuissoz}
  et~al.}{2008}]{VuissozEtal08}
{Vuissoz} C.,  et~al., 2008, \mn@doi [\aap] {10.1051/0004-6361:200809866},
  \href {http://adsabs.harvard.edu/abs/2008A%26A...488..481V} {488, 481}

\bibitem[\protect\citeauthoryear{{Walsh}, {Carswell}  \& {Weymann}}{{Walsh}
  et~al.}{1979}]{WalshEtal79}
{Walsh} D.,  {Carswell} R.~F.,   {Weymann} R.~J.,  1979, \mn@doi [\nat]
  {10.1038/279381a0}, \href {http://adsabs.harvard.edu/abs/1979Natur.279..381W}
  {279, 381}

\bibitem[\protect\citeauthoryear{{Warren} \& {Dye}}{{Warren} \&
  {Dye}}{2003}]{WarrenDye03}
{Warren} S.~J.,  {Dye} S.,  2003, \mn@doi [\apj] {10.1086/375132}, \href
  {http://adsabs.harvard.edu/cgi-bin/nph-bib_query?bibcode=2003ApJ...590..673W&db_key=AST}
  {590, 673}

\bibitem[\protect\citeauthoryear{{Weinberg}, {Mortonson}, {Eisenstein},
  {Hirata}, {Riess}  \& {Rozo}}{{Weinberg} et~al.}{2013}]{WeinbergEtal13}
{Weinberg} D.~H.,  {Mortonson} M.~J.,  {Eisenstein} D.~J.,  {Hirata} C.,
  {Riess} A.~G.,   {Rozo} E.,  2013, \mn@doi [\physrep]
  {10.1016/j.physrep.2013.05.001}, \href
  {http://adsabs.harvard.edu/abs/2013PhR...530...87W} {530, 87}

\bibitem[\protect\citeauthoryear{{Williams}, {Momcheva}, {Keeton}, {Zabludoff}
  \& {Leh{\'a}r}}{{Williams} et~al.}{2006}]{WilliamsEtal06}
{Williams} K.~A.,  {Momcheva} I.,  {Keeton} C.~R.,  {Zabludoff} A.~I.,
  {Leh{\'a}r} J.,  2006, \mn@doi [\apj] {10.1086/504788}, \href
  {http://adsabs.harvard.edu/abs/2006ApJ...646...85W} {646, 85}

\bibitem[\protect\citeauthoryear{{Wisotzki}, {Koehler}, {Kayser}  \&
  {Reimers}}{{Wisotzki} et~al.}{1993}]{WisotzkiEtal93}
{Wisotzki} L.,  {Koehler} T.,  {Kayser} R.,   {Reimers} D.,  1993, \aap, \href
  {http://adsabs.harvard.edu/abs/1993A%26A...278L..15W} {278, L15}

\bibitem[\protect\citeauthoryear{{Wisotzki}, {Christlieb}, {Bade}, {Beckmann},
  {K{\"o}hler}, {Vanelle}  \& {Reimers}}{{Wisotzki}
  et~al.}{2000}]{WisotzkiEtal00}
{Wisotzki} L.,  {Christlieb} N.,  {Bade} N.,  {Beckmann} V.,  {K{\"o}hler} T.,
  {Vanelle} C.,   {Reimers} D.,  2000, \aap, \href
  {http://adsabs.harvard.edu/abs/2000A%26A...358...77W} {358, 77}

\bibitem[\protect\citeauthoryear{{Wisotzki}, {Schechter}, {Bradt},
  {Heinm{\"u}ller}  \& {Reimers}}{{Wisotzki} et~al.}{2002}]{WisotzkiEtal02}
{Wisotzki} L.,  {Schechter} P.~L.,  {Bradt} H.~V.,  {Heinm{\"u}ller} J.,
  {Reimers} D.,  2002, \mn@doi [\aap] {10.1051/0004-6361:20021213}, \href
  {http://adsabs.harvard.edu/abs/2002A%26A...395...17W} {395, 17}

\bibitem[\protect\citeauthoryear{{Wong} et~al.,}{{Wong}
  et~al.}{2017}]{WongEtal16}
{Wong} K.~C.,  et~al., 2017, MNRAS in press, ArXiv e-prints (1607.01403), \href
  {http://adsabs.harvard.edu/abs/2016arXiv160701403W} {}

\bibitem[\protect\citeauthoryear{{Wucknitz}}{{Wucknitz}}{2002}]{Wucknitz02}
{Wucknitz} O.,  2002, \mn@doi [\mnras] {10.1046/j.1365-8711.2002.05426.x},
  \href {http://adsabs.harvard.edu/abs/2002MNRAS.332..951W} {332, 951}

\bibitem[\protect\citeauthoryear{{Wucknitz}, {Biggs}  \& {Browne}}{{Wucknitz}
  et~al.}{2004}]{WucknitzEtal04}
{Wucknitz} O.,  {Biggs} A.~D.,   {Browne} I.~W.~A.,  2004, \mn@doi [\mnras]
  {10.1111/j.1365-2966.2004.07514.x}, \href
  {http://adsabs.harvard.edu/abs/2004MNRAS.349...14W} {349, 14}

\bibitem[\protect\citeauthoryear{{Xu}, {Sluse}, {Schneider}, {Springel},
  {Vogelsberger}, {Nelson}  \& {Hernquist}}{{Xu} et~al.}{2016}]{XuEtal16}
{Xu} D.,  {Sluse} D.,  {Schneider} P.,  {Springel} V.,  {Vogelsberger} M.,
  {Nelson} D.,   {Hernquist} L.,  2016, \mn@doi [\mnras]
  {10.1093/mnras/stv2708}, \href
  {http://adsabs.harvard.edu/abs/2016MNRAS.456..739X} {456, 739}

\bibitem[\protect\citeauthoryear{{van Dokkum} \& {Conroy}}{{van Dokkum} \&
  {Conroy}}{2010}]{vanDokkumConroy10}
{van Dokkum} P.~G.,  {Conroy} C.,  2010, \mn@doi [\nat] {10.1038/nature09578},
  \href {http://adsabs.harvard.edu/abs/2010Natur.468..940V} {468, 940}

\bibitem[\protect\citeauthoryear{{von der Linden} et~al.,}{{von der Linden}
  et~al.}{2014}]{vonderLindenEtal14}
{von der Linden} A.,  et~al., 2014, \mn@doi [\mnras] {10.1093/mnras/stt1945},
  \href {http://adsabs.harvard.edu/abs/2014MNRAS.439....2V} {439, 2}

\makeatother
\end{thebibliography}
\bibliographystyle{mnras}


\bsp	
\label{lastpage}
\end{document}